\documentclass[sn-basic, linktocpage, pdflatex]{sn-jnl}

\usepackage{hyperref}
\hypersetup{pdfproducer={}, pdfcreator={Me}}
\usepackage{graphicx}%
\usepackage{multirow}%
\usepackage{amsmath,amssymb,amsfonts}%
\usepackage{mathrsfs}%
\usepackage[title]{appendix}%
\usepackage{xcolor}%
\usepackage{textcomp}%
\usepackage{manyfoot}%
\usepackage{booktabs}%
\usepackage{algorithm}%
\usepackage{algorithmicx}%
\usepackage{algpseudocode}%
\usepackage{listings}%
\usepackage{siunitx}%
\usepackage{array, makecell}%
\usepackage{comment}%
\usepackage{tabularx}%
\usepackage{anyfontsize}%
\usepackage{orcidlink}

\usepackage{lineno}

\usepackage{bm}%

\newcommand{\va}{V_{\rm A}}
\newcommand{\vsw}{V_{\rm SW}}
\def\kms{\hbox{km$\;$s$^{-1}$}} 

\setcitestyle{authoryear,open={(},close={)}}

\begin{document}

\title[Switchback Formation]{Magnetic switchback formation: a review of proposed mechanisms}


\author*[1]{\fnm{Peter} \sur{Wyper} \orcidlink{0000-0002-6442-7818}}
\email{peter.f.wyper@durham.ac.uk}
\affil*[1]{\orgdiv{Department of Mathematical Sciences}, \orgname{Durham University}, \orgaddress{\street{Stockton Road}, \city{Durham}, \postcode{DH1 3LE}, \country{UK}}}

\author[2]{\fnm{Jonathan} \sur{Squire} \orcidlink{0000-0001-8479-962X}}
\email{jonathan.squire@otago.ac.nz}
\affil[2]{\orgdiv{Department of Physics}, \orgname{University of Otago}, \orgaddress{\street{PO Box 56}, \city{Dunedin}, \postcode{9016}, \country{NZ}}}

\author[3,4]{\fnm{Etienne} \sur{Pariat} \orcidlink{0000-0002-2900-0608}}
 \email{etienne.pariat@lpp.polytechnique.fr}
\affil[3]{\orgdiv{French-Spanish Laboratory for Astrophysics in Canarias, CNRS, Instituto de Astrofísica de Canarias}, \orgaddress{\postcode{38205}, \city{La Laguna},  \state{Tenerife}, \country{Spain}}}
\affil[4]{\orgdiv{Sorbonne Universit\'e, \'Ecole polytechnique, Institut Polytechnique de Paris, Universit\'e Paris Saclay, Observatoire de Paris-PSL, CNRS, Laboratoire de Physique des Plasmas}, \orgaddress{\postcode{75005}, \city{Paris}, \country{France}}}

\author[5]{\fnm{Oleksiy V.} \sur{Agapitov} \orcidlink{0000-0001-6427-1596}}
\email{agapitov@ssl.berkeley.edu}
\affil[5]{\orgdiv{SSL}, \orgname{University of California Berkeley}, \orgaddress{\street{7 Gauss Way}, \city{Berkeley}, \state{CA}, \postcode{94720}, \country{USA}}}

\author[6]{\fnm{Jim F.} \sur{Drake} \orcidlink{0000-0002-9150-1841}}
\email{drake@umd.edu}
\affil[6]{\orgdiv{Department of Physics}, \orgname{University of Maryland}, \orgaddress{\street{4150 Campus Drive}, \city{College Park}, \state{MD}, \postcode{20742}, \country{USA}}}

\author[7]{\fnm{Norbert} \sur{Magyar} \orcidlink{0000-0001-5731-8173}}
\email{n.magyar@stariongroup.eu}
\affil[7]{\orgdiv{Starion Group S.A.}, \orgaddress{\street{Rue des Etoiles 140}, \city{Libin}, \postcode{6890}, \country{Belgium}}}

\author[8]
{\fnm{William H.} \sur{Matthaeus} \orcidlink{0000-0002-0117-4059}}
\email{whm@udel.edu}
\affil[8]{\orgdiv{Bartol Research Institute and Department of Physics and Astronomy, University of Delaware}, \orgaddress{
\street{104 The Green}, 
\city{Newark}, \state{Delaware} \postcode{19716}, \country{USA}}}

\author[9]{\fnm{Lorenzo} \sur{Matteini} \orcidlink{0000-0002-6276-7771}}
\email{l.matteini@imperial.ac.uk}
\affil[9]{\orgdiv{Physics Department, Imperial College London}, \orgaddress{\street{South Kensington}, \city{London}, \postcode{SW7 2BW}, \country{UK}}}

\author[10]{\fnm{David} \sur{Ruffolo} \orcidlink{0000-0003-3414-9666}}
\email{david.ruf@mahidol.ac.th}
\affil[10]{\orgdiv{Department of Physics, Faculty of Science, Mahidol University}, \orgaddress{ \street{272 Rama VI Rd.}, \city{Bangkok}, \postcode{10400}, \country{Thailand}}}

\author[11]{\fnm{Victor} \sur{Réville} \orcidlink{0000-0002-2916-3837}} 
\email{victor.reville@utoulouse.fr}
\affil[11]{\orgdiv{Institut de Recherche en Astrophysique et Planétologie (IRAP), Université de Toulouse, CNRS, CNES}, \orgaddress{\city{Toulouse}, \country{France}}}

\author[12]{\fnm{Chen} \sur{Shi} \orcidlink{0000-0002-2582-7085}}
\email{chenshi@auburn.edu}
\affil[12]{\orgdiv{Department of Physics}, \orgname{Auburn University}, \orgaddress{\city{Auburn}, \state{Alabama}, \postcode{36849}, \country{USA}}}

\author[13]{\fnm{Munehito} \sur{Shoda} \orcidlink{0000-0002-7136-8190}}
\email{shoda@eps.s.u-tokyo.ac.jp}
\affil[13]{\orgdiv{Department of Earth and Planetary Science, School of Science}, \orgname{The University of Tokyo}, \orgaddress{\street{7-3-1 Hongo, Bunkyo-ku}, \city{Tokyo}, \postcode{1130033}, \country{Japan}}}

\author[14]{\fnm{Marc} \sur{Swisdak} \orcidlink{0000-0002-5435-3544}}
\email{swisdak@umd.edu}
\affil[14]{\orgdiv{IREAP}, \orgname{University of Maryland}, \orgaddress{\street{8279 Paint Branch Drive}, \city{College Park}, \state{MD}, \postcode{20740}, \country{USA}}}

\author[15]{\fnm{Marco} \sur{Velli} \orcidlink{0000-0002-2381-3106}}
\email{mvelli@ucla.edu}
\affil[15]{\orgname{Earth, Planetary and Space Sciences, University of California, Los Angeles,
\postcode{91109},\country{USA}}}

\author[16]{\fnm{Mojtaba} \sur{Akhavan-Tafti} \orcidlink{0000-0003-3721-2114}}
\email{akhavant@umich.edu}
\affil[16]{\orgdiv{Department of Climate and Space Sciences and Engineering}, \orgname{University of Michigan}, \orgaddress{\street{2455 Hayward St.}, \city{Ann Arbor}, \postcode{MI 48109}, \country{USA}}}

\author[17]{\fnm{Bahaeddine} \sur{Gannouni} \orcidlink{0000-0002-1711-1802}}
\email{bgannouni@irap.omp.eu}
\affil[17]{\orgdiv{IRAP, CNRS, Observatoire Midi-Pyrénées, Universit\'e Toulouse III - Paul Sabatier}, \orgaddress{ \street{9 Avenue du Colonel Roche}, \postcode{31400}, \city{Toulouse}, \country{France}}}

\author[18]{\fnm{Roberto} \sur{Lionello} \orcidlink{0000-0001-9231-045X}}
\email{lionel@predsci.com}
\affil[18]{\orgdiv{Predictive Science Inc.}, \orgaddress{\street{9990 Mesa Rim Rd., Suite 170}, \city{San Diego}, \state{CA}, \postcode{92121}, \country{USA}}}

\author[19,20,21]{\fnm{Maria S.} \sur{Madjarska} \orcidlink{0000-0001-9806-2485}}
\email{madjarska@mps.mpg.de}
\affil[19]{Max Planck Institute for Solar System Research, Justus-von-Liebig-Weg 3, 37077, G\"ottingen, Germany}
\affil[20]{Korea Astronomy and Space Science Institute, 34055, Daejeon, Republic of Korea}
\affil[21]{Space Research and Technology Institute, Bulgarian Academy of Sciences, Acad. G. Bonchev Str., Bl. 1, 1113, Sofia, Bulgaria}

\author[22]{\fnm{Mathew J.} \sur{Owens} \orcidlink{0000-0003-2061-2453}}
\email{m.j.owens@reading.ac.uk}
\affil[22]{\orgdiv{University of Reading}, \orgaddress{\street{Earley Gate}, \city{Reading}, \state{Berkshire} \postcode{RG6  6BB}, \country{UK}}}

\author[23]{\fnm{Nour E.} \sur{Raouafi\thanks{Now know as Nour E. Rawafi}} \orcidlink{0000-0003-2409-3742}}
\email{Nour.Rawafi@jhuapl.edu}
\affil[23]{\orgdiv{Johns Hopkins Applied Physics Laboratory}, \orgaddress{\street{11100 Johns Hopkins Road}, \city{Laurel}, \state{MD} \postcode{20723}, \country{USA}}}

\author[24]{\fnm{Alphonse C.} \sur{Sterling} \orcidlink{0000-0003-1281-897X}}
\email{alphonse.sterling@nasa.gov}
\affil[24]{\orgname{NASA Marshall Space Flight Center}, \city{Huntsville}, \state{AL} \postcode{35812}, \country{USA}}

\author[25]{\fnm{Durgesh} \sur{Tripathi} \orcidlink{0000-0003-1689-6254}}
\email{durgesh@iucaa.in}
\affil[25]{\orgdiv{Inter-University Centre for Astronomy and Astrophysics}, \orgaddress{\street{Post Bag 4}, \city{Ganeshkhind}, \state{Pune} \postcode{411007}, \country{India}}}


\abstract{Magnetic switchbacks are large amplitude deflections of the magnetic field within the solar wind. They are Alfv\'{e}nic in character and so are associated with a spike in velocity and a generally small variation in local plasma density. Early orbits of Parker Solar Probe revealed that the solar wind near the Sun is dominated by these structures, and therefore, they may be playing an important role in the energy budget and acceleration of the young solar wind. In this review, we present an overview of different mechanisms that have been proposed for how switchbacks could be formed. We group the mechanisms by whether they predominantly act in the {low solar atmosphere} or within the solar wind (in situ). We focus on mechanisms that can create reversals of the ambient magnetic field direction and, thus, account for the most extreme perturbations. The general consensus is that mechanisms in the lower solar atmosphere do not form such reversals on their own but provide the seed perturbations, flows, or particle beams necessary for in situ mechanisms to create switchbacks within the solar wind. Switchback observations thus likely contain an imprint of the coronal source of the seed perturbation or flow, which is evolved further locally by one of several plausible in situ mechanisms. We discuss the strengths and weaknesses of each mechanism and outline future observational and theoretical tests that could help differentiate between them.}



\maketitle
\tableofcontents
 
\section{Introduction}\label{sec: intro}

Since the discovery of the ubiquitous presence of magnetic switchbacks in the near-Sun solar wind \citep{Bale2019,Kasper19} by Parker Solar Probe \citep[PSP,][]{Fox2016,Raouafi2023}, there have been many mechanisms put forward to explain how they form. The purpose of this article is to provide an overview of the different mechanisms that have gained the most traction throughout the community, providing a critical evaluation of the validity of each mechanism and observed properties with which each mechanism is consistent. To do so, first requires a discussion of which observable features should be reproduced in order for a magnetic switchback generation mechanism to be considered as successful. This is to say, what precisely do we consider to be a switchback?

The term magnetic switchback was first used by \citet{2004JGRA..109.3104Y} to refer to a short-term reversal in sign of the radially directed solar wind magnetic field in Ulysses data \citep[see ][for a historical discussion]{Velli2025}. With the launch of PSP, such reversals were found to be much more frequent nearer the Sun \citep{Bale2021}. Each reversal was also associated with a jet or spike in radial plasma velocity, suggesting that the structures were Alfv\'{e}nic in nature \citep{Kasper19}. The direction of electron strahl and cross helicity measurements revealed that these structures were local folds of the magnetic field lines \citep{Kasper19,McManus20}, ruling out heliospheric current sheet crossings. Schematic depictions quickly began to depict switchbacks as ``S-shaped" bends in field lines \citep[e.g.,][]{Kasper19,Zank2020}, just like a switchback in a trail or road from which their name was derived.

The definition of a switchback has evolved since these early results, with different authors using different definitions and with more properties being added as distinguishing characteristics. A full review is given in \citet{Badman2025}, but we highlight here that many authors have relaxed the constraint that the typically radial magnetic field should reverse sign, i.e. that the deflection angle should exceed $90^\circ$ \citep[e.g.,][]{Horbury20,Zank2020,Laker21}, with deflection angles as low as $25^\circ$ even being considered \citep[e.g.,][]{DudokdeWit2020}. The lower cut-off for what constitutes a switchback above the background of waves and turbulent fluctuations within the solar wind is, therefore, not currently well defined. However, all mechanisms should be able to reproduce the magnetic field reversals that were so striking in those first results from PSP. Therefore, in this article, for the purposes of evaluating different mechanisms, we will adopt the definition that a \emph{switchback} is a perturbation of the solar wind plasma at radial distances as sampled by PSP ($\gtrsim 10$ solar radii) with the following key properties:
\begin{itemize}
\item A deflection of greater than $90^\circ$ away from the background magnetic field direction and back, {such that the magnetic field involves a locally Sunward directed section consistent with electron strahl and cross-helicity measurements.}
\item A high correlation between magnetic field and velocity variations. As a consequence, reversals in the background magnetic field direction ($B_R$ when the Parker Spiral is negligible)
are associated with outward radial bulk velocity enhancements \cite[e.g.,][]{Gosling2009,Matteini2014}. 
\end{itemize}
We emphasize that this is a stricter definition than adopted in other reviews in this collection \citep[e.g.][]{Badman2025}. Throughout this review, magnetic field deflections of less than $90^\circ$ will be referred to as simply \emph{magnetic deflections} or \emph{Alfv\'{e}nic magnetic deflections}, depending on the correlation with plasma velocity. Deflections of greater than $90^\circ$, but without a correlated velocity enhancement, will be referred to as \emph{magnetic inversions} or \emph{reversals}. Further features of switchbacks that are generally observed are that they:
\begin{itemize}
    \item Have an approximately constant magnetic field amplitude $B=|\bm{B}|$, meaning that the magnetic field vector $\bm{B}$ evolves on a sphere throughout the structure. This feature is known as ``spherical polarisation.'' 
    \item Have little variation in proton density within the switchback (i.e. they are approximately incompressible).
    \item Appear as patches in in situ data with multiple deflections grouped together. This produces a ``spikiness" within each patch. 
    \item Are deflected in all directions, but show a slight bias towards the plane of the solar rotation, i.e., towards the $T$ direction in RTN coordinates \citep[e.g.,][]{Laker2022}.
\end{itemize}
The reader should consult \citet[][]{Badman2025} for a complete list of switchback properties, for a complementary view of switchback definition and for the state-of-the-art of switchback observations. An observational overview of potential solar sources is also given in \citet{Tripathi2025}. 

As will be discussed in this article, some mechanisms are better at reproducing the observed properties listed above than other mechanisms. Some mechanisms have also become outdated as new observations have become available. As of the time of writing, a key new observable gained from PSP's recent sampling of solar wind below the Alfv\'{e}n point is that in the sub-Alfv\'{e}nic wind, there are apparently fewer switchbacks as they are defined above \citep{Bandyopadhyay22,PecoraEA22sb,Akhavan-Tafti2024}. For instance, \citet{Bandyopadhyay22} measured Alfv\'{e}nic deflections only up to $\approx 50^\circ$ in periods of locally sub-Alfv\'{e}nic wind from encounters 8 and 9. However, these results remain controversial because the velocity enhancements within switchbacks, which are comparable to the Alfv\'en speed, naturally make them locally super-Alfv\'{e}nic in regions of sub-Alfv\'{e}nic wind. This complicates the identification of the transition from super to sub-Alfv\'{e}nic background wind \citep[see][ for examples and further discussion]{Sioulas2024,Badman2025}. The outcome of this controversy, particularly in the wake of observations from future encounters, will have important implications for our understanding of the type(s) of switchback formation mechanisms that operate in the solar corona.

In the following, we break our discussion  into two  sections, grouping those mechanisms into {those acting predominantly in the  \emph{lower solar atmosphere}}
(Sect.~\ref{sec: ex situ}), and those that operate \emph{in situ} in the solar wind at or near solar radial distances accessible by spacecraft (Sect.~\ref{sec: in situ}). The mechanisms -- particularly those across the different groups -- are certainly not mutually exclusive. Indeed, all of the \emph{in situ} mechanisms rely on features sourced at low altitudes, such as fluctuations, waves,  or shear flows between different streams. Likewise, it is plausible that multiple different low-coronal or \emph{in situ} mechanisms could be generating switchbacks simultaneously. With this in mind, the most likely possibility seems to be that one or more of the low-coronal mechanisms pairs with one or more of the \emph{in situ} mechanisms to generate switchbacks.

In the following, each subsection of Sect.~\ref{sec: ex situ} or Sect.~\ref{sec: in situ} is used to describe a mechanism that has been previously proposed and studied in the literature. For each, we provide an overview and explanation of the mechanism, a list of physical conditions required in order for it to operate, then a description of its observable signatures (i.e., predictions or post-dictions), and potential advantages and limitations in describing current observations. We conclude in Sect.~\ref{sec: conclusions} with an overview, as well as some suggestions for future observational and theoretical studies that may be used to distinguish between ideas.

\subsection*{Notation}
Some common notations used throughout the article are as follows: The magnetic field is $\bm{B}$ with magnitude $B=|\bm{B}|$, the plasma's bulk velocity is $\bm{u}$,  its mass density is $\rho$, and its thermal pressure is $P$. In the CGS-Gaussian units, the ratio of thermal to magnetic pressure is $\beta=8\pi P/B^2$. The background solar wind magnetic field is $\bm{B}_0$, and the background solar wind speed is $\vsw$.  Where appropriate, fluctuations are denoted with $\delta \bm{u}$ and $\delta \bm{B}$ (i.e., $\delta \bm{B} = \bm{B}-\bm{B}_0$). The Alfv\'en speed is defined via the background field as $\va=B_0/\sqrt{4\pi \rho}$. $R$ is the heliocentric radius and  $R_\odot$ is the solar radius, such that the solar surface is at $R=R_\odot$. The local Alfvén radius, i.e. the radius within the considered open flux tube or region where $\vsw(R_{\rm A})=\va(R_{\rm A})$, is denoted  $R_{\rm A}$. The switchback component of the magnetic field is denoted as $B_R$ (or $B_r$) (for the radial component) or $\delta B_\|=\delta \bm{B}\cdot \bm{B}_0/|\bm{B}_0|$. Finally, the deflection angle relative to the unperturbed background magnetic field direction is given by $\theta_B= \cos^{-1} (\bm{B}\cdot \bm{B}_0/(B B_0))$, which reduces to $\theta_{BR}= \cos^{-1} (B_R/B_0)$ when the ambient field is purely radial and the perturbation is spherically polarised ($B=B_0$).

\section{Mechanisms Operating in the Lower Solar Atmosphere}
\label{sec: ex situ}

\subsection{Direct Injection from Convective Motions}
\label{sec: swirls}

\paragraph{Overview of the Mechanism}
Convective motions in the photosphere directly impart waves and flows into the chromosphere, which itself is highly dynamic and abounds with complex magnetic structure and dynamics \citep[e.g.,][]{2014A&ARv..22...78W,Tziotziou23}. In this environment, reversals in the radial magnetic field component are expected to form, which could potentially then maintain their field reversal and be propagated directly into the corona and later into the solar wind as switchbacks.\\

\begin{figure*}
\centering
\includegraphics[width=0.9\textwidth]{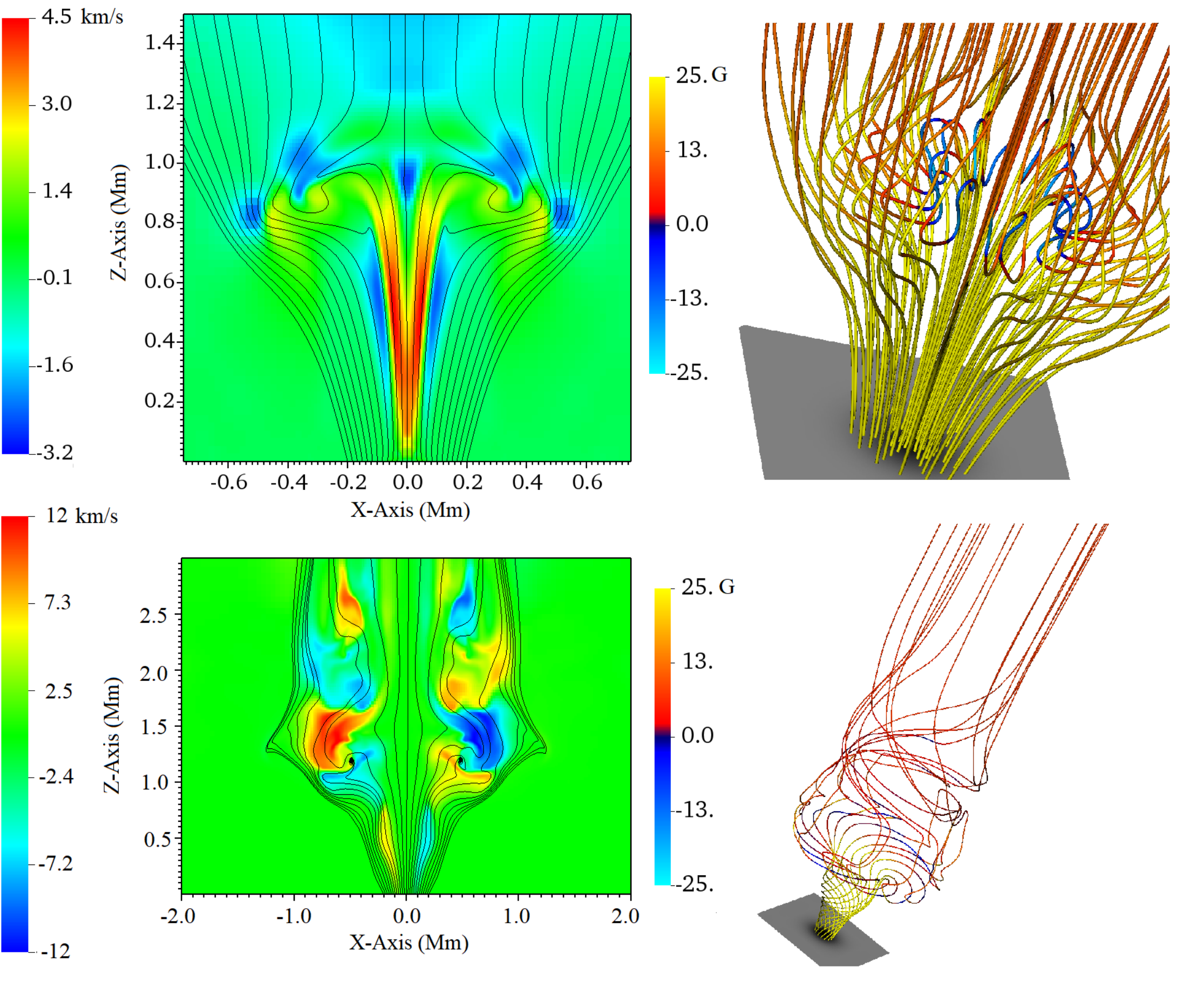}
\caption{Snapshots showing the dynamics in a lower solar atmospheric flux tube resulting from strong upflows at its base (top row) and vortex flows (bottom row). Both rows show a slice through the 3D simulation domain (on the left) depicting the magnetic field lines and vertical (top) or LOS (bottom) velocity, respectively, in \kms, as well as a 3D rendering of magnetic field lines traced from the base of the flux tube (on the right), with field lines colored according to the vertical (Z-Axis) magnetic field strength, in Gauss. Figure reproduced with permission from \citet{2021ApJ...911...75M}, copyright by AAS} 
\label{Fig_Norbert1}
\end{figure*}

\paragraph{Explanation \& Previous Work}
Phenomena of particular interest for magnetic fold generation in the lower solar atmosphere are vortex flows \citep[see the review of ][]{Tziotziou23} and strong (i.e., $\gtrapprox$3~\kms) upflows \citep[e.g.,][]{2019SoPh..294...18M}, in addition to magnetic reconnection \citep[e.g.,][]{2019ARA&A..57..189C}. Such flow dynamics can lead to strong nonlinearities via local wave generation and steepening or instabilities, processes that are thought to be relevant for switchback generation \citep{2023SSRv..219....8R}. The generation of switchbacks by vortex flows and upflows in the lower solar atmosphere was investigated via 3D MHD numerical simulations by \citet{2021ApJ...911...75M}. They show the development of strong localized magnetic folds in the lower solar atmosphere as a result of such flows. Applied vortical flows induce large-amplitude torsional Alfvénic waves, which steepen in their upward propagation, somewhat in analogy with the expansion-driven steepening of Alfvén waves in situ in the solar wind (see Sect.~\ref{sec: in situ} and in particular Sect.~\ref{sec: expanding aws}). On the other hand, applied strong upflows lead to roll-ups at the edge of the flows, reminiscent of Rayleigh-Taylor or Kelvin-Helmholtz type instabilities. The switchbacks resulting from these two formation mechanisms and the underlying magnetic field lines are depicted in Fig.~\ref{Fig_Norbert1}. 

\begin{figure*}
\centering
\includegraphics[width=\textwidth]{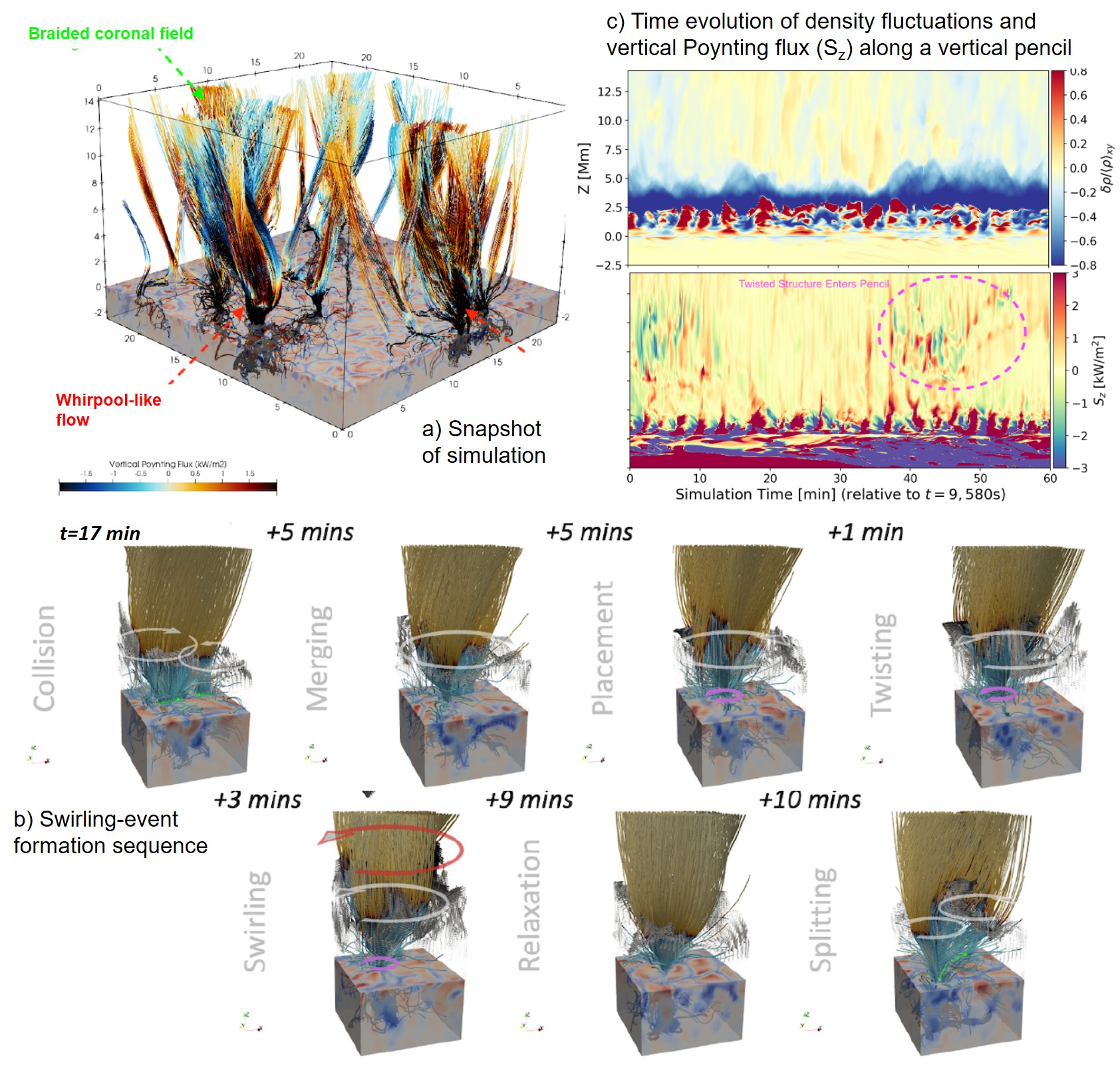}
\caption{Twisting and formation of torsional Alfvén waves by photospheric flows. (a) Snapshot of the BIFROST simulation domain, highlighting the formation of braided coronal structures (with field lines color coded by the vertical Poynting flux, $S_z$) thanks to shuffling of photospheric magnetic concentration. (b) Sequence of snapshots illustrating the merging of two magnetic concentrations (lasting $\sim 30$ minutes) inducing the formation of an impulsive swirling event (of $\sim 10$ min duration). (c) Time-distance diagrams of the relative density fluctuations and of $S_z$ along a vertical pencil, illustrating the passage of vertically propagating torsional wave pulse. Figure reproduced with permission from \citet{Finley22}, copyright by ESO} 
\label{fig:Finley}
\end{figure*}

In a complementary approach, \citet{Finley22} carried out a comprehensive 3D simulation of a patch ($24\times24 \, \text{Mm}^2$) of a coronal hole, void of significant large-scale magnetic bipoles, in which the low atmosphere and the corona are driven by the subphotospheric convective motions. In this simulation, \citet{Finley22} found that the shuffling of small-scale magnetic concentrations leads to the braiding of the coronal magnetic field (see Fig.~\ref{fig:Finley}). The magnetic concentrations, which expand with height to form the magnetic funnel network, are regularly buffeted by the convective motions, inducing the launch of torsional Alfvén waves up and along the magnetic funnels. The convective motions often drive the magnetic concentrations into complex intersections between multiple granules and create whirlpool-like downflows. Whirlpool-like photospheric flows twist the magnetic field too quickly for it to remain in equilibrium with its surroundings. This causes the field to dramatically rotate/swirl, forming impulsive swirling events (as illustrated in panel (b) of Fig.~\ref{fig:Finley}, see also e.g. \cite{Iijima2017}). These swirling events are associated with significant vertical Poynting flux, which is favorably transmitted toward the upper corona. While this mechanism does not directly form magnetic field inversions, it induces magnetic deflections, which can propagate toward the solar wind and serve as seeds for in situ mechanisms (see Sect.~ \ref{sec: in situ}).

\paragraph{Requirements}
In the simulations conducted by \citet{2021ApJ...911...75M}, vortices, swirls, or strong upflows in the vicinity of intense flux tubes (e.g., photospheric bright points) are required for the generation of magnetic reversals or switchbacks within the chromosphere. The amplitude of these inducing flows is generally larger (3--6~\kms) \citep{2019SoPh..294...18M,2020ApJ...898..137S} than root-mean-square amplitudes typically encountered in the photosphere (around 1~\kms); thus the flows assumed in these simulations would be sporadic events. However, for the torsional wave generation and braiding demonstrated by \citet{Finley22} gentler chromospheric swirling flows as are produced naturally by convective motions are sufficient. Simulations of magnetic reconnection in the chromosphere as a result of convective motions \citep[e.g.,][]{Hansteen2019,Peter2019} show that reconnection in this region also generates further strong flows and waves, which may also form chromospheric magnetic reversals as these flows and waves evolve. If we suppose that the switchbacks observed in situ in the solar wind could originate in the chromosphere, it would imply, firstly, that these crossed into the corona, and secondly, that they could propagate or be advected into the solar wind. The numerical simulations of magnetic reversal generation in the chromosphere also show that these reversals are unable to enter the coronal part of the simulation or that they enter as waves that present minimal magnetic field deflections of a few degrees (discussed further below). Notwithstanding this apparent inability, once magnetic reversals are in the low corona, they would then have to propagate out into the solar wind without much deformation in order to be considered of low-atmospheric origin. \\

\paragraph{Observable Signatures}
Magnetic reversals generated in the chromosphere through nonlinearities and instabilities, as modeled by \citet{2021ApJ...911...75M}, are in a highly dynamic state initially and do not possess the characteristic properties of switchbacks observed in the solar wind, e.g., nearly constant magnetic field magnitude, high Alfvénicity, small density perturbations. However, this would not be at odds with observations since these primordial switchbacks would undergo significant changes in their propagation. It is known that perturbations or waves with variations in their total magnetic field strength ($B \neq$\,const.) evolve towards the more nonlinearly stable constant magnetic field configuration \citep[e.g.,][see also Sect.~\ref{sec: in situ}]{1998JGR...103..335V,2005ESASP.592..785L,Shoda2021}.
\par 
Given that the simulated chromospheric magnetic reversals present strong density perturbations and Doppler shifts, these should be observable through remote sensing, i.e., spectroscopic or imaging observations in various relevant lines of chromospheric emission, if large enough to be spatially resolved by the existing instrumentation. Although many types of transient events are observed through chromospheric lines \citep[e.g.,][]{2012RSPTA.370.3129R,2012SSRv..169..181T}, it is unclear whether any of these could be related to the generation of switchbacks.

\paragraph{Advantages and Limitations}
One of the advantages of switchback generation via flows and waves in the chromosphere generated by convective motions is that the convective motions required are ever-present and thus could represent a continuous source of switchbacks. However, a deal-breaking limitation for this generation mechanism is the inability of such strong magnetic folds to enter the corona, as shown by \citet{2021ApJ...911...75M}. There are two principal reasons why these magnetic reversals struggle to enter the corona. On one hand, if the reversals were generated by fluid instabilities, e.g., Rayleigh-Taylor or Kelvin-Helmholtz, they would be localized within the chromosphere once formed. An additional process would be required to advect or transfer these magnetic structures across the transition region and into the corona, which would be challenging given the orders of magnitude difference between chromospheric and coronal densities and temperatures. On the other hand, if the switchbacks are steepened, large-amplitude torsional or transverse waves, these can partially propagate through the transition region and into the corona but suffer strong reflections \citep{2022Physi...4.1050C}. Moreover, as a result of the large jump in Alfvén speed in the corona, even waves with strong deflections become long-wavelength waves with weak deflections (visible in the bottom-right of Fig.~\ref{Fig_Norbert1}).

Nevertheless, let us suppose that a chromospheric-origin switchback or magnetic reversal enters the corona and continues propagating away from the Sun. \citet{Tenerani2020} studied this problem in a uniform medium and showed that constant magnetic field propagating Alfv\'en wave switchbacks could survive long enough to be measured within the solar wind with their lifetime being dependent upon the efficiency of the parametric decay instability. However, \citet{2021ApJ...914....8M} have repeated these simulations, including density stratification and magnetic field expansion, showing a much higher rate of deformation of propagating switchbacks from other effects not present in a homogeneous background plasma. Therefore, it is also questionable whether magnetic reversals of chromospheric origin would be able to enter the corona and whether these would survive out into the solar wind. 

However, the Alfvén waves with lower amplitude deflections that are launched as these structures enter the corona could potentially be later amplified by mechanisms acting within the solar wind (Sect.~\ref{sec: in situ}), embedding the spatial structure of the bright points from which they originated. Similarly, as discussed by \citet{Finley22}, swirling events can be induced through the interaction of magnetic field concentrations that can generate magnetic deflections, which can propagate in the upper corona and later in the solar wind. Future observations will help determine whether the statistical properties (e.g. size, frequency, distribution) of the torsional waves induced by the swirling events are consistent with the in situ properties of switchbacks.

\subsection{Open-field Reconnection and Quasi-2D Turbulence}
\label{sec:turbulence}

\paragraph{Overview of the Mechanism}
As mentioned in the previous section, buffeting motions in the photosphere launch a near-continual flux of transverse fluctuations into the chromosphere, a fraction of which propagate through the transition region into the low-density corona. Here, the interaction of these fluctuations can lead to non-WKB
reflections, turbulence, and subsequent coronal heating \citep[e.g.,][]{MattEA99-ch}. {Alternatively, it has also been suggested that turbulence originating within the magnetic carpet may also be advected through the chromosphere and into the corona \citep[e.g.][]{Zank2021}.} Notably, in the open field of coronal holes, {either} scenario also leads to component reconnection between adjacent open flux tubes \citep[e.g.][]{RappazzoEA12}. This kind of component reconnection is a basic feature of quasi-two-dimensional MHD-scale turbulence, as described, e.g., by 
\citet{ServidioEA10-recon}. Such turbulence will increase in amplitude in the expanding corona \citep{ChhiberEA19-2} and, therefore, could contribute to switchback formation.

\begin{figure*}
\centering
\includegraphics[width=1.0\textwidth]{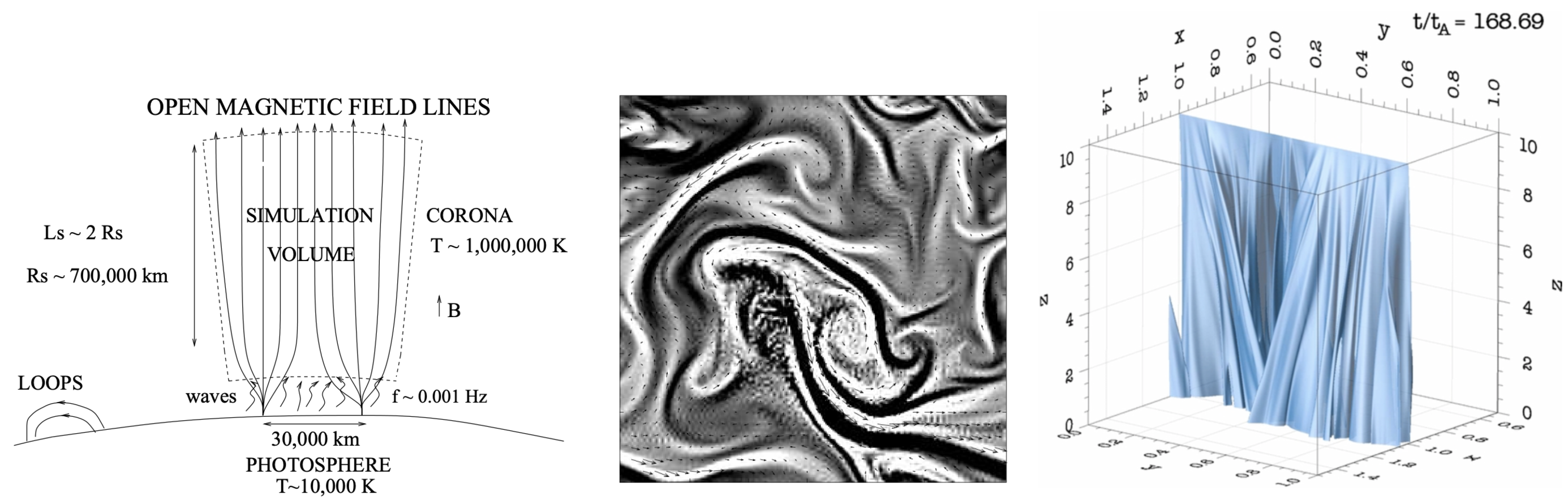}
\caption{Left and middle: schematic of the simulation setup and turbulent current density and flows produced in the open field as a result of lower boundary driving, \citet{Dmitruk2002}. Right: iso-surface showing the complex shape taken by the open-closed boundary following turbulent component reconnection in the closed field region \citep{RappazzoEA12}. Both figures are reproduced with permission from the authors, copyright by AAS} 
\label{fig:turb}
\end{figure*}

\paragraph{Explanation and Previous Work}
In the solar photosphere, the ``magnetic carpet'' consists of a vast number of bipolar regions, which are mostly smaller loops of magnetic field with one footpoint having radial-outward field direction and the other radial-inward \citep{SchrijverEA98-carpet}. Closely following the discovery of this important magnetic feature of the deep solar atmosphere, it was pointed out
\citep{PriestEA02} that there are several ways in which the magnetic carpet can contribute to heating the overlying coronal plasma and, in coronal holes, potentially structuring and accelerating the solar wind. These include several types of reconnection, such as null point reconnection (see Sect.~\ref{sec:interchange}), component reconnection at separators and quasi-separatrix layers, and component reconnection associated with magnetic braiding from footpoint motions. Other heating mechanisms include wave generation and dissipation and the formation of dissipative current sheets between flux tubes.
It is noteworthy that coronal holes cannot be distinguished from quiet-Sun regions at the height level of the magnetic carpet, mainly because the number of closed loops overwhelms the open flux regions. Coronal holes become distinctively clear with increasing altitude.

Plasma jets and wave fluctuations are ubiquitous in the chromosphere, both from reconnection events in the magnetic carpet and from buffeting photospheric motions. While a significant proportion is reflected at the transition region, a fraction escapes into the low corona, e.g. \cite{Morton2025}. In the closed field of the quiet Sun, simulations suggest that the induced component reconnection events are widely distributed throughout large-scale coronal loops and are thought to be associated with, for example, nanoflare events 
\citep[e.g.,][]{EinaudiVelli94,DmitrukGomez97,RappazzoEA08,Klimchuk2015,Viall2017}. In the open field of coronal holes, reflections of the waves caused by density and/or wave speed gradients in the open flux tubes produce the necessary counter-propagating waves required to excite the same anisotropic turbulence, leading to a vast number of current sheets and component reconnection events. This has been shown to lead to heating at the right altitudes that may be sufficient to power the fast solar wind \citep{LeamonEA00,MattEA99-ch,Dmitruk2002}. Figure \ref{fig:turb} (left and middle panels) shows an example of plasma flows and multiple current sheets based on this mechanism \citep{Dmitruk2002}. It should be noted that the plasma jets are mainly transverse to radial, given that the magnetic field is principally radial in coronal holes, while the current sheets are stretched out along the radial direction. The fluctuations induced by this component reconnection are, therefore, predominantly transverse to the radial direction and so do not produce reversals of the radial magnetic field component (switchbacks). 

{An alternative scenario for the origin of turbulence and heating of the open field and corona has been proposed by \citet{Zank2018,Zank2021}. They argue that turbulence originating within the magnetic carpet is not susceptible to the same reflection issues as pure waves, and can be advected through the chromosphere, across the transition region and into the corona. In this scenario, the turbulence is injected into the corona directly rather than forming dynamically within it, but is similarly quasi-2D in nature, with relatively small magnetic field deflections.}

{Regardless of its origins, it is expected that in the low corona a reservoir of quasi-2D turbulent fluctuations are present that could}
subsequently grow in amplitude via {in situ} effects (see \S \ref{sec: in situ}). Furthermore, the turbulent fluctuations are expected to be structured on the supergranular scale as a result of the flux tubes from which the perturbations originate (e.g., Fig. \ref{fig:turb}; left panel).

Finally, it is worth noting that component reconnection is not limited to purely open or closed field. \citet{RappazzoEA12} demonstrated in a simulation study that such a quasi-2D turbulent evolution occurring in the closed field adjacent to an open-closed separatrix boundary can induce reconnection across the boundary, i.e., interchange reconnection, leading to a highly corrugated separatrix boundary surface (Fig. \ref{fig:turb}; right panel). Due to the driving scales involved, such interchange reconnection is likely prevalent on the flanks of helmet streamers and pseudostreamers. Still, it is perhaps less likely on small-scale open-closed boundaries such as those associated with coronal jets, coronal bright points, and jetlets, which are the topics of the next sections and in which reconnection around the null point typically dominates. This purely component interchange may, however, contribute to the slow solar wind in the context of pseudo and helmet streamers, but again, as the fluctuations are mainly transverse to the magnetic field, switchbacks are not produced directly by this process. \\

\paragraph{Requirements}
The key requirements for this mechanism are
\begin{itemize}
\item Alv\'{e}nic fluctuations {or quasi-2D turbulence} entering open flux tubes in the low corona from below.
\item Gradients in the wave speed along the flux tubes leading to wave reflections that set up counter-propagating waves and turbulence.
\item To fully develop into switchbacks a secondary mechanism, {see Sect.~\ref{sec: in situ},}
is then required to amplify the radial magnetic field perturbations to the point that switchbacks form.
\end{itemize}

\paragraph{Observational Signatures}
A key expectation of the quasi-2D turbulence occurring in the low corona is the appearance after propagation of turbulent fluctuations further out in the solar wind directly measurable by PSP. In this respect, there is ample evidence of turbulence in the solar wind data \citep[e.g.,][]{Chen2020}. Furthermore, in the scenario that the turbulence is driven by wave reflections, an expectation is the presence of counter-propagating waves. This is again seen in data from PSP; for example, \citet{Chen2020} showed that the turbulence in the first two PSP encounters was highly Alfv\'{e}nic, and although dominated by outward propagating waves, inward propagating waves were also present. {However, \citet{Zank2021} in their review of PSP data interpret e.g. cross helicity measurements as being more consistent with advected quasi-2D turbulence. Finally,} \citet{Shi2021} also noted in certain encounters that, closest to the Sun, the magnetic energy of the fluctuations at times was greater than their kinetic energy, which they took to imply that the fluctuations had a magnetic origin in the low corona. 

Regarding switchbacks, the deflections of this mechanism are modest and clearly not enough to suggest that quasi-2D turbulence and component reconnection form switchbacks in the low corona. An expectation of any mechanisms that act to amplify these fluctuations is that they must, therefore, grow with distance from the Sun. Finally, as highlighted in the previous section, the dynamics and amplification of waves and turbulence in the low corona are expected to be closely correlated with the local expansion of flux tubes in this region. Regions of rapid expansion associated with supergranular scales should have higher amplitude fluctuations, which ultimately could lead to higher amplitude switchbacks at Parker Solar Probe \citep[e.g.,][]{Bale2021,Fargette2021}.\\

\paragraph{Advantages and Limitations}
The main advantages of this mechanism are that wave perturbations are plentiful in the lower corona, and wave speeds in the open field naturally drop with height due to expansion. Thus, the presence of quasi-2D turbulence and component reconnection in the low corona is highly likely. Indeed, it is one of the key fundamental mechanisms by which the solar wind is thought to be heated and accelerated. Furthermore, the perturbations required (transverse fluctuations) are quite general and could equally act on waves produced by surface motions such as those in Sect.~\ref{sec: swirls} but also on waves launched by multiple small-scale reconnection events as will be described in Sects.~\ref{sec:interchange} and \ref{sec:untwisting}. A further advantage is that the turbulence is potentially structured by the over-expansion of flux tubes on the supergranular scale, which may provide an explanation for the typical sizes of switchback patches \citep[e.g.,][]{Bale2021}. 

However, one major limitation is that this mechanism does not produce reversals of the radial magnetic field directly. Rather, it must rely upon some other mechanism to amplify the fluctuations during propagation. Such mechanisms will be discussed in Sect.~\ref{sec: in situ}.

\subsection{Interchange Reconnection} 
\label{sec:interchange}

\paragraph{Overview of the Mechanism}
Interchange reconnection refers to the reconnection between open and closed magnetic field lines in the solar corona. As such, interchange reconnection can occur at any open-closed separatrix boundary, including the flanks of helmet streamers and pseudostreamers, as well as at much smaller closed-field regions dotted around within coronal holes. Due to the prevalence of switchbacks in wind originating well away from large-scale coronal hole boundaries,
we focus the majority of our discussion on these smaller-scale regions dominated by null point reconnection. In that context, one conceptual idea that has been proposed is that interchange reconnection creates switchbacks {directly in some form}, through {for example} the formation and ejection of highly kinked U-shaped field lines \citep[``U-loops"; e.g.,][]{Fisk2020}, {a distribution of deflections potentially up to and including U-loops \citep{Zank2020}} or {as} magnetic islands, formed within the current layer \citep{Drake2021}. Another idea is that the perturbations caused by interchange reconnection could become seeds for in situ mechanisms within the solar wind to take over \citep[e.g.,][see also Sect. \ref{sec: in situ} for a discussion of the various mechanisms]{Wyper2022}. Both ideas are discussed below. We focus here on structures launched into the solar wind by the reconnection process itself, predominantly in the context of continually driven interchange reconnection. A discussion of the impulsive release of twist that is previously stored away from the interchange reconnection site, in the form of coronal jets and jetlets, is deferred to Sect.~\ref{sec:untwisting}.\\

\begin{figure*}
\centering
\includegraphics[width=1.0\textwidth]{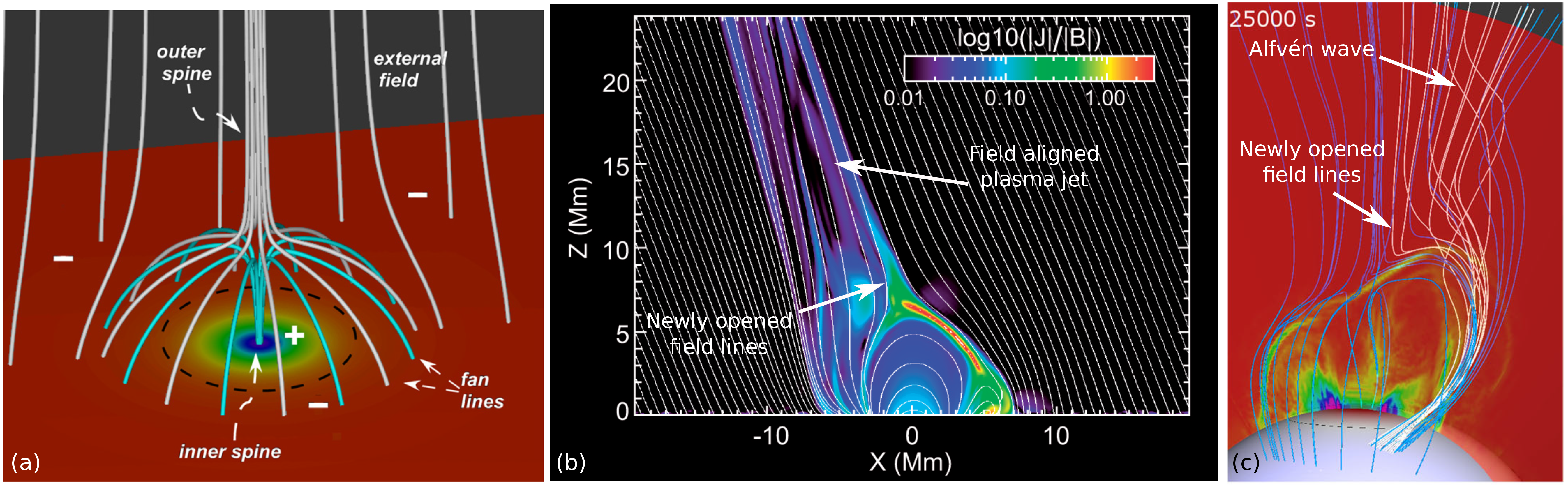}
\caption{
(a) The magnetic topology of a 3D coronal magnetic null point \citep{Pariat2009} reproduced with permission of the AAS. (b) Newly opened white field lines and field-aligned plasma jet in the flux emergence simulation of \citet{MorenoInsertis2008} adapted and reproduced with permission of the AAS. (c) Field line kinks propagating Alfv\'{e}nically along the open field  \citep[reproduced with permission from][copyright by Springer]{Lynch2014} } 
\label{fig:int_fls}
\end{figure*}

\begin{figure*}
\centering
\includegraphics[width=0.9\textwidth]{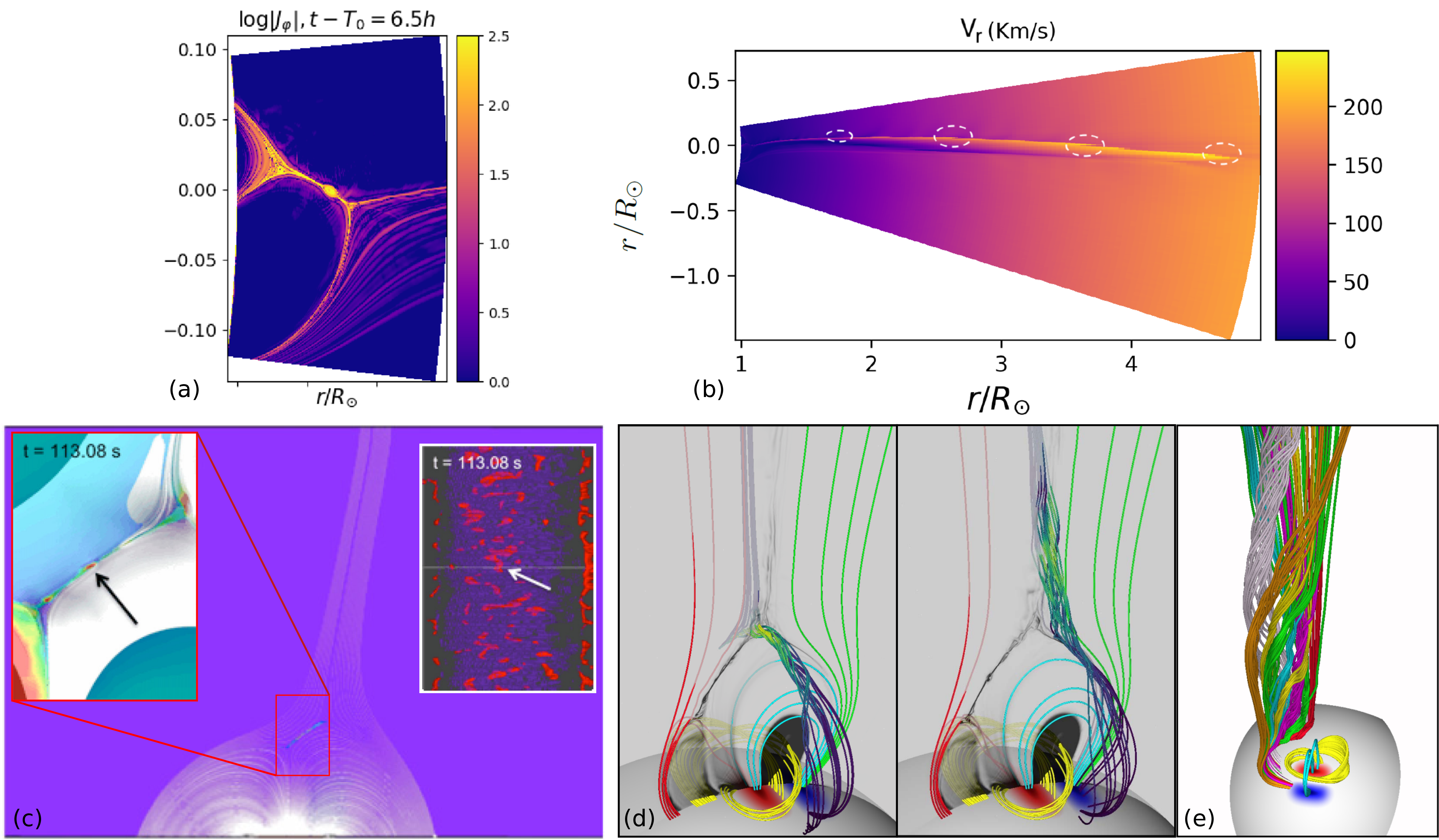}
\caption{(a) Plasmoid in the 2D flux emergence simulation of \cite{Gannouni2023}. (b) Quasi-periodic modulations to the jet/microstream are produced. (c) 3D plasmoid flux ropes and associated enhanced density in the interchange reconnection simulation of \citet{Edmondson2010} reproduced with permission of the AAS. (d) A single plasmoid flux rope ejection launches torsional Alfv\'{e}nic waves in the simulation of \citet{Wyper2022}. (f) The combined effect of multiple plasmoid flux rope ejections.  Note that the twist in each is of the same handedness.  All figures are reproduced with permission from the authors, copyright by AAS} 
\label{fig:ropes}
\end{figure*}

\paragraph{Explanation \& Previous Work}
In this scenario, a closed-field region surrounded by an open field is assumed to be either pre-existing or formed by flux emergence. The magnetic structure typically takes the form of a 3D magnetic null point with a separatrix surface (known as the fan plane) that divides the open from the closed magnetic field, Fig. \ref{fig:int_fls}(a) \citep[e.g.,][]{Pariat2009}. These magnetic structures are under continual stress as magnetic flux on the solar surface emerges, cancels, and moves around. Furthermore, changes in the inclination of the open field \citep[e.g.,][]{Fisk2020}, erupting structures in the closed field \citep[e.g.,][]{Sterling2015,Nobrega-Siverio2023} and even p-mode driving \citep{Kumar2022} could in theory also apply stress to the null point system. Regardless of how it is created, this stress focuses around the null point, forming a current sheet (for a review of current sheet formation at null points, see e.g., \citealt{Priest2009}). Within this current sheet, interchange reconnection is induced, which produces highly kinked field lines and plasmoid flux ropes that could potentially produce switchbacks. We will use the term ``plasmoid flux ropes" to refer to the small-scale flux ropes formed within reconnection layers and to differentiate them from the large-scale flux ropes associated with eruptions. Given the significant interest in interchange reconnection and the varying levels of detail used when invoking it, we will sequentially build up the picture of the interchange reconnection process and its imprint on the solar wind. We will start with 2D laminar reconnection and build up to 3D, plasmoid-dominated interchange reconnection.

Some of the earliest simulation work on interchange reconnection was in the context of flux emergence \citep{Shibata1994,Yokoyama1995,Yokoyama1996}. These ground-breaking simulations showed that as flux emerges into the open field of a coronal hole, a current layer forms at the boundary between the open and closed field. The onset of interchange reconnection in this current layer then forms the null point topology. This early work was mostly in 2D or 2.5D, but later \citet{MorenoInsertis2008} extended this concept to a localised 3D emergence, finding similar results in the early stages of the emergence. Figure \ref{fig:int_fls}(b) shows the current sheet and field lines in this simulation in a plane of symmetry where the reconnection is essentially anti-parallel with a smooth, laminar current layer. In all these studies, U-shaped kinked field lines form in the current layer.
However, it is universally observed that the U-shape \emph{does not survive} the ejection from the current layer. Newly ejected field lines {crash} into older ones in the outflow region, and the reversal in the radial field component is destroyed. However, what is produced is a hot, field-aligned plasma jet launched out along the open field from one end of the current layer. Such jets have been hypothesised as the source of microstreams or even the solar wind itself \citep[e.g.,][discussed further below]{Neugebauer12,Bale2023,Raouafi2023}.

Another strand of work on interchange reconnection has been motivated by understanding the source of the slow solar wind. The focus of these works has been on large-scale null point topologies relevant to pseudostreamers \citep[][]{Tripathi2025}. These works typically involved the injection of shear into the closed field region as an energisation mechanism, but which is also released during the interchange process. For example, \citet{Edmondson2010b} considered a 3D null point topology where interchange reconnection was induced by broad rotational surface motions. The closed-field shear these motions create is imparted to the opening field lines during the 3D interchange process, which then propagates away along open field lines as an Alfv\'{e}n wave \citep{Lynch2014}. This wave behaviour is driven by the difference in the shear component between the open and closed field regions \citep[e.g.,][]{Shibata86,Karpen1998} and was not present in the previously mentioned 2D or 2.5D simulations with uniform guide fields. In general, the bigger the difference in shear component, the more non-linear the wave and, typically, the more explosive the reconnection. In these slow-wind motivated simulations, the difference, and, therefore, the wave, is rather modest, but as will be discussed in more detail in the next section, many coronal jet models also follow this general principle.

Another key difference from 2D and 2.5D simulations is that in 3D the current sheet spans not just the null point, but also spreads over the fan plane either side of it. The reconnection process occurs continually within this layer, and although it looks like 2D reconnection in the plane of symmetry (Fig. \ref{fig:int_fls}(b)), the process is inherently three-dimensional. Where the current layer extends down the sides of the separatrix dome, the reconnection process locally has an effective guide field component \citep[e.g. see, for instance ][for details]{Priest2009}. The upshot of this 3D region is that newly reconnected field lines end up with a range of kinks depending upon where on the surface (relative to the symmetry plane) they connect to.

However, even with extra closed-field shear and the broader range of field line kinks afforded by the 3D geometry, any initial reversal in the radial field direction is still destroyed as the waves are launched (see Fig. \ref{fig:int_fls}c). This is a general feature of simulations of the interchange process on these large scales, with a number of subsequent works finding similar evolutions \citep[e.g.,][]{Aslanyan2022,Aslanyan2021,Scott2021,Pellegrin-Frachon2023}. The same evolution has also been shown on much smaller-scale closed-field regions more relevant to coronal bright points \citep{2019LRSP...16....2M} and the moving magnetic features observed at the base of plumes \citep{Wyper18a,Nobrega-Siverio2023}. Overall, in simulations where the interchange reconnection process involves a \emph{smooth, laminar current sheet}, a fast field-aligned flow accompanied by Alfv\'{e}nically propagating field line kinks with modest ($< 90^\circ$) deflections are produced. 

However, current layers are also expected to be highly unstable to the plasmoid instability in the solar corona. Early stability analysis of resistive current sheets showed that Sweet-Parker type current sheets are subject to various instabilities, including a long wavelength tearing mode \citep{Furth1963}. This instability was considered to be slow, however, in comparison with observed solar eruptive processes such as flares, as its growth rate scaled as $S^{-1/2}$ where $S=a\va/\eta$ is the Lundquist number, with $\va$ the Alfv\'{e}n speed in the inflow region, $\eta$ the plasma resistivity and $a$ the thickness of the current sheet. Although tearing-induced bursty reconnection was observed in numerical simulations with (relatively) low Lundquist numbers \citep{Biskamp1986, Shibata1992a}, it was first discarded as a relevant process in a $S \sim 10^{13}$ solar corona. Yet, \citet{TajimaShibata2002}, \citet{Loureiro2007}, and \citet{Bhattacharjee2009} showed that, through a renormalization of the Lundquist number $S=L\va/\eta$,  where $L$ is the finite length of the current sheet, the instability could achieve infinitely fast growth, with a positive dependence of the growth rate on the Lundquist number. Finally, \citet{PucciVelli2014} proposed that the tearing mode will most likely be triggered in forming current sheets that achieve the critical aspect ratio $a/L = S^{-1/3}$, where the transition between slow and fast growth occurs (see also \citealt{Uzdensky2016}, as well as \citealt{Tenerani2016} for an overview of the `ideal' tearing process). This has been confirmed by numerous numerical simulations \citep[e.g.,][]{Tenerani2015, Reville2020ApJL}, where the Lundquist number lies between $10^4$ and $10^6$ in 2D. In the ideal tearing theory, the growth rate of the instability becomes independent of the Lundquist number for $S \geq 10^8$, which suggests that numerical simulations can reproduce reconnection rates and timescales that are not too far from realistic regimes. 

In the context of such bursty, 2.5D interchange reconnection involving plasmoids, \citet{Drake2021} conducted an idealised particle-in-cell (PIC) simulation and found that a plasmoid survived ejection from the current layer. They, therefore, suggested such plasmoids could be a source of switchbacks. However, their idealised setup lacked a continuous open-field region, which likely contributed to the plasmoid's survival. \citet{Gannouni2023} showed that an emerging bipole in a continuous unipolar region triggered tearing-induced reconnection during the current sheet formation as soon as $a/L$ reaches $S^{-1/3}$. Plasmoids are created and ejected on both ends of the current sheet (Fig.~\ref{fig:ropes}). In contrast to the findings of \citet{Drake2021}, these plasmoids are destroyed at the pseudostreamer current sheet endpoints, launching slow magneto-acoustic jets in the fan's vicinity (see also \citealt{Yang2013,Yang2015}). During the flux emergence phase, the current sheet is subject to several cycles of disruption and reformation, which leads to a periodic release of jets and wave trains in the corona and solar wind. \citet{Gannouni2023} found typical periods around 19 minutes that are essentially related to the ideal timescale $t_A = L/\va$ of the emerging bipole.

Due to the high computational demands, comparatively fewer 3D simulations have reached Lundquist numbers where the plasmoid instability operates. For instance, \citet{Edmondson2010} conducted an early 3D numerical experiment where an extended 2D null line was subject to external stress (similar to the motions supposed by e.g., \citealt{Fisk2020}). Once formed, the current sheet quickly fragmented, producing short, highly twisted flux ropes with enhanced density (Fig. \ref{fig:ropes}: bottom left and inserts), which were less inflated in size than in 2D. However, the authors did not focus on what this structure launched along the open field. This work was later extended to cases with different guide fields \citep{Edmondson2017}. These works provided a bridge to the full 3D evolution but still remain 2.5D in their large-scale field structure. \citet{Wyper2014a,Wyper2014b} were the first to study the plasmoid instability in the context of a localised current sheet formed at a 3D magnetic null point. They found that the onset of the instability occurs above a similar Lundquist number threshold to the 2D case. The current layer was localised to a patch (instead of an infinite sheet), which led to a key new finding that the twist within the plasmoid flux ropes, which is initially confined within the fragmented reconnection region, spread out along the field lines as \emph{torsional Alfv\'{e}nic waves}. Building on this work, \citet{Wyper2022} conducted a three-dimensional simulation of interchange reconnection occurring at a 3D null point atop a separatrix dome (recall that this is the magnetic configuration expected in most magnetic bright points and jets, Fig. \ref{fig:int_fls}(a) see also \citet{Tripathi2025}). Their simulation included an isothermal solar wind and followed the structures ejected from the interchange current layer at high resolution out to 4 solar radii. They demonstrated that, as flux ropes are ejected from the current layer, they align to the open field, destroying any reversal of the radial magnetic field component initially present in the flux rope. It was shown that the ejection of these twisted structures launches a continual stream of torsional Alfv\'{e}nic waves (see Fig. \ref{fig:ropes}(d)), which due to the asymmetry of their system all had the same sense of rotation, (Fig. \ref{fig:ropes}(e)). Following in behind these waves is the super-sonic field-aligned reconnection outflow seen in 2D simulations. This finding that the field line kinks/twist propagates away ahead of the field-aligned flow of ejected plasma has also been noted in a number of other contexts \citep[e.g.,][see also the next section]{Lynch2014,Wyper2016,Karpen2017}.

In summary, the studies described above show that when the interchange process is \emph{3D} and \emph{intermittent/bursty}, the field-aligned plasma jet becomes modulated by the plasmoid ejection and that torsional Alfv\'{e}nic waves are launched in addition to the Alfv\'{e}nic propagation of field line kinks associated with the release of closed-field shear. \\

\begin{figure*}
\centering
\includegraphics[width=0.95\textwidth]{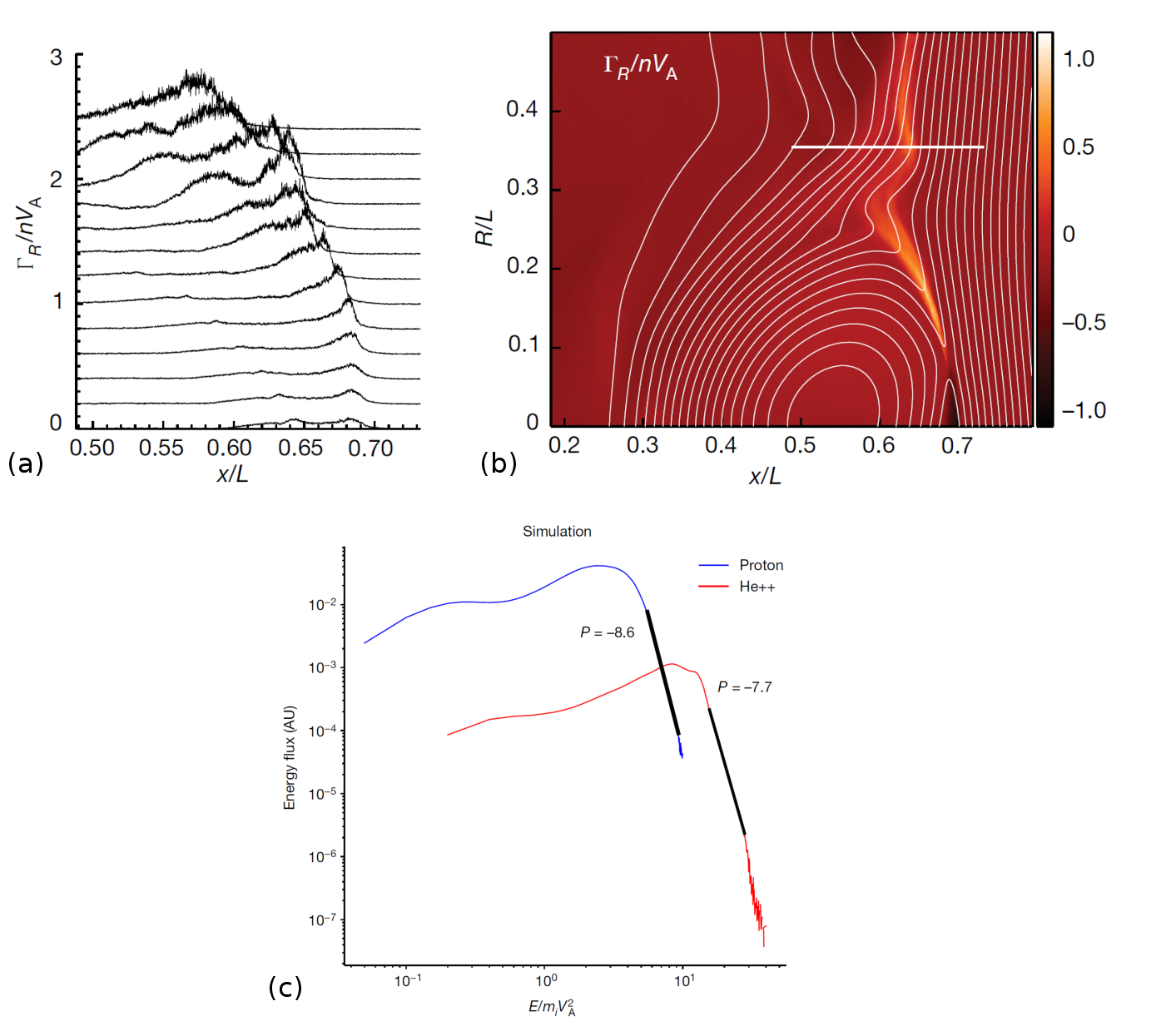}
\caption{Observational predictions for the intermittent interchange model from the 2.5D kinetic simulation of \citet{Bale2023}. (a) Time evolution of the radial velocity. (b) Radial velocity and the position of the line sample. (c) Spectra of the energy flux of protons and Helium ions in this simulation. Figure reproduced with permission from \citet{Bale2023}, copyright by Springer} 
\label{fig:int_obs1}
\end{figure*}

\begin{figure*}
\centering
\includegraphics[width=0.95\textwidth]{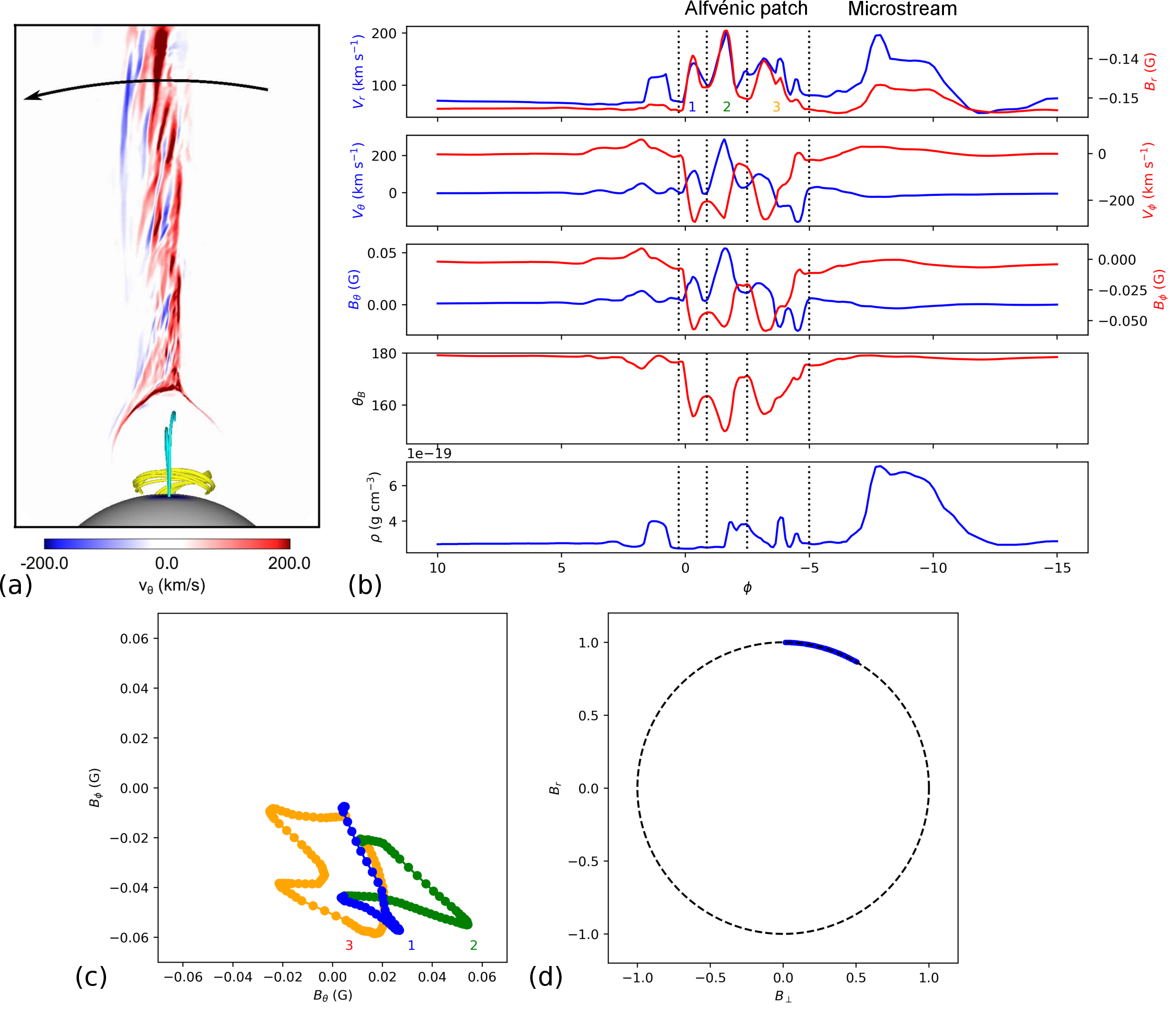}
\caption{Observational predictions for the intermittent interchange model from the 3D simulation of \citet{Wyper2022}, reproduced with permission from ESO. (a) The out-of-plane velocity shows the velocity reversals associated with torsional Alfv\'{e}n waves. The black arrow shows the path of the sample. (b) Density, field deflection ($\theta_B$), velocity, and magnetic field components along the sample. Three large deflections within the patch are highlighted. (c) Hodogram showing the characteristic change in field direction within each deflection. (d) $B_r/\langle B\rangle$ vs $B_\perp/\langle B\rangle$ (where $\langle B\rangle$ is the average magnetic field strength within the sample) showing the arc polarisation of the waves} 
\label{fig:int_obs2}
\end{figure*}

\paragraph{Requirements}
The two key requirements for inducing null point interchange reconnection are:
\begin{itemize}
\item A localised closed magnetic field region at the base of a coronal hole. This implies the closed and open field regions are separated by at least one coronal null point and its associated separatrix surface. 
\item A source of stress to form a current layer at the null point and induce interchange reconnection. 
\end{itemize}

Such closed-field magnetic regions associated with coronal null points are extremely prevalent and are detectable in magnetograms as small patches of minority polarity within the majority polarity of the coronal hole \citep{Shimijo2009,Platten2014,2017A&A...606A..46G}. In EUV, these features are associated with coronal bright points varying in size from tens of Mm to a fraction of an Mm in the fast-changing magnetic field at the base of plumes \citep[e.g.,][]{Zhang2012,2017A&A...606A..46G,Kumar2019,Madjarska2024} (see also \citet{Tripathi2025}).

As for the second requirement, as discussed, there are many potential sources of stress for these magnetic systems. This includes surface motions of magnetic features, flux emergence, flux cancellation, flux divergence or convergence, wave driving, motion or eruption of small-scale filaments, and the systematic inclination of the open field by, for example, the differential motion of the open field. Which are most prevalent is likely to depend on the scale of the closed-field region. 

Finally, for intermittent reconnection dynamics driven by the formation of plasmoid flux ropes, a related requirement is that the stress is maintained long enough for the current sheet at the null point to thin and lengthen to the point that the plasmoid instability sets in. However, given the high conductivity of the solar corona, it is expected that this criterion is quickly met for current sheets formed by many of the stressing mechanisms above. Indeed, an order of magnitude analysis by \citet{Bale2023} suggests that the null point current sheets are actually thin to the point that the reconnection occurs on kinetic scales. \\

\paragraph{Observable Signatures}
Regarding remote sensing observables, we point the reader to \citet{Tripathi2025}, where numerous observations attributed to interchange reconnection are described. The most directly observable interchange reconnection events are energetic jet-like events (coronal jets, jetlets), which are the topic of the next section. However, quasi-steady continual reconnection at larger structures has also been observed in the form of inflows/outflows, \citep[e.g.,][]{Masson2014}. It is expected that such quasi-continual reconnection also occurs at smaller scales. For instance, \citet{TriNS_2021,UpeT_2022} recently noted a systematic correlation between blue-shifted flows in the transition region and surface magnetic field strength, which they interpreted as evidence of widespread interchange reconnection and heating. For further details, see Sect.~5 of \citet{Tripathi2025}.

Turning to in situ observables, in Figs.~\ref{fig:int_obs1} and \ref{fig:int_obs2}, we show the expected in situ observable signatures obtained from the simulation studies of \citet{Bale2023} and \citet{Wyper2022}. {The general picture presented by both simulations is that the in situ signatures are present in the post-reconnection open field lines that map back to nearby the closed field region, appearing as patches when sampled in situ.} The key expected observation features of the interchange reconnection process, as highlighted by these simulations, are
\begin{itemize}
\item An asymmetric patch of arc-polarised Alfv\'{e}nic deflections.
\item Each deflection within a patch follows a similar direction.
\item A field-aligned quasi-periodic outward plasma jet/microstream, also appearing in situ as an asymmetric patch.
\item A power law ion energy spectra above 100 keV when the reconnection process transitions to the collisionless regime.
\end{itemize}

Figure \ref{fig:int_obs1}(a) shows the time evolution of the radial velocity flux in the cut shown in panel (b), taken from \citet{Bale2023}. The most recently reconnected field lines exhibit the fastest field-aligned flows at the position of the sample, with older field lines exhibiting slower upflows. This leads to an asymmetric patch across the sample, which also exhibits some fluctuations (spikiness) due to the modulation of the plasma jet by plasmoids formed in the current layer. Such patches have been suggested as potential sources of microstreams measured in situ by PSP. The power-law dependence of the ion energy spectra is shown in Fig. \ref{fig:int_obs1}(c). By varying the guide field in their simulation \citet{Bale2023} managed to match the power law indices to those observed in spectra by PSP.

The Alfv\'{e}nic patch associated with the intermittent ejection of plasmoid flux ropes from \citet{Wyper2022} is highlighted in Fig. \ref{fig:int_obs2}(b), where the in-phase relationship between velocity and the magnetic field is apparent. The arc polarisation is shown in Fig. \ref{fig:int_obs2}(d). The deflections from the ambient magnetic field direction, $\theta_B= \cos^{-1} (\bm{B}\cdot \bm{B}_0/(B B_0))$, peak around $30^\circ$ in this sample  with a maximum of $45^\circ$ found in the simulation overall. The asymmetry of the patch is evident by the sharper deflection as the most recently launched waves are entered (around $\phi=0.0$), with a tailing off upon leaving the patch (around $\phi=-5$). The adjacent denser asymmetric patch of field-aligned flow is older still and appears next to the Alfv\'{e}nic patch due to the lateral motion of the newly opened field lines, which at this height makes the waves and field-aligned flow move diagonally. Above it on the same flux tubes are further torsional Alfv\'{e}nic waves. 

A final prediction from the model of \citet{Wyper2022} is that the Alfv\'{e}nic deflections each deflect towards a similar direction. This follows from the torsional waves having the same sense of rotation and creates a characteristic out-of-phase relationship in some cases (although not all) between the two lateral field components (Fig. \ref{fig:int_obs2}(a)) and a near linear back-and-forth evolution in the hodogram (Fig. \ref{fig:int_obs2}(c)). The three largest deflections within the patch are shown in different colours. Similar coherent deflection patterns were found by \citet{Laker2023} in PSP data. It should be noted that the preferential sense of rotation of the waves is linked with the shear injected into the closed field in this simulation. However, as discussed, shear in the closed field is a typical occurrence in the solar corona.

Returning now to observations, while interchange reconnection is expected to be triggered in magnetic structures of many scales, in situ data, in particular switchback patches, show a modulation at granular and supergranular scales \citep{Bale2021, Fargette2021, Shi2022}. However, it should be noted that multiple assumptions are made when conducting the back mapping that implies these scales. Nevertheless, these scales are reminiscent of coronal bright points and solar plumes, which are thought to involve interchange reconnection \citep[see, e.g.,][]{Kumar2022}. One idea is that the modulation of switchback patches could be understood as a distance between the coronal bright points/the null point dome fan surfaces where perturbations (Alfv\'{e}n waves or fast flows) are the strongest \citep{Bale2023}. This would, however, rely on near-continual, low-energy interchange reconnection at the coronal bright points, which has not been unambiguously observed (see \citet{Tripathi2025} for a discussion). This is in contrast to the idea that the patches reflect the over-expansion of ``funnels" as mentioned in Sect.~\ref{sec:turbulence}, which sit between the coronal bright points. Waves from interchange reconnection at smaller scales (e.g., at the base of plumes) may locally have a patch structure within the wind but collectively would also have higher amplifications near the centres of the funnels where the expansion is high. Whatever sets the modulation distance, it is, however, clear from simulations that evolutionary processes within the solar wind are needed to form the $B_r$ reversals typically observed.

Looking to larger scales and further out in the heliosphere, long-duration heliospheric magnetic field inversions ($>$ 1~hour) that survive out to 1 AU have been ballistically mapped back to the Sun to show that they originate close to the large-scale coronal separatrix layers of streamers and pseudostreamers \citep{owensSolarOriginHeliospheric2013}. On these larger scales, it has been suggested that ``U-loops" opened near the Alfv\'{e}n surface maintain their $B_r$ reversal \citep{owensSolarOriginHeliospheric2013}, although this has not been demonstrated in simulations to date. There remains debate about whether these inversions are remnants of the U-shaped magnetic flux tubes formed in the corona or form later in the heliosphere via solar wind stream shear associated with the coronal interchange reconnection (see Sect.~\ref{sec: stream shear}). However, these inversions form only a subset of most switchbacks, with the majority being measured deep within coronal holes away from the large-scale coronal open-closed boundaries. \\

\paragraph{Advantages and Limitations}
The key advantages of this mechanism are that
\begin{itemize}
\item Small-scale regions capable of supporting interchange reconnection are extremely common in coronal holes. Interchange reconnection is, therefore, potentially prevalent enough to explain the prevalence of switchbacks.

\item Plasmoid ejection launches spherically polarised Alfv\'{e}nic waves into the solar wind, which could act as seed fluctuations for Alfv\'{e}n wave growth mechanisms further out. 
\item The reconnection jet launches a modulated field-aligned flow (microstreams) into the solar wind, which could set the right conditions for flow stream instabilities further out.  
\item The in situ patch structure of both the Alfv\'{e}nic fluctuations and field-aligned flows is consistent with PSP observations of switchback patches. 
\item The patch structure and size could be related to supergranular structures such as coronal bright points.
\item The intermittency generated by plasmoid ejection creates a ``spikiness" within the patches that is consistent with PSP patch observations. 
\item Ion spectra in the collisionless regime match those observed by PSP \citep{Bale2023}. 
\end{itemize}

Taken together, the expected ubiquitous interchange reconnection at very small scales, which produces jets of hot plasma that can lead to shear flows further out, alongside the launching of Alfv\'{e}n waves, has led some authors to argue that interchange reconnection provides all the required properties needed for in situ mechanisms to produce switchbacks further out \citep[e.g.,][]{Raouafi2023}. However, some {strong} limitations still remain, including:

\begin{itemize}
\item Simulations show that switchbacks (in the sense of $B_r$ reversals) are not created directly in the corona. U-shaped field lines and $B_r$ reversals within plasmoid flux ropes are not expected to survive ejection from the current sheet. Additional mechanisms are, therefore, required to produce larger deflections. However, as noted above, some of the ingredients required by these mechanisms are generated by the reconnection itself, e.g., the spherically-polarized Alfv\'en waves and flow shear.

\item In MHD simulations, the size/width of modelled current sheets are significantly larger than those estimated from solar observations. In kinetic simulations, the opposite is true, and in that case, the current sheet is also then not self-consistently dynamically driven by the large-scale MHD magnetic environment. This presents an issue about the unknown scale dependence when extrapolating to solar parameters. 

\item Interchange reconnection occurs along separatrices. Models of switchbacks that develop along the large-scale separatrices of streamers and pseudostreamers, i.e., those which require relatively long current sheets along those separatrices, would have a spatial distribution limited to the imprint of these separatrices in the heliosphere: the S-web \citep{Antiochos2011}. Switchback formation models solely focusing on interchange reconnection involving large-scale structures would thus mainly produce switchbacks in the slow solar wind, which is not observed. Interchange reconnection, as a formation mechanism of switchbacks, must, therefore, develop mostly at small-scale regions within coronal holes. This implies that the length and extent of the current sheet that generates the plasmoids cannot exceed some small length scale. So far, most 3D global MHD models (i.e. in spherical geometry including a solar wind) focusing on the interchange reconnection process within coronal holes rely on relatively large-scale structures, i.e., large anemone domes, which extend over several to tens of latitudinal/longitudinal degrees. The scale dependence when modeling smaller structures needs to be studied.
\end{itemize}

In summary, simulations to date show that the interchange reconnection process imparts bursty patches of flows and waves to the solar wind but that the interchange reconnection process itself does not produce radial field reversals directly. However, both the flows and waves are ideal candidates for evolutionary processes within the solar wind to take over and turn these seed perturbations into switchbacks further out (see Sect.~\ref{sec: in situ}). 



\subsection{Solar Jets and Untwisting Magnetic Waves}\label{sec:untwisting}

\paragraph{Overview of the Mechanism}
Magnetic twist/magnetic helicity is present in numerous solar structures of the low solar atmosphere \citep[see][in this collection]{Tripathi2025}. Such magnetic twist helicity is distributed over the volume of the structures, although not necessarily uniformly. When such a closed twisted magnetic structure reconnects with a surrounding non-twisted field, thanks to interchange magnetic reconnection, the newly reconnected field lines are then strongly out of equilibrium, with a twisted section on one part and a straight one on the other part. The Lorentz forces act in order to distribute the twist over the length of the field line, which actually corresponds to the generation of a non-linear Alfvénic wave that propagates upward along the open section. 

If the open field line is connected to the solar wind, this untwisting magnetic wave shall propagate up, first in the low $\beta$ solar corona, eventually reaching the high $\beta$ region ($\beta >1$) and the super-Alfvénic wind. The untwisting magnetic wave, being by nature Alfvénic, may directly explain switchback properties.

While interchange magnetic reconnection is essential to this mechanism, it is, however, largely agnostic to the precise reconnection dynamics. Unlike the mechanism discussed in Sect.~\ref{sec:interchange}, interchange reconnection here only serves as a way to form new field lines. The majority of the twist/helicity distributed over the closed volume does not transit through the reconnection site as the closed field is opened by interchange reconnection. Once the field line is opened, the twist/helicity then propagates away on the newly reconnected field lines. 

\begin{figure*}
\centering
\includegraphics[width=0.95\textwidth]{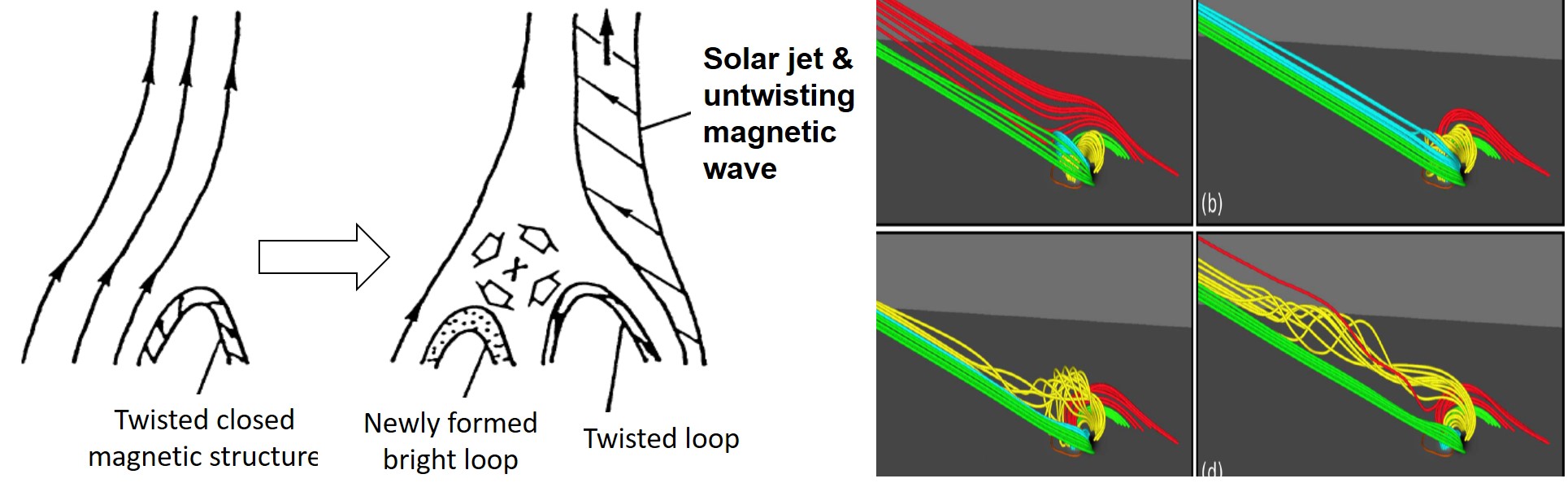}
\caption{Left: cartoon of the untwisting magnetic wave mechanism. Adapted from \citet{Schmieder95} and reproduced with permission from Springer Nature. Right: snapshots of a 3D MHD numerical simulation of the generation of an untwisting magnetic wave. The initially closed yellow twisted flux rope experiences interchange reconnection. As the closed twisted field lines sequentially reconnect, they untwist, inducing upward propagating Alfvénic waves. Adapted from figure 8 of \citet{Wyper19}. Figure reproduced with permission by the authors, copyright by Oxford University Press}
\label{fig:unstiwtingmech}
\end{figure*}

\paragraph{Explanation \& Previous Work}
The untwisting wave mechanism is an essential part of our actual understanding of the generation of jet-like events in the solar corona. Jet-like events/jets are {very common features of the solar atmosphere, e.g., coronal jets, macro-spicules, jetlets, chromospheric jets, and spicules. The interested reader is encouraged to consult \citet{Tripathi2025} in this article collection for a complete description of the different types, scales, and properties of jet-like solar events. Observationally, jet events very often display helical structure and/or twisting motions (see reviews of \cite{Raouafi16,Shen21,Tziotziou23}, and see also Sect.~5 of \citet{Tripathi2025}). The (pre-)eruptive structure frequently shows the existence of mini-filament structures, which are thought to be the signature of twisted magnetic flux tubes \citep[e.g.,][]{Kayshap13,Moore15,Baikie22,Sterling22}. Modelling of coronal bright points, which are commonly associated with jet events, regularly reveals the presence or formation of twisted magnetic flux ropes \citep[e.g.,][]{Galsgaard19,Madjarska20,Madjarska22}.}
Signatures of a helical magnetic structure are thus present in a noticeable sample of all jet-like phenomena despite their very different environments and distinct scales \citep[cf.][]{Tripathi2025}. Several observational studies have thus naturally proposed {low solar atmosphere}
formation scenarios in which switchbacks are induced by jet-like events from the low solar atmosphere \citep[e.g.,][]{Neugebauer95,Neugebauer12,Sterling20}.

From the modeling side, jets have benefited from a long history of numerical efforts \citep[see reviews of][]{Raouafi16,Shen21}. Numerical models, mostly within the MHD framework because of the scales and the properties of these events, have aimed at understanding multiple observational features of these events, such as their magnetic topology, their pre-eruptive dynamics, their trigger, their (thermo-)dynamics and flow properties, their different emission features, and their link with particle acceleration. Of particular interest for switchback formation are the studies that have focused on the propagation of jets toward the interplanetary medium, and in particular their untwisting magnetic waves.

In their seminal 2D numerical simulations, \citet{Shibata86} already conceptualised the propagation of jets thanks to the magnetic twist relaxation in open flux tubes. The concept was later cartooned in 3D by \citet{Schmieder95,Canfield96,Jibben04} to explain the dynamics of observed coronal jets and chromospheric surges. \citet{Pariat2009} performed the first 3D MHD simulation of the self-consistent impulsive generation of untwisting magnetic waves during the onset of a solar jet.  
Using different configurations (with the emergence of a twisted flux rope, with the shear-driven formation of a flux rope, in open or closed magnetic domains), several studies have provided evidence that jet-like events are driven/accelerated, 
at least partly, by propagating nonlinear Alfv\'enic waves due to the untwisting post-reconnection loop \citep[e.g.,][]{Torok09,Pariat10,Pariat15,Archontis13,MorenoInsertis13,Lee15,Wyper18a,Wyper19,Doyle19}. The properties of the untwisting wavefront were more specifically studied by \citet{Pariat16}, which showed that its upward wave packets consisted of non-linear (partly compressible) Alfvénic waves. \citet{Uritsky17} later characterized the different substructures (wavefront, shear Alfvén turbulence, shear and compressible turbulence, and dense jet) associated with this untwisting mechanism.

\begin{figure*}
\centering
\includegraphics[width=0.95\textwidth]{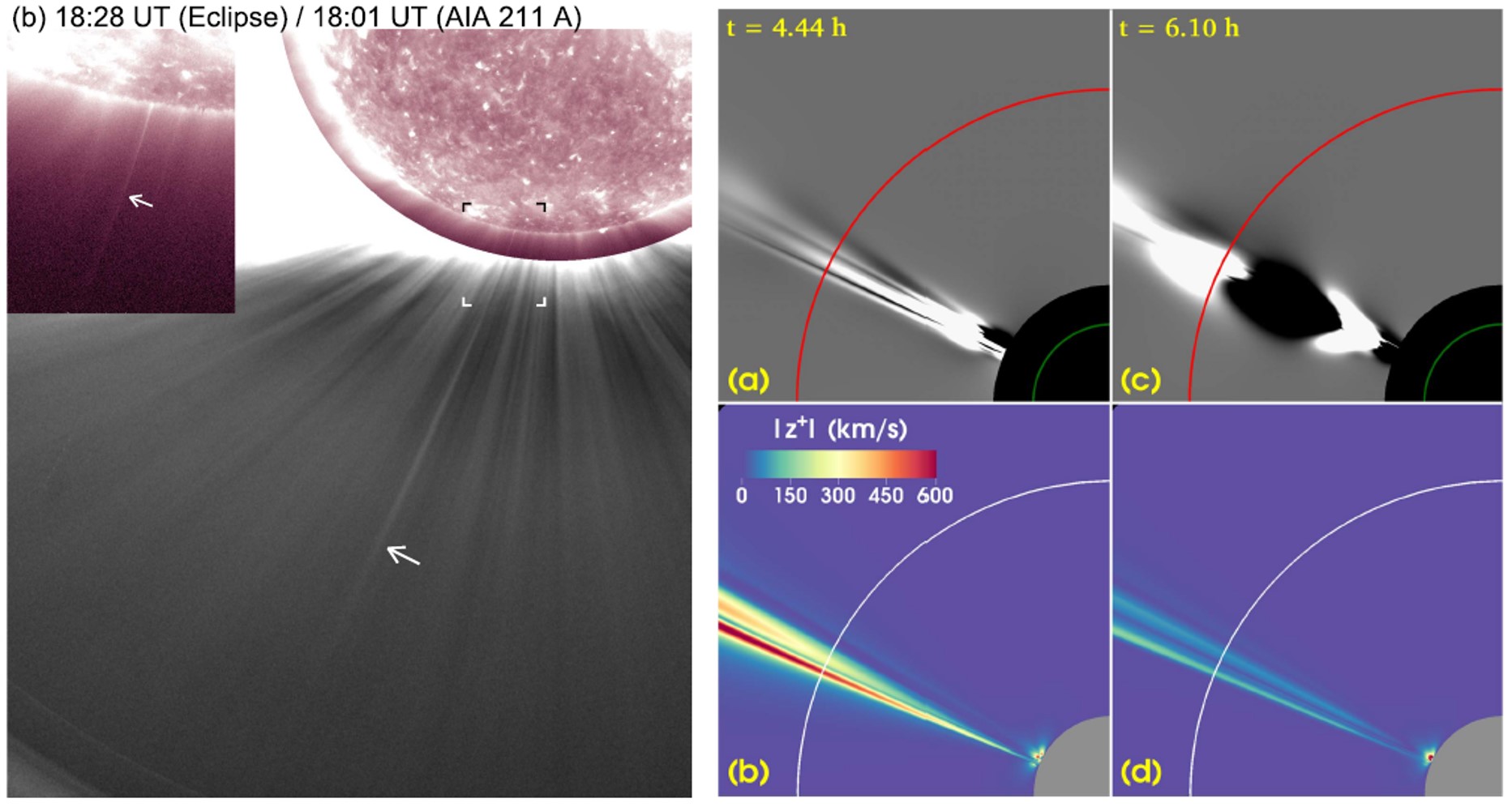}
\caption{Left: EUV and Eclipse white-light observation of the emission associated with a jet propagating toward the upper corona. Figure reproduced with permission from \citet{Hanaoka18}, copyright by AAS. Right: snapshots of a 3D MHD numerical simulation of the propagation of jets. Synthetic running-difference image of polarization brightness (top panels) and magnitude of Elsässer variable (bottom panels). Figures reproduced with permission from \citet{Lionello16}, copyright by AAS}
\label{fig:unstiwtingpropag}
\end{figure*}

The propagation of the untwisting wave over several solar radii, and hence its eventual signature in situ, has been the focus of only a few studies. \citet{Lionello16}, \citet{Karpen2017}, and \citet{Szente17} have produced the first 3D MHD simulations of jets propagating over tens of solar radii (respectively 20, 9, and 24 $R_\odot$) in stratified atmospheres including a solar wind (cf. Fig.~\ref{fig:unstiwtingpropag}). Although the simulation of \citet{Karpen2017} was later extended to 60 solar radii in \citet{Roberts18}. In \citet{Lionello16}, the spherically symmetric, steady-state solar wind results from the relaxation of an initial thermodynamic solar wind solution \citep[see][]{Lionello13}. In  \citet{Szente17}, the solar wind is heated by low-frequency Alfvén wave turbulence \citep{vanderHolst14}, while \citet{Karpen2017} relies on an isothermal Parker solar wind solution. In \citet{Lionello16}, a series of jets results from flux emergence of a twisted flux rope, in \citet{Karpen2017} and \citet{Szente17}, the jets are self-consistently induced following energy/helicity accumulation by boundary shearing motions \citep[as in][]{Pariat2009}. Despite these differences in the simulation frameworks, all nonetheless demonstrated that the untwisting wave does propagate far up in the solar wind and that it dominates the energy budgets. While the simulations of \citet{Karpen2017} were performed in a domain in which the plasma $\beta$ was always small (hence the wind was sub-Alfvénic), more recently \citet{Touresse2024} has produced new simulations of jet propagation including a super-Alfvénic domain and showed that the untwisting wave mechanism was present for all of the atmospheric profiles that were studied.  

\paragraph{Requirements}
The requirements for the untwisting wave mechanism are: 
\begin{itemize}
\item Interchange reconnection between twisted closed field and untwisted open field, hence sharing the similar requirements of the previous section.
\item Storage of magnetic twist/helicity in the closed magnetic structure involved in interchange reconnection. This requirement is a key element of the mechanism that distinguishes it from the pure interchange reconnection mechanism. The properties of the produced untwisting Alfvénic wave are indeed linked with the properties of the helicity-carrying structure rather than with the properties of the reconnection region. 
\item Newly reconnected untwisting field lines shall be connected to the open field of the solar wind.
\end{itemize}

\begin{figure*}
\centering
\includegraphics[width=0.95\textwidth] {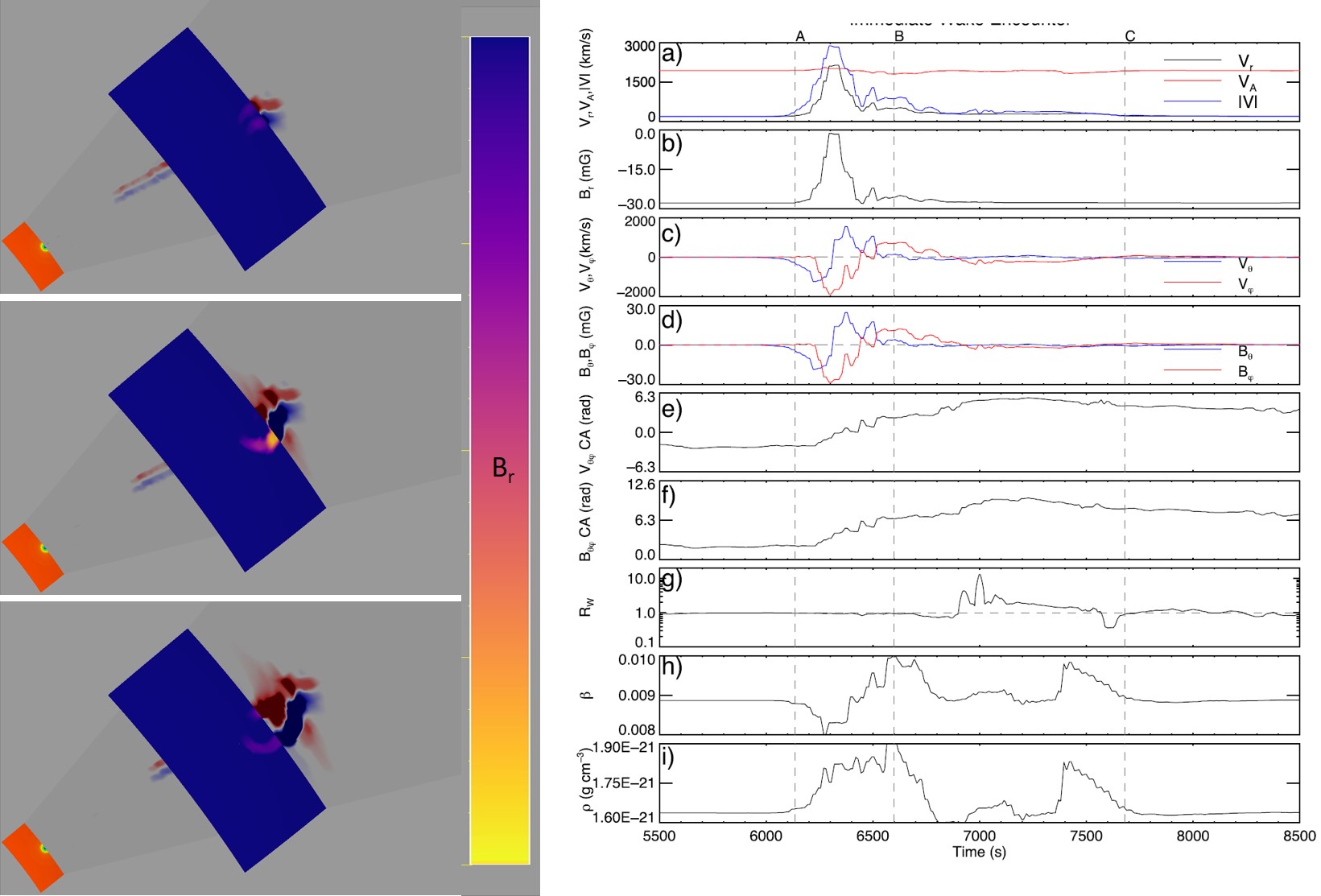}
\caption{Left: Snapshots of the propagation of the untwisting magnetic wave. The vertical cut presents the azimuthal velocity in Doppler-shift coded colour. The horizontal cut at $R=4R_\odot$ displays the distribution of the radial magnetic field $B_r$. Adapted from \citet{Touresse2024}.
Right: Time profile of different quantities simulating encounter with an in situ instrument. From top to bottom: (a) radial velocity, Alfvén speed, and total velocity, (b) radial magnetic field, (c) transverse velocity components, (d) transverse magnetic field components, (e) clock angle of the transverse velocity vector, (f) clock angle of the transverse magnetic field vector, (g) Walén ratio, (h) ratio of gas to magnetic pressure, and (i) mass density. Figure reproduced with permission from \citet{Roberts18}, copyright by AAS}
\label{fig:unstiwtingsynthobs}
\end{figure*}

\paragraph{Observable Signatures}

The untwisting wave mechanism is inherently linked with the formation of jet-like structures in the low solar atmosphere. Jet-like events are ubiquitous on the Sun over a broad range of scales and are transient and impulsive. They do not correspond to a constant, ever-present structure of the solar atmosphere. Overall, jet-like events can thus explain the ubiquitous while transient nature of switchbacks \citep[see the discussion in][]{Badman2025}. Additionally, multiple lines of observational evidence show that jet-like events are associated with helical motions and rotation of the plasma. There are also numerous observational indications that some large-scale jets have been observed to propagate over several solar radii (see Fig.~\ref{fig:unstiwtingpropag}, and see also, e.g., \citealt{Patsourakos08,Nistico09,Hanaoka18}). 

Typical large-scale jets, usually observed in EUV and X-rays, do not account for the prevalence of switchbacks. These typical jets are not numerous enough and are randomly distributed across the solar disk. However, jets at smaller scales, often referred to as ``jetlets", are more numerous and are localised predominantly to the inter-granular network so may be more relevant to the formation and clustering of magnetic switchbacks in PSP data and potentially even the formation of the solar wind itself \cite[see][]{Raouafi2014,Panesar2018,Kumar2023,Uritsky2023,Raouafi2023,LeeJ24}. It is important to note that the jetting observed on all scales is believed to operate under the same physical process, namely, an untwisting magnetic wave permitted by interchange reconnection. Small-scale jets are expected to exhibit properties similar to those found in larger jets. Hence, projecting the well-known properties of larger jets is likely relevant to the topic of magnetic switchbacks.

There is little doubt that magnetic untwisting waves do occur at the Sun at different scales and eventually induce a detectable signature in situ. The question is, rather, whether the observable signature of the untwisting mechanism corresponds to switchbacks or not. Observationally, it is not/has not been possible to demonstrate a one-to-one correspondence between a given solar jet and a given switchback. The possible correspondence between them has rather been addressed statistically \citep[e.g.,]{Badman2025,Kumar2023}.

In terms of expected in situ observables, building on \citet{Karpen2017}, \citet{Roberts18} and \citet{Touresse2024} produced synthetic signatures of the untwisting mechanism. Figure \ref{fig:unstiwtingsynthobs} shows that the untwisting wavefront does correspond to a local Alfvénic magnetic deflection, with a decrease of the radial magnetic field intensity $|B_r|$, while the norm of the magnetic field $|B|$ remained roughly constant, and that the untwisting wave was also associated with enhanced radial velocity. 

So far, all 3D MHD simulations confirm that the untwisting wave mechanism can indeed lead to the generation and propagation of non-linear Alfvénic waves from the low atmosphere up into the high-$\beta$, super-Alfvénic solar wind. This non-linear Alfvénic wave mechanism could explain some of the observed Alfvénic deflections. However, while propagating in the low-beta corona ($\beta <1$), the simulations show that the dominating magnetic forces inhibit the propagation of full magnetic reversals. While  U-shaped loops are present near the solar surface in the models - in the twisted helical progenitor - these magnetic inversions do not survive propagation. Immediately after reconnection, the Lorentz force unbends the magnetic field. Looking at the propagation of the untwisting magnetic wave in different atmospheric $\beta$ profiles, \citet{Touresse2024} observed that ``U-loops"/magnetic inversions were never present when the Alfvénic wave was below the Alfvénic surface. Thus, the untwisting magnetic wave mechanism seems unable to explain full reversal switchbacks directly. In any case, the magnetic untwisting mechanism may provide the seeds needed for in situ mechanisms to produce switchbacks (see Sect.~\ref{sec: expanding aws}).

\paragraph{Advantages \& Limitations}
The advantages of this mechanism are that
\begin{itemize}
\item Ubiquitous and intermittent solar jet-like events can explain ubiquitous and intermittent switchbacks. The largest-scale jets observed in EUV are likely too infrequent to explain the occurrence of switchback \citep{Huang2023}. The latest results indicate that microstream/switchback periodicities (and compositions) appear to be more consistent with solar coronal jetlets \citep{Raouafi2023,Kumar2023} and/or spicules \citep{LeeJ24}.  
\item Jet generation is naturally associated with the production of an Alfvénic wave/Alfvénic magnetic deflection, straightforwardly explaining the Alfvénic nature of switchbacks \citep{Pariat15,Pariat16,Wyper18a}.
\item Numerical simulations show that untwisting magnetic waves can propagate over several solar radii toward the super-Alfvénic region \citep{Lionello16,Karpen2017,Szente17}. The propagation of the untwisting jet is also associated with strong velocity and magnetic shear \citep{Touresse2024}. 
\item The magnetic untwisting wave is not purely Alfvénic and can thus account for some deviations from Alfvénicity as observed with switchbacks \citep{Larosa2021}.

\item Simulations \citep[e.g.,][]{Touresse2024} tend to show no ``U-loop"/magnetic inversion in the sub-Alfvénic wind, possibly explaining the seldom presence of switchbacks in such medium compared to super-Alfvénic wind \citep{BandyopadhyayEA22sub,Akhavan-Tafti2024,Sioulas2024}.
\item While the untwisting wave seems unable to directly induce switchbacks  \citep{Roberts18,Touresse2024}, the propagating Alfvénic deflections induced by the untwisting mechanism may constitute the seeds for the solar wind in situ processes forming switchbacks. 
\end{itemize}

Some of the limitations of this mechanism to explain switchbacks and of the current models are:
\begin{itemize}
\item The low-plasma-beta environment of the corona does not permit magnetic inversion initially present in the newly reconnected field lines to ``survive" within the low solar corona. This mechanism cannot directly lead to switchbacks \citep{Touresse2024}.
\item  It is still unclear which type of jets,
which spatial scale, is the one that could lead to switchbacks, even though evidence is pointing toward jetlets \citep{Raouafi2023,Uritsky2023,Kumar23} or spicules \citep{LeeJ24}. While larger-scale jets can indeed propagate far in the solar atmosphere, one does not know exactly what their signature is: do they correspond to a specific class of switchbacks? Furthermore, it is unclear whether the untwisting wave of the smaller-scale events can reach the upper coronal layers/inner heliosphere/super-Alfvénic wind.
\item So far, the propagation of the untwisting jet has been studied mostly in MHD models using relatively simple solar wind backgrounds \citep[with the notable exception of][]{Szente17}. Studying the jet propagation in more complex and realistic models
would further allow the determination of how the properties and dynamics of the untwisting wave is impacted when reaching the super-Alfv\'{e}nic wind.

\end{itemize}

\section{Mechanisms to Generate Switchbacks in the Solar Wind}\label{sec: in situ}

\subsection{Alfv\'{e}n Wave and Turbulence Growth Through Expansion}\label{sec: expanding aws}

\paragraph{Overview of the Mechanism}
In a constant density and pressure  background plasma, any 
\emph{spherically polarized} perturbation --- i.e., one that  satisfies $B=|\bm{B}|={\rm const.}$ and $\delta \bm{u}=\delta \bm{B}/\sqrt{4\pi \rho} $ in a background field $\bm{B}_0$ (with $\delta \bm{B} = \bm{B}-\bm{B}_0$, see notations in Sect.~\ref{sec: intro}) ---
propagates unchanged at the Alfv\'en speed $\va$  as a nonlinear 
solution to the compressible MHD or drift-kinetic equations. 
{Moreover, \citet{Barnes1974} showed that when such perturbations vary predominantly along a single direction and rapidly compared to a smooth background $\bm{B}_0$, their evolution is governed by the same characteristics as for linear Alfvén waves, even when $|\delta\bm{B}| \sim |\bm{B}_0|$ (a similar result, from \citealt{Hollweg1974}, relaxes some assumptions on the perturbations
for a specific solar-wind-like background).  Thus, within this class of constant-$|\bm{B}|$ Alfvénic structures, the leading-order dynamics are essentially amplitude independent, allowing $|\delta \bm B|/B_0 \gtrsim 1$ so long as the constant-$|\bm{B}|$ condition is maintained.} 

{The constant-$|\bm B|$ requirement places stringent restrictions on 
the form of allowable $\delta \bm B$ perturbations. One specific case is 
circularly polarized waves, in which there are no perturbations to the parallel magnetic field ($\delta B_\|$), while the perpendicular components trace out a circle as the wave propagates, keeping $|\bm B|$ constant. 
However, the $\nabla\cdot \bm B=0$ constraint disallows circularly 
polarized fluctuations from having any structure perpendicular to $\bm B_0$, making them a poor fit to observations \citep{Goldstein1974}. With general perpendicular structure, $|\bm B|$
can remain constant if (in contrast to linear Alfv\'en waves)  the fluctuations also involve perturbations to $\delta B_\|$, as well as developing non-monochromatic (i.e., non-sinusoidal) structure in at least some components.\footnote{Some authors distinguish spherically and circularly polarized waves, with spherically polarized waves being defined as having $\delta B_\|\neq 0$. We instead define circularly polarized fluctuations to be a subset of the more general spherically polarized (constant $|\bm B|$) condition.} In this case, regions with large perpendicular field perturbations can have $\delta \bm{B}\cdot \bm B_0<0$ to compensate, thereby keeping $|\bm B|$ constant. At sufficiently large amplitudes, i.e., if $|\delta B_\||\gtrsim B_0$,
these parallel perturbations reverse the background field. Thus, large-amplitude, spherically polarized perturbations naturally create switchbacks as part of their constraint to maintain constant $|\bm B|$.
}

There also exists a natural source of  {large-amplitude spherically polarized} perturbations in the solar wind. As the plasma and magnetic field expand in the corona and solar wind, flux and mass conservation imply that the background Alfv\'en speed decreases with increasing radius $R$. This decrease in propagation speed causes the relative amplitude of  {Alfv\'enic perturbations $|\delta \bm{B}|/B_0$} to increase in the same way as predicted by linear (WKB; Wentzel–Kramers–Brillouin) theory. Thus, if the amplitude of such perturbations at  low altitudes is sufficiently large (but still $\ll1$) and $\va$ decreases sufficiently rapidly with $R$ (beyond some radius where there is a maximum in $\va$), $|\delta \bm{B}|/B_0$ can approach (or exceed) $\simeq1$ by the Alfv\'en radius or higher altitudes. So long as they remain spherically polarized as this occurs, the evolution of their amplitude with radius is identical to that of linear waves from WKB theory \cite{Hollweg1974}, {allowing perturbations to develop with $|\delta \bm B|\gtrsim B_0 $ and $\delta B_\|\gtrsim B_0 $, which  reverse the background field}. Thus, perturbations with {$|\delta \bm B| \ll B_0$} in the low corona can naturally evolve into switchbacks further out {by staying spherically polarized  as they grow.}

Of course, there exist many complicating factors, particularly the influence of turbulence and other dissipation processes on the ideal evolution processes described above. 

\paragraph{Previous Work and Results}

\begin{figure}
    \centering
    \includegraphics[width=\columnwidth]{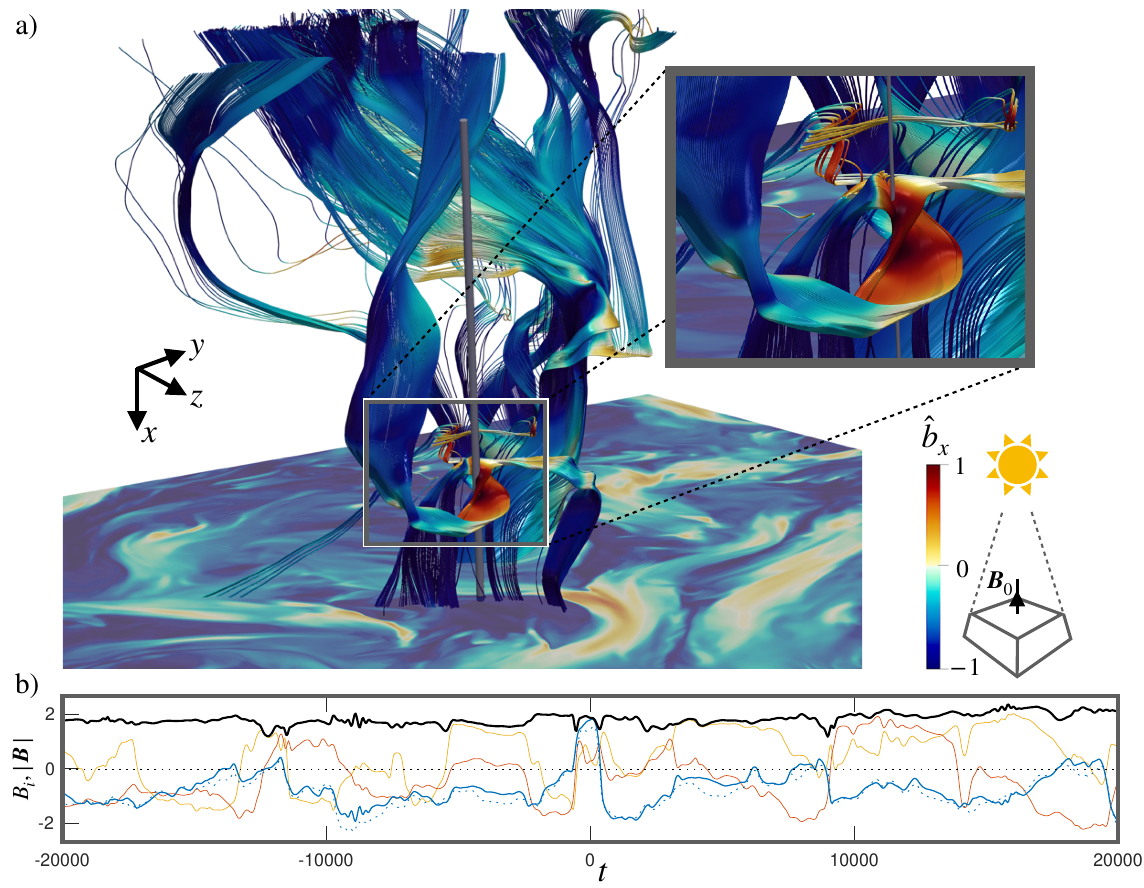}
    \caption{Switchback formation in 3-D reflection-driven MHD turbulence in the expanding box model 
    from \citet{Squire2020}, reproduced with permission of the AAS. The simulation is initialized with random perpendicular fluctuations with $\delta \bm{u}
    \propto\delta\bm{B}$ on top of the mean magnetic field $\bm{B}_0=-1\hat{\bm{x}}$, then evolves as the box expands, which causes the normalized fluctuation amplitude $\delta B_\perp/B_0$ to grow. The upper panel shows a visualization of the magnetic field line structure, colored by the normalized radial field component $B_x/B_0$. The lower panel plots the field components 
    along the grey line visualized in the top panel, which is intended to approximate a spacecraft trajectory in an outflowing wind. The blue, yellow, and red lines show $B_x/B_0$, $B_y/|B_0|$, and $B_z/B_0$, respectively, while the black line shows $|\bm{B}|/B_0$. We see large amplitude constant-$B$ perturbations, which have completely reversed the field around the point labelled $t=0$ (the origin $t$-axis scale is arbitrary, while its scale is tuned to approximately match PSP scales at $R\approx 35 R_\odot$). }
    \label{fig:aws-squire}
\end{figure}

\begin{figure}
    \centering
    \includegraphics[width=\columnwidth]{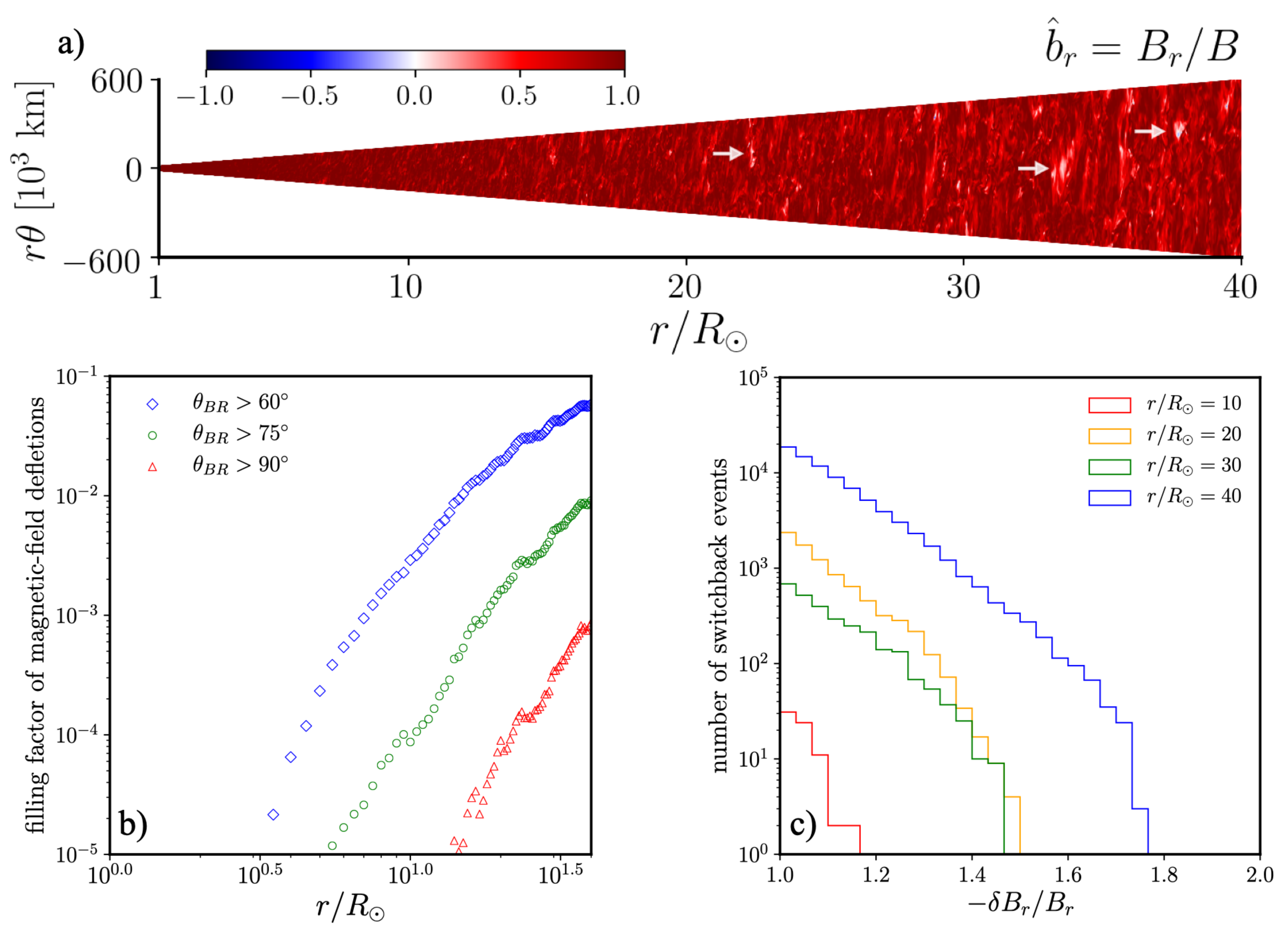}
    \caption{Formation of switchbacks in a radially extended 3-D flux tube with fluctuations driven by random motions at its base. Reproduced with permission of the AAS. Panel (a) shows the radial magnetic field at one time along a 2-D slice through the 3-D domain. Regions, where the field has reversed (switchbacks), are highlighted with arrows. Panels (b)-(c) show the switchback fractions versus heliocentric radius $R$ (denoted $r$ on the panel). Panel (b) shows the fractional filling factor of regions with deflections $\theta_{BR}=\cos^{-1}(B_R/B_0)$ larger than a specified threshold, as labelled. Panel (c) shows 
    the cumulative histogram of the number of switchback deflections at various radii, counted via the number of grid cells with  $\delta B_R/B_0$ larger than the relevant point on the $x$ axis (such that $-\delta B_R/B_R=1$ signifies a $90^\circ$ deflection). We see the continuous growth of switchbacks with $r/R_\odot$, although their volume filling fraction remains lower than that observed by PSP. Figures reproduced with permission  from \citet{Shoda2021}, copyright by AAS}
    \label{fig:aws-shoda}
\end{figure}

The theoretical basis for the physics described above was worked out in the early 1970's in seminal papers of \citet{Voelk1973a}, \citet{Hollweg1974}, and  \citet{Barnes1974}, among others. These works considered the evolution of linear or spherically polarized Alfv\'enic perturbations as they propagated out from the Sun, without focusing specifically on field reversals or switchbacks (although this was recognized as being possible). The direct relationship to switchback observations from Parker Solar Probe (PSP) was made in \citet{Squire2020}, who used numerical simulations with the ``expanding box model'' \citep{Grappin1993}, which captures a small patch of plasma as it flows outwards in the super-Alfv\'enic wind. Their results showed that spherically polarized field reversals (switchbacks) can grow from smaller-amplitude initial conditions (without switchbacks) as a result of the expansion-induced wave growth, while also demonstrating that the fluctuations becomes spherically polarized (constant $|B|$) even with complex, 3-D structures and a well-developed turbulent spectrum (see Fig.~\ref{fig:aws-squire}). These results required starting from waves with already relatively large amplitudes, an issue that was rectified in \citet{Shoda2021} with simulations of the small spherical wedge of the solar wind (i.e., a radial flux tube), which accelerates out from the low corona to $40$ solar radii. They also observed the formation of spherically polarized, radially elongated field reversals, with the relative fraction of these switchbacks growing out to the end of the simulation domain at $R\sim 40 R_\odot$ (see Fig.~\ref{fig:aws-shoda}).

The development of small-amplitude Alfv\'en waves into magnetic switchbacks can be interpreted as follows. Initially, the relative amplitude of Alfv\'en waves, $|\delta \bm{B}|/B_0$, increases with wave propagation due to the WKB effect. Subsequently, as demonstrated by \citet{Vasquez98}, a second-order effect (magnetic pressure) generates a magnetic field disturbance parallel to the background field, $\delta B_{\parallel} = -\delta B_{\perp}^2/(2B_0)$, as required in order to maintain approximate spherical polarization of the fluctuations. Hence, when the amplitude is sufficiently large, $\delta B_{\parallel}$ can grow sufficiently so that $\delta B_\| \sim B_0$, indicating the occurrence of a switchback. The expression $\delta B_{\parallel} = -\delta B_{\perp}^2/(2B_0)$ is an approximation valid when $\delta B_\perp \lesssim B_0$ and $\ell_{\perp}\gg \ell_\|$ (see \citealt{Mallet2021} and below for more general requirements; here $\ell_{\perp}$ and $\ell_\|$ refer to the scales perpendicular and parallel to the magnetic field). The process also requires compressible motions in order to develop. Therefore, both compressibility and the three-dimensional nature of disturbances must be considered for switchback formation, as done by \citet{Squire2020} and \citet{Shoda2021}.

\begin{figure}
    \centering
    \includegraphics[width=\columnwidth]{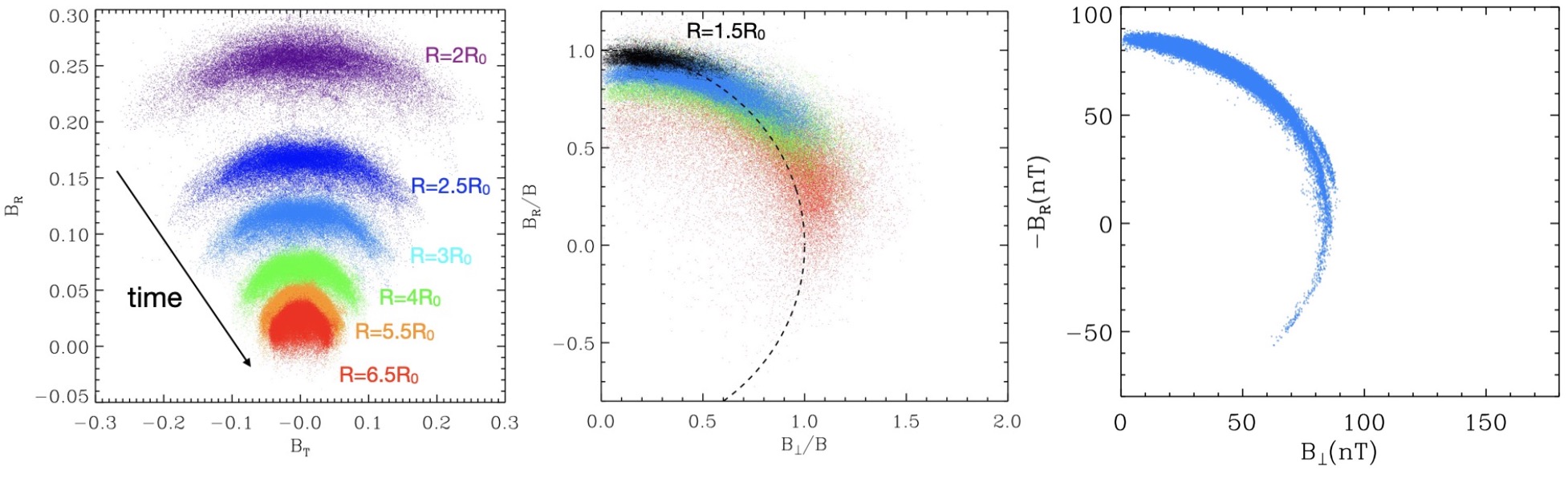}
    \caption{Evolution of spherical polarisation in an expanding hybrid simulation \citep{Matteini2024}. Left panel: Projection of the magnetic fluctuations in the ($B_T$, $B_R$) plane where T is one of the transverse directions, for different radial distances encoded by different colors. A distinct arc-like polarization is visible since the initial stage, evolving toward the 2D projection of a spherical surface at larger distances. Middle panel: projection in the plane ($B_\perp$,$B_\parallel$), where fluctuations are normalized to the local average magnitude $\langle B \rangle$ for each $R$; a subset of distances from the left panel is shown, with the same color code. Black data refer to $R=1.5R_0$. Right panel: PSP data from first perihelion. Figure reproduced with permission from \citet{Matteini2024}, copyright by AIP publishing}
    \label{fig:hybrid_matteini}
\end{figure}

Various additional aspects of the expansion and propagation process have also been considered. \citet{Mallet2021} derived  a simplified
equation to capture the evolution of 1-D spherically polarized solutions, which 
also predicts various correlations related to the compressible features of switchbacks. 
\citet{Johnston2022} and \citet{Squire2022a} studied the influence of the Parker spiral on the fluctuations 
as they expand, finding, surprisingly, that it often helps switchback formation, as well 
as introducing asymmetries into the fluctuation statistics. 
Finally, the picture above has also been generalized beyond the MHD model by \citet{Matteini2024} using 2D hybrid-kinetic simulations (kinetic ions, fluid electrons) with the background magnetic field $B_{0}$ out of the simulation plane (see Fig.~\ref{fig:hybrid_matteini}). As $|\delta \bm{B}|/B_0$ grows due to the expansion of the volume with radial distance, the simulations show the emergence of spherical polarisation in the magnetic fluctuations. $\delta B_\|$ fluctuations, which are initially absent in the simulation, form and grow with radial distance. The level of $\delta B_\|$ follows well the relation mentioned above: $\delta B_{\parallel} = -\delta B_{\perp}^2/(2B_0)$, leading also to local field reversals when $\delta B_\|\sim B_0$. The right panel of Fig. \ref{fig:hybrid_matteini} shows the comparison with PSP data near the Sun. 

\paragraph{Requirements}
Most importantly, the mechanism requires a source of Alfv\'enic perturbations propagating outwards from the low corona. These must grow in amplitude sufficiently (to $|\delta \bm{B}|/B_0\simeq1 $) to form a spherically polarized Alfv\'enic structure with a parallel field perturbation comparable to the mean field, viz., a switchback. This implies that there exists some minimum required fluctuation amplitude at the source, which can, in principle, be computed from WKB theory (neglecting other destructive effects) for a given background field and density profile. Reflection heating models, which involve similar calculations, generally show fluctuations reaching amplitudes $|\delta \bm{B}|/B_0\simeq1 $ around the Alfv\'en radius \citep[e.g.,][]{Cranmer2009a,ChhiberEA19-1}, suggesting that switchbacks should indeed form around this point.

A further requirement arises from the  $\nabla \cdot \bm{B}=0$ constraint. Specifically $0=\nabla\cdot \delta \bm{B}\sim  \delta B_\|/\ell_\| + \delta B_\perp/\ell_\perp$, 
implies that the maximum $\delta B_\|$ expected is $\delta B_\| \sim \ell_\|/\ell_\perp \delta B_\perp$. (Note that this estimate
only provides an estimate of the maximum $\delta B_\|$ because it is possible that $\nabla_\perp \cdot \delta \bm{B}_\perp=0$, and indeed
this is the case for a linear Alfv\'en wave; see \citealt{Mallet2021}). In other words, 
the requirement  $\delta B/B_0\gtrsim 1$ is not in itself sufficient; in order to reverse the field ($\delta B_\|\sim  B_0$), 
the fluctuations must also be at least modestly elongated along
the background magnetic field $\ell_\|\gtrsim \ell_\perp$ (as is indeed observed; \citealp{Horbury20}). 
Given that expansion tends to elongate structures in the opposite 
way (towards $\ell_\|\lesssim  \ell_\perp$) this requirement suggests 
that the original low-coronal source of Alfv\'enic perturbations
is required to produce highly anisotropic structures ($\ell_\perp\ll  \ell_\|$)
near the coronal base \citep{Mallet2021}.

As well as the linear (WKB) effects on the Alfv\'enic
perturbations, as invoked in the arguments above (amplitude growth and stretching in the perpendicular direction), nonlinear effects may influence the perturbations as they propagate outwards. Such effects are certainly needed to accommodate 
 important physical effects and match observations of in situ heating \citep{ZhouMatt90a,Totten1995}. Of 
particular interest are turbulence and parametric decay, both of which could 
 decrease the perturbation amplitude below WKB expectations, probably making it harder to form switchbacks via expansion. 
Turbulence can develop through reflection of outwards propagating waves
from the background gradient (\cite{Velli1989}, \cite{MattEA99-ch}, see Sect.~\ref{sec:turbulence}). This causes energy contained in Alfv\'enic fluctuations to be converted to heat by nonlinear interactions. 
(Driving of turbulence by shear or other effects could also \emph{create} larger fluctuations, but this is a separate effect; see Sect.~\ref{sec: KH}). If this 
happened sufficiently rapidly at low altitudes, the fluctuations will never reach $|\delta \bm{B}|/B_0\sim 1$ or form switchbacks. However, in the standard 
phenomenology of reflection-driven turbulence, based on the ideas of \citet{Velli1989}, \citet{ZhouMatt89-nonwkb}, and \citet{Dmitruk2002}, the fluctuation amplitude $|\delta \bm{B}|/B_0$ exhibits 
substantial growth  at $R<R_{\rm A}$, and in this sense, the qualitative prediction of WKB theory remains largely unchanged, even as reflection effects, turbulent heating, 
and dynamics of cross helicity and residual energy evolution become distinctly different (see, e.g., \citealt{Cranmer2005a,Verdini2007,Chandran2009,ChhiberEA19-2}).

Parametric decay is an instability of the nonlinear Alfv\'en-wave solution \citep[e.g.,][]{Goldstein1974, Derby1978, Jayanti1993}, and could, in principle, break up switchback progenitors before they reached $|\delta \bm{B}|\sim B_0$. Indeed, in idealised setups, 
numerical simulations of parametric decay in low-$\beta$ plasma result in the reversal of normalized cross helicity, completely disrupting the Alfv\'enicity characteristic of switchbacks \citep{DelZanna2001, Shoda2018a}. However, in reality, it appears to be suppressed for more complex, 3-D structures \citep{Cohen1974a, DelZanna2001, Tenerani2020}, as well as by expansion \citep{Tenerani2013, DelZanna2015, Shoda2018b, Reville2018} and the non-monochromatic nature of the parent wave \citep{Malara1996, Malara2000}. This could result in a reduced saturation level without disrupting Alfv\'enicity, thus seeding a low level of turbulence that could cause heating and fluctuation decay, with similar effects to reflection.
The presence of switchbacks, therefore, does not necessarily contradict the solar wind formation scenario driven by parametric decay, as suggested in the literature \citep{Suzuki2005, Shoda2019, AsgariTarghi2021}.

We may conclude, therefore, that even with relatively pessimistic assumptions about the efficiency of turbulent damping and/or parametric decay during propagation, the amplitude-growth mechanism 
for promoting switchback formation never requires $|\delta \bm{B}|\sim B_0$ in the low corona. It is also
worth noting that enhancements of turbulence might \emph{help} the formation 
of switchbacks, 
by creating more perpendicular structures, 
thus 
interfering with the WKB tendency to develop pancake-like  ($\ell_\perp<\ell_\|$) structures with small $\delta B_\|/\delta B_\perp$.

Finally, we consider more specific switchback properties 
beyond the field-reversal requirement. Various studies have discussed potential signatures in compressible features, the sharpness of 
the reversals, or asymmetries compared to the Parker spiral direction \citep[e.g.,][]{Farrell2020,Fargette2021}.
Although the actual field reversals form at large altitudes in this scenario, this does not exclude
signatures of a low-altitude source (e.g., reconnection) from being observed at higher altitudes if individual progenitor  ``events'' maintain their structure as they propagate outwards. For example, it may be possible to measure asymmetrical or helical fluctuations related to interchange reconnection. However, various compressible features \citep{Mallet2021,Johnston2022}, sharp boundaries \citep{Squire2022c,Mallet2023}, and asymmetries \citep{Johnston2022,Squire2022a} can naturally arise during propagation/expansion, so one must be cognisant of propagation effects before concluding that such signatures could provide useful information about the coronal source.

\paragraph{Observable Signatures}

The basic observable signatures of the Alfv\'enic expansion mechanism are that switchbacks 
should be outwards propagating, spherically polarized Alfv\'enic structures, which are at least modestly radially elongated and involve $\delta B_\perp$ perturbations that are similar in magnitude to $B_0$. The model predicts that switchbacks grow in amplitude (and therefore prevalence) until around the Alfv\'en radius, and therefore, field reversals do not exist in the low corona.

Beyond the Alfv\'en radius, assuming turbulence is strong and driven primarily by reflection,
models predict that the fluctuation amplitude, and thus the switchback occurrence rate, remains approximately constant with distance from the Sun so long as the fluctuations maintain high relative cross helicity (imbalance). This is because in super-Alfv\'enic regions, the WKB growth and turbulent decay balance such that $\delta B_\perp/B_0 \sim {\rm const.}$ (cf.~sub-Alfv\'enic regions, where $\delta B_\perp/B_0$ grows with $R$; \citealp{Verdini2007,Chandran2009,ChhiberEA19-2}), while the strong nonlinear interactions should maintain approximately constant and  predominantly perpendicular anisotropy $\ell_\perp\lesssim \ell_\|$. At larger radii, the turbulence becomes balanced and such switchbacks cannot exist in the same form (they are no longer approximate nonlinear solutions of ideal MHD), although the system may evolve towards other large-amplitude magnetically dominated states \citep{Tu1991,Bruno2013,Meyrand2023}.

\paragraph{Advantages \& Limitations}

This mechanism is agnostic to any specific coronal/solar-wind conditions or the mechanism that itself generates Alfv\'enic perturbations in the low corona in the first place. 
In that sense, it is a particularly simple hypothesis: we know that the Alfv\'en speed
decreases significantly with altitude, and this will cause perturbations to grow in amplitude; so long as the low-coronal perturbations are large enough and the fluctuations  
are not significantly dissipated (e.g., by turbulence or parametric decay), switchbacks are a 
natural outcome.

A possible issue with the simulations of both \citet{Shoda2021} and \citet{Squire2020} is that the fraction of switchbacks -- as measured e.g., by the simulation volume occupied by reversals of the background field -- is lower than observed (up to $\simeq 3\%$ in  \citealt{Squire2020} or $\simeq 0.1\%$ in \citealt{Shoda2021}, compared to $\simeq 6\%$ from early PSP encounters). 
This may signal that impulsive events such as interchange reconnection (Sect.~\ref{sec:interchange}) or jets (Sect.~\ref{sec:untwisting}), rather than continuous random driving of Alfv\'en waves, are needed to explain observations. Similarly, time series from PSP usually appear
``spikier'' than those produced by 
the numerical simulations. However, results are certainly also affected by numerical resolution, and using a similar set-up to \citet{Squire2020}, \citet{Johnston2022} showed that switchback fractions of $\simeq 6\%$ occur 
when starting from larger amplitudes and reaching the strong turbulence regime. 
Another aspect that is not obviously reproduced by expansion alone (i.e., without invoking other low-coronal or evolution physics) is the very sharp boundaries of observed switchbacks (e.g., \citealt{Bale2021}). Whether these evolve naturally in MHD, given sufficient resolution, remains unclear, although there are promising signs from a simplified model of the expansion process \citep{Squire2022c}. Other possibilities could include kinetic effects during propagation \citep{Mallet2023,Tenerani2023}, or the properties of the source itself (i.e., various forms 
of impulsive events may create sharp boundaries that then survive during propagation).

\subsection{Magnetic Field Distortion by Shear}
\label{sec: stream shear}

\paragraph{Overview of the Mechanism}
In the ideal MHD regime, the magnetic field is frozen in to the plasma and thus is advected by the plasma flow. 
As a result, {perpendicular velocity shears or variation in the Alfv\'en speed} will distort magnetic field lines that are not parallel to the flow lines, potentially causing the field to locally reverse. 
{This holds for both the mean magnetic field \citep{schwadron2021switchbacks}}, and for the magnetic field perturbations in large-amplitude Alfv\'enic fluctuations, i.e., an Alfv\'en wave propagating through a transverse velocity shear or density gradient can roll up into kinked magnetic field lines, resulting in a polarity reversal. {
\citet{schwadron2021switchbacks} studied the first scenario (shearing of the mean field), while \citet{landi2006heliospheric} demonstrated this using a narrow shear (a jet) imposed on a monochromatic AW. \citet{toth2023theory} generalized the mechanism, showing that  large-scale transverse shear of the radial Alfv\'en-wave propagation speed \(\bar{v} \equiv V_{\rm SW} + \va\) --- arising from gradients in \(V_{\rm SW}\), \(B_0\), or \(\rho\) --- can twist circularly polarized AWs and make switchbacks}

\begin{figure}
    \centering
    \includegraphics[width=0.8\textwidth]{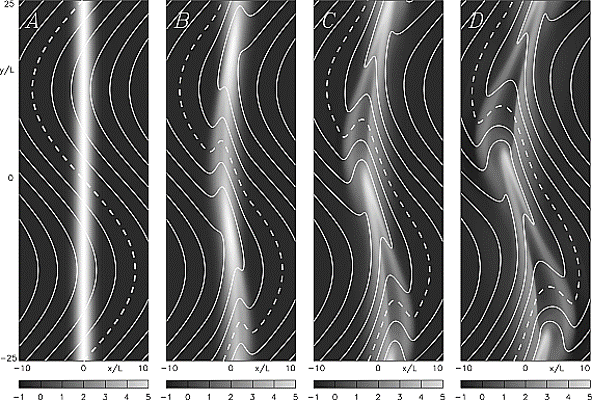}
    \caption{Snapshots of a 2D simulation with a plasma jet on top of a monochromatic Alfv\'en wave. The white lines are magnetic field lines, and the grey scale shows the flow speed. Adapted from \citep{landi2006heliospheric} and reproduced with permission, copyright by Wiley \& Sons}
    \label{fig:landi2006_fig01}
\end{figure}

\paragraph{Explanation \& Previous Numerical Results}

This reversal  process via velocity shear is seen to act for large-scale magnetic-field inversions in the heliosphere. In near-Earth space, ion charge state and compositional data suggest that magnetic field inversions seen in the heliosphere are not directly created by interchange reconnection in the corona but are instead subsequently generated by the solar wind speed shear along the flux tube \citep{owensGenerationInvertedHeliospheric2018, owensSignaturesCoronalLoop2020}. This ``kinematic'' shear effect on the heliospheric magnetic field has been employed \citep{lockwoodExcessOpenSolar2009} to describe the apparent increase in total heliospheric flux observed at increasing radial distance from the Sun \citep{owensEstimatingTotalHeliospheric2008}. {However, these large-scale 
reversals could not reasonably be termed switchbacks based on the definitions adopted for this review.}


{Figure~\ref{fig:landi2006_fig01} shows the narrow-shear Alfv\'en-wave  variant of \citet{landi2006heliospheric}: a jet imposed on a standing, monochromatic AW distorts field lines and can reverse \(B_0\) near the jet center (local polarity inversions). In this setup the effective shear length is comparable to, or smaller than, the transverse  width over which the field lines oscillate \(w\sim (B_\perp/B_0)\lambda_r/\pi\), where $\lambda_r$ is the parallel wavelength of the Alfv\'en wave. Extending this idea to  more realistic situations, \citet{toth2023theory} vary the shear lengthscale \(\lambda_y\) and show three regimes in 2D ideal MHD: (i) \(\lambda_y\!\gg\! w\):  \(B_R\) folds to form switchbacks with near-Alfv\'enic correlation; (ii) \(\lambda_y\!\sim\! w\): complex structures with intermittent reversals; (iii) \(\lambda_y\!\ll\! w\): strong distortion but no switchbacks, unless the guide field is very weak. They further verify that it is specifically the shear of \(\bar{v}=V_{\rm SW}+\va\) that matters (the same reversals arise whether the shear is introduced via \(V_{\rm SW}\), \(B_0\), or \(\rho\)), and that the mechanism persists in a radially expanding wedge, as more relevant to the heliosphere.}

\begin{figure}
    \centering
    \includegraphics[width=0.8\textwidth]{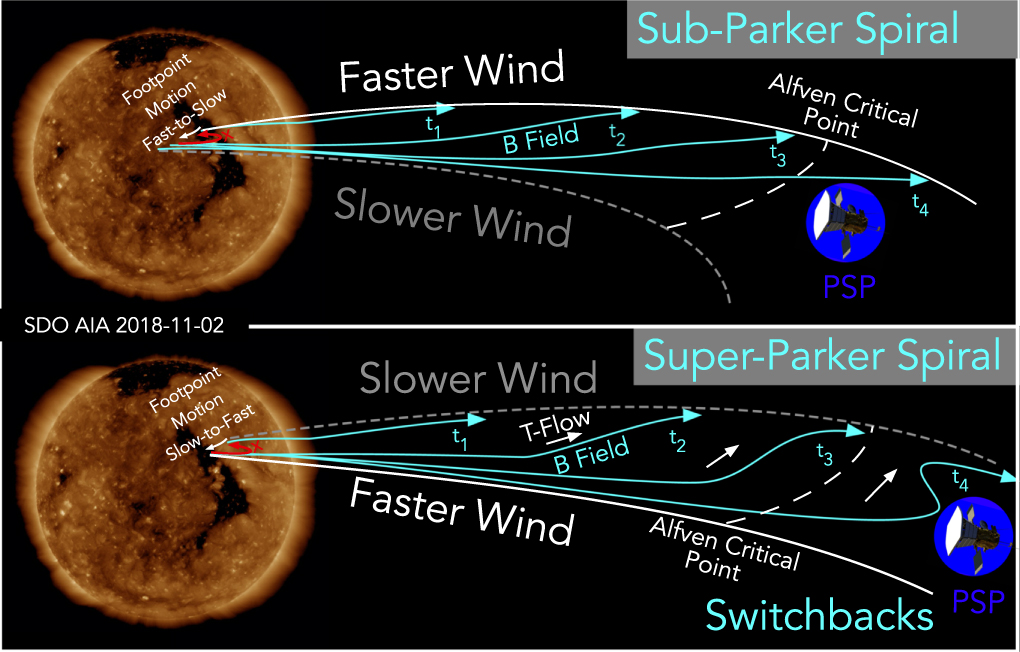}
    \caption{Illustration of the sub-Parker spiral (top) and super-Parker spiral (bottom) formed due to the motion of the magnetic footpoint of the solar wind. Figure reproduced with permission from \citep{schwadron2021switchbacks}, copyright by AAS}
    \label{fig:schwadron2021fig01}
\end{figure}

{\citet{schwadron2021switchbacks} give a geometric, footpoint-motion picture for how perpendicular flows could arise to shear the mean field (as opposed to Alfv\'en waves). This is, in essence, a small-scale analogue of the large-scale magnetic-inversion process mentioned above: motion from fast\(\to\)slow sources produces sub-Parker fields in rarefactions, while the opposite (slow\(\to\)fast) produces super-Parker fields (see Fig.~\ref{fig:schwadron2021fig01}). Above the Alfvén point, contraction between adjacent flows can reverse \(B_0\) and yield switchbacks, while below it, one-sided, co-rotation–oriented tangential flows are expected. In support, they argue that switchbacks occur preferentially outside rarefactions and emphasize their association with large, one-sided transverse flows.}

\paragraph{Requirements}

Transverse gradients of the Alfv\'en wave speed are required for the Alfv\'en-wave-shear mechanism of \citet{toth2023theory} to generate switchbacks from Alfv\'en waves and velocity shear. 
Consider the outward propagating Alfv\'en wave, whose speed is ${V}_{\rm SW}+\va$.
If any of $V_{\rm SW},\,{B}_0$, or $\rho$  vary along the transverse (to $\bm{B}_0$) direction across 
scales larger than the wave {(i.e., the \(\lambda_y\!\gg\! w\) regime)}, the wavefront will be distorted, and switchbacks will be generated. 

In the super-Parker spiral mechanism of \citet{schwadron2021switchbacks}, a gradient of the solar wind speed is necessary. In addition, the photosphere must undergo  small-scale horizontal motion relative to the solar wind, so that the magnetic footpoints move between source regions of solar wind streams with different speeds.

\begin{figure}
    \centering
    \includegraphics[width=0.9\linewidth]{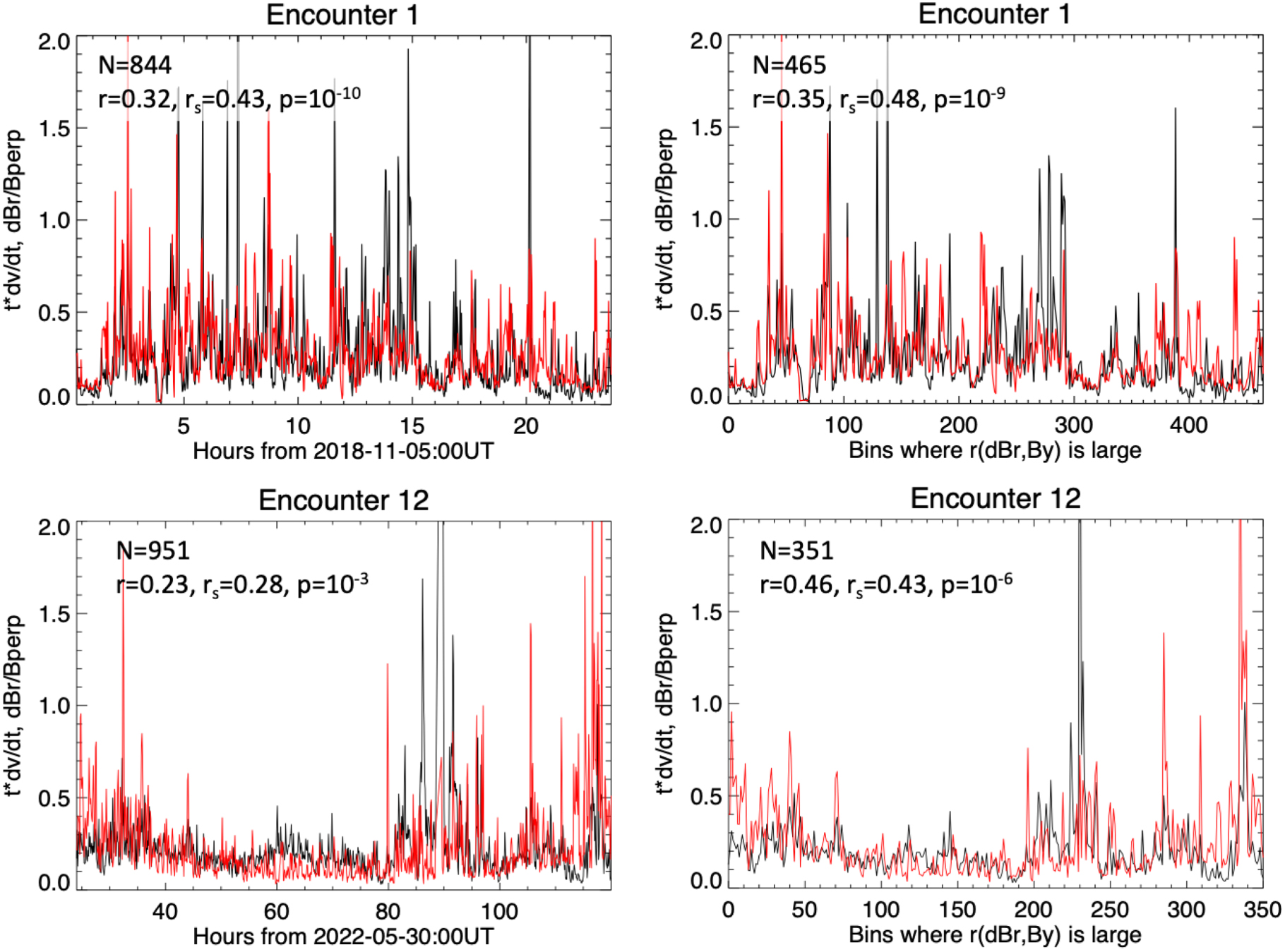}
    \caption{PSP observations in encounters 1 (top) and 12 (bottom). The black lines show $\delta B_R/B_\perp$ where $\delta B_R$ (denoted $dB_r$ in the figure) is the perturbation of radial magnetic field, $B_\perp$ is the magnitude of the perpendicular magnetic field. The red lines show the transverse gradient of the wave speed $d \bar{v}/(R d\alpha)$ multiplied by the wave propagation time $t= D/\bar{v}$. The left column shows the original data, and the right column only shows intervals with a high correlation between the radial magnetic field perturbation $\delta B_R$ and one of the transverse components of the magnetic field $B_y$. In each subplot, $r$ and $r_s$ stand for Pearson's and Spearman's correlation coefficients between the black and red curves, and $p$ is the $p$-value. Figure reproduced with permission from \citet{toth2023theory} }
    \label{fig:toth2023_fig06}
\end{figure}

\paragraph{Observable Signatures}
If a switchback is generated by the shear of wave speed, i.e. roll-up of the transverse component of the magnetic field due to the velocity shear, one expects \citep{toth2023theory}
\begin{equation}\label{eq:toth2023_shear}
    \frac{\left<\left| \delta B_R \right| \right>}{\left<B_\perp \right>} \approx \frac{D}{\left< \bar{v} \right>} \left< \left| \frac{d \bar{v} }{R d\alpha} \right| \right>.
\end{equation}
Here, $\left< \right>$ stands for average over a certain time period,  $D$ is the total distance that the wave travels, and $d\bar{v}/(R d\alpha)$ is the transverse gradient of the wave speed  ($\alpha$ is the transverse angular coordinate in radians).
In Fig.~\ref{fig:toth2023_fig06}, a comparison between the left-hand-side and right-hand-side of  equation \eqref{eq:toth2023_shear} is shown for two intervals of PSP observations. 
The top panels show observations in encounter 1 and the bottom panels show observations in encounter 12, while the left columns show all the data points and the right columns  show only data points with high correlation between $\delta B_R$ and one of the transverse components of magnetic field $B_y$, an indicator of the shearing process.
One can see that there is a {strong} correlation between the two curves, especially for the right column.
{This correlation with Eq.~\eqref{eq:toth2023_shear} supports the phase–speed–shear picture, though it is not necessarily unique --- for example, spherically polarized states  with large radial deflections will generically produce a correlation between $\delta B_R$ and the local radial velocity and hence with its shear.}

{\paragraph{Advantages \& Limitations}
Some advantages of the Alfv\'en-wave shear mechanism include: 
\begin{itemize}
  \item It is naturally Alfvénic: the fluctuations remain approximately  Alfv\'en-wave like, with \(\delta\boldsymbol u_\perp\)–\(\delta\boldsymbol B_\perp\) correlations preserved. Likewise, the  simulations explicitly show near-Alfvénic relationships in all components through a switchback. 
  \item Predictive observable: Eq.~\eqref{eq:toth2023_shear} links \(\langle|\delta B_R|\rangle/B_\perp\) to the path‐integrated transverse shear of \(\bar v\); PSP intervals satisfy this relation with high significance. 
  \item Generality: any cross-field gradients in \(V_{\rm SW}\), \(B_0\), or \(\rho\) can supply the required shear; the mechanism operates in planar boxes and/or in an expanding solar-wind wedge. 
\end{itemize}
While some  potential limitations  are that: 
\begin{itemize}
  \item The shearing mechanism does not by itself produce a spherically polarized wave with constant $|\bm B|$. Indeed, figure 5 of \citep{toth2023theory} shows structures that with modest small-scale variation in $|\bm B|$ and greater variation on larger scales, with the larger scales in total pressure balance. 
  \item {As so-far studied, the simulations in \citet{toth2023theory} assume a source of circularly polarized fluctuations close to the Sun, with the incoming Alfvén wave varying only along the radial direction (no intrinsic perpendicular structure on scales comparable to the parallel wavelength; see \S\ref{sec: expanding aws}). Given the strong dependence of the results on the ratio $\lambda_y/w$, it will be important in future work to extend the theory to waves that themselves possess perpendicular structure and to assess how this modifies the switchback properties. In more general cases where $|\boldsymbol B|$ is not initially constant, additional processes would be required to produce the very nearly constant-$|\boldsymbol B|$ switchbacks observed at PSP.}
\end{itemize}
An advantage of the super-Parker spiral mechanism
is that it could provide a simple geometric link between footpoint motion, stream shear, and the observed {one-sided, co-rotation–oriented} tangential flows. It is likewise consistent with the relative lack of switchbacks inside rarefactions.
However, an important  limitation is that by itself, the geometry does not enforce spherical polarization (constant \(|\boldsymbol B|\)) or Alfvénicity, the latter of which is argued to arise generically from large-scale evolution rather than being intrinsic to the mechanism. Explaining the observed spiky, nearly constant-\(|\boldsymbol B|\) events may therefore require additional Alfvénic structure or constraints beyond just shearing of a large-scale field.}

\subsection{WKB Growth of Interchange-Generated Fast Waves}
\label{sec: wkb fast waves}

\paragraph{Overview of the Mechanism}
Compressive fluctuations launched from low altitudes grow in amplitude as they propagate 
outwards due to the background variation in magnetic field and density. This 
physics is similar to the WKB growth of Alfv\'en waves discussed above (Sect.~\ref{sec: expanding aws}),
and, in the absence of damping via collisionless effects or turbulence,  could lead to large-amplitude 
{shock-like} perturbations around the Alfv\'en radius $R_{\rm A}$. \citet{Zank2020} propose that compressive, fast-magnetosonic structures are released by interchange reconnection events at around $6R_\odot$, growing in amplitude
as they propagate outwards to then produce switchbacks. 

\paragraph{Previous Work and Results}

The idea is motivated by the significant parallel magnetic field and velocity perturbations seen in switchbacks, which are argued to look similar to linear fast-magnetosonic modes. In addition, given an origin in violent interchange reconnection, compressive features are a natural expectation. \citet{Zank2020} propose a phenomenological model for the types of structures that could be released via reconnection, then apply WKB theory to study the change in structure and relative amplitude with radius, showing this peaks around $50 R_\odot$ for a model solar wind \citep{2020ApJ...901..102A}. They fit various PSP time series with the expected model structures, showing that most fields (e.g., velocity, magnetic, density and temperature fluctuations) can be fit for the selected events. This analysis was extended by \citep{2021ApJ...917..110L}, who use MCMC (Markov chain Monte Carlo) to fit 96 observed switchbacks, finding that with a six-parameter linear fast-wave model, the parameters of the switchback (density, velocity, and magnetic fluctuations) can be well fit for around half of the observed events.

\paragraph{Requirements}

The mechanism requires impulsive reconnection events that release strong compressive 
waves outwards. It also requires that, as they propagate outwards and grow to 
larger amplitudes, these structures do not dissipate significantly, as might occur 
{due to collisionless damping}, the formation of shocks, or through mixing via turbulence. 
This propagation process should be studied in 3-D numerical simulations 
to both understand
how well the linear 
theory of \citet{Zank2020} applies as structures evolve towards nonlinear amplitudes and to
study their dissipation {by collisionless damping mechanisms}.

\paragraph{Observable Signatures}

The most fundamental observational signature is that the 
switchbacks should be compressive fast-magnetosonic structures.
Given an angle of propagation, this fixes the relationship between the different fluctuations
(magnetic field, velocity, density, and temperature; see Appendix of \citealt{Zank2020}), although the waveform itself 
remains general. \citet{Zank2020} suggest that double- or multiple-humped structures should be
most common as a consequence of the reconnection's structure. {In linear MHD, fast magnetosonic modes are compressive: they involve coupled perturbations in density, temperature, and magnetic-field strength. In a low-$\beta$ plasma, quasi-parallel fast modes tend toward an almost acoustic polarization with relatively weak magnetic perturbations, whereas more oblique or perpendicular fast modes display larger $|\delta\bm{B}|$ and total-pressure variations. Thus, events with sizeable $\delta B_\parallel$ should also display  large $|B|$, density, and temperature fluctuations.}


Another potential signature of the mechanism could be compositional if the initial 
structure  entrains plasma that is hotter than the surroundings.

\paragraph{Advantages \& Limitations}

{If waves remain undamped as they propagate, the WKB growth} mechanism could  produce large-amplitude fluctuations, including in the radial 
field, as observed. Similar to the Alfv\'enic mechanism, at lower altitudes, perturbations are
expected to be smaller, fitting with recent observations showing a lack of 
switchbacks inside the Alfv\'en radius (see Sect.~\ref{sec: conclusions}). {However, as a compressive 
perturbation, fast-mode structures  perturb both the magnetic pressure and 
total pressure, 
 making it difficult to explain observations of   nonlinear Alfvénic kinks with constant $|\delta\bm{B}|$ and  pressure.
These features make fast-mode structures poor candidates for the many PSP switchbacks that are very nearly spherical-polarized and exhibit almost constant $|B|$ and weak density variations, 
although there are a subset of events that do show  compressive signatures (e.g., \citealp{Farrell2020,Laker2023}). Whether such compressive events constitute a dynamically important subset of switchbacks remains to be assessed.}

Another limitation is the relatively large heights so far assumed {in \citet{Zank2020}'s WKB calculations} for where the interchange reconnection takes place ($\sim 6R_\odot$), which limits the applicability of the mechanism to flux tubes adjacent to the largest closed-field structures, i.e. helmet streamers and pseudostreamers. With switchbacks being observed in solar wind originating far from coronal hole boundaries, future work should also explore how well the mechanism performs for interchange reconnection occurring at much lower heights in, for example, coronal bright points and plume bases \citep[see Sect.~5 in][]{Tripathi2025}.

\subsection{Nonlinear Effects of the Kelvin-Helmholtz Instability and Shear Driving} \label{sec: KH}

\begin{figure*}
\begin{centering}
\includegraphics[width=.90\columnwidth]
{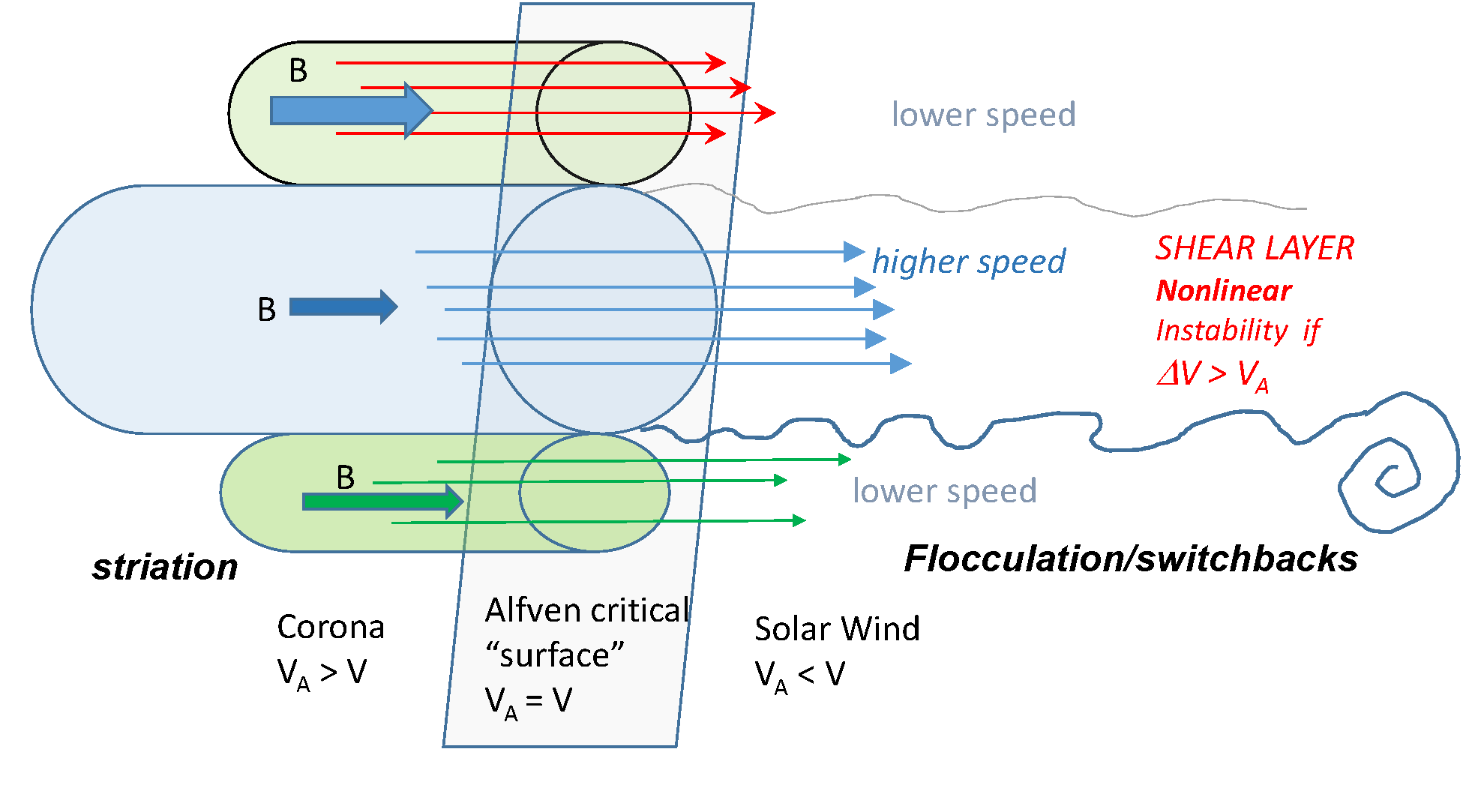}
\caption{Sketch showing the hypothesis that switchbacks are formed by shear-driven dynamics outside the Alfvén critical zone, see text for details. Figure reproduced and modified with permission from \citet[][]{RuffoloEA20}, copyright by AAS}
    \label{fig:shear}
    \end{centering}
\end{figure*}

\paragraph{Overview of the Mechanism}
Motivated by  prior remote  observations  of  a  transition  from  striated solar coronal structures  to  more  isotropic  ``flocculated''  fluctuations \citep{DeForestEA16}, it was proposed that the dynamics of the inner solar wind just outside the Alfv\'en critical zone is driven by the relative shear between adjacent coronal magnetic flux tubes \citep{RuffoloEA20}. This model suggests that large amplitude flow contrasts are magnetically constrained at lower altitudes, but shear-driven dynamics are triggered as such constraints are released above the Alfv\'en critical zone (see Fig.~\ref{fig:shear}), leading to enhanced turbulence. This has been seen in global magnetohydrodynamic (MHD) simulations that include self-consistent turbulent transport \citep{ChhiberEA19-1}. This dynamical evolution could account for features observed by PSP, including magnetic switchbacks and large transverse velocities that saturate near the local Alfv\'en speed \citep{RuffoloEA20}.  

Supporting evidence includes a comparison with a high Mach number 3D compressible MHD simulation of nonlinear shear-driven turbulence, which can reproduce several observed diagnostics, including characteristic  distributions  of  fluctuations that are qualitatively similar to PSP observations. The concurrence of evidence from remote sensing observations, {\it in situ} measurements, and both global and local simulations supports the idea that the dynamics just above the Alfv\'en critical zone boost low-frequency plasma turbulence to the level routinely observed throughout the explored solar system.  

\paragraph{Explanation}

The velocity of the solar wind is observed to be highly inhomogeneous, including streams of different speeds, so velocity shear has long been considered a mechanism for providing energy to solar wind turbulence. Specifically, the Kelvin-Helmholtz instability was proposed as a mechanism for converting velocity shear into turbulent fluctuations \citep{Sturrock66,Coleman68,Roberts92}.
However, other work argued that Kelvin-Helmholtz activity would be damped or too weak to explain the observed level of turbulence \citep{Parker64,Bavassano78}. In contrast with the flanks of Earth's magnetosphere, where strong and regular Kelvin-Helmholtz structures can be observed \citep{Kivelson95}, regularly spaced Kelvin-Helmholtz structures were not reported in the solar wind until Solar Orbiter observations at a distance of 0.69 au from the Sun \citep{Kieokaew21}, suggesting that the Kelvin-Helmholtz instability is localized to certain regions inside 1 au, and rapidly produces a turbulent mixing layer \citep{Rogers92} that no longer contains regularly spaced structures.

Direct imaging by \citet{DeForestEA16} has uncovered a transition in solar wind density structure from elongated striae to relatively isotropic flocculae. They interpreted this as a signature of the onset of shear-driven turbulent activity some 20-80 solar radii from the photosphere outside the Alfv\'en critical ``surface,'' where the Alfv\'en speed of magnetic fluctuations equals the solar wind speed. Indeed, given the highly dynamic and turbulent nature of both the solar wind and corona, these boundaries are almost certainly better described as an Alfv\'en critical {\it zone} \citep{DeForestEA18,Cranmer23}, which is fragmented and intermittent according to global MHD simulations that include turbulence transport \citep{ChhiberEA22-MNRAS}.

The Alfv\'en critical zone coincides with where the magnetic field ceases to be a dominant constraint on transverse motions, allowing coronal structures with differences (shear) in the outflow velocity to develop the Kelvin-Helmholtz instability. This interpretation is supported by results from turbulence-driven global simulations of the solar wind \citep{ChhiberEA18-global-floc}. The presence of velocity shears is also strongly suggested by coronal imaging at lower altitudes \citep{DeForestEA18}. 

Using a high Mach number 3D MHD simulation, \citet{RuffoloEA20} presented this hypothesis that shear-driven dynamics outside the Alfv\'en critical zone, including Kelvin-Helmholtz dynamics that evolve into mixing layers, could be responsible for the generation of switchbacks and the enhancing of solar-wind turbulence. This hypothesis is summarized in Fig.~\ref{fig:shear}. In the corona, the strong magnetic field regulates the dynamics of the nascent solar wind. Each flux tube may contain differing radial speeds and different radial field strengths due to processes at lower altitudes. Beyond the Alfv\'en critical zone, the magnetic field is no longer capable of constraining the dynamics, and the energy in the velocity contrasts becomes available to drive nonlinear magnetized Kelvin-Helmholtz-like dynamics, including magnetic field amplification and directional change, with associated deflection of velocities into the transverse directions. This may explain the transition from striated to flocculated structures in {\it STEREO} images \citep{DeForestEA16}. \citet{RuffoloEA20} highlighted characteristics of Parker Solar Probe data that are consistent with this picture of how shear-driven dynamics at and above the Alfv\'en critical zone generates switchbacks and boosts low-frequency turbulence to the levels observed throughout the heliosphere.

\paragraph{Requirements}
A general requirement for the Kelvin-Helmholtz instability is a spatially sharp change in the flow speed $V$ by a magnitude $|\Delta V|$ that exceeds the local Alfv\'en speed $\va$; otherwise, the instability is suppressed by magnetic tension \citep{Chandrasekhar81}.
According to \citet{RuffoloEA20}, in PSP measurements of the solar wind during initial solar encounters, the increment of the solar wind radial velocity $\Delta V_R$ (over time lags comparable to the correlation scale) frequently exceeded the Alfv\'en speed, as required. 
Note that the Kelvin-Helmholtz instability 
was also considered in greater detail by \citet{LauLiu80} and \citet{MiuraPritchett82}, who refined instability criteria for particular assumptions corresponding to regular velocity structures; in light of the chaotic nature of the solar wind, we consider the Chandrasekhar criterion as more generally applicable.

\paragraph{Observable Signatures}

\begin{figure}
    \centering
    \includegraphics[width=.45\columnwidth]{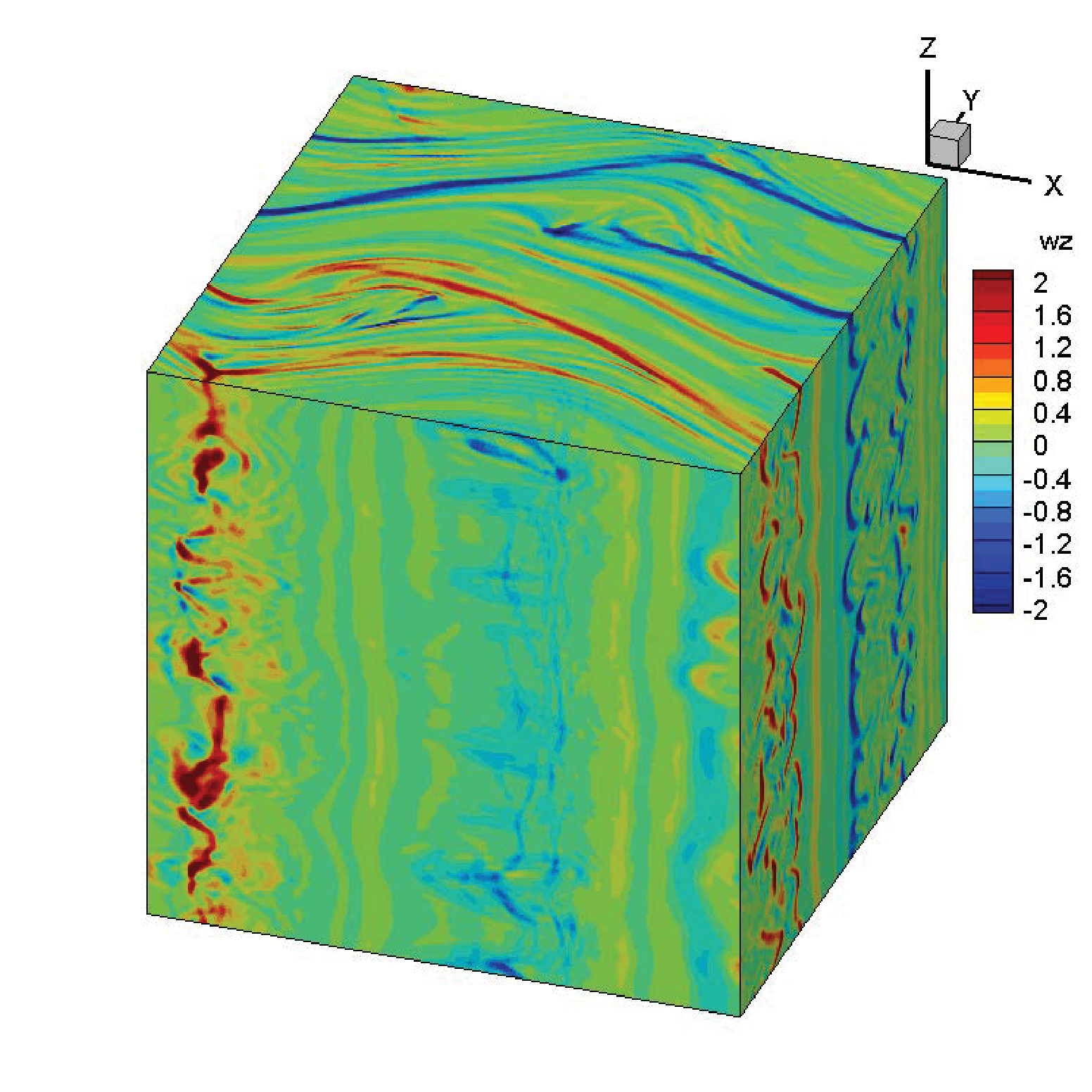} \includegraphics[width=.45\columnwidth]{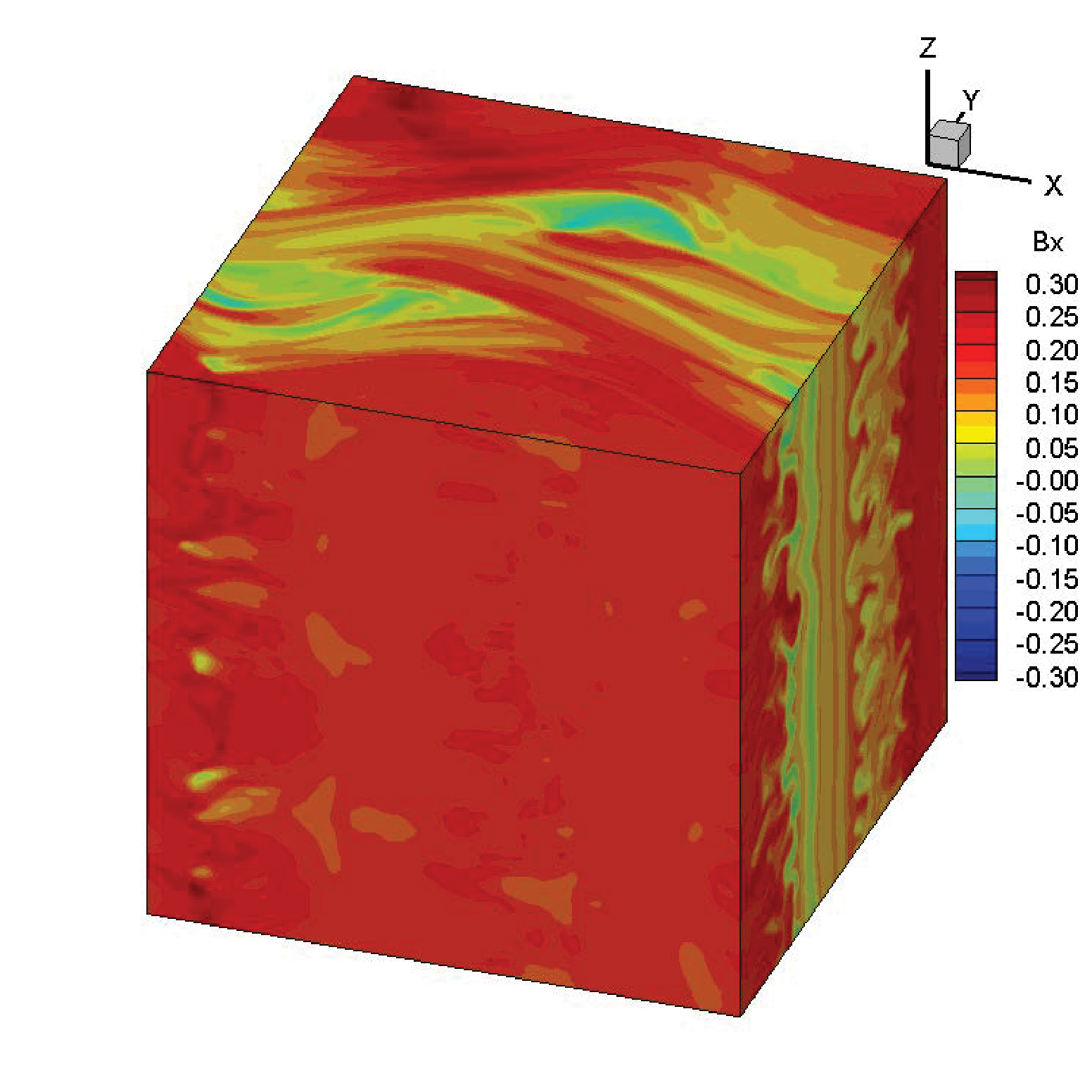}
    \caption{(Left) Vorticity in the $z$-direction, $\omega_z$, and (right) magnetic field in the $x$-direction, $B_x$ (shown by color scales), from a 3D compressible MHD simulation that started from a planar sheared configuration as in Fig.~\ref{fig:shear}.  
    Here, the vortex rollup is well developed, producing two regions with switchbacks (reversals in the sign of $B_x$), which recur intermittently throughout the simulation run. Figure reproduced with permission from \citet[][]{RuffoloEA20}, copyright by AAS
            }
    \label{fig:shearsim}
\end{figure}

Observations that are indicative of, or consistent with, the hypothesis that shear-driven dynamics near the Alfv\'en critical zone generates switchbacks and enhances turbulence include the following (arranged approximately in chronological order):
\begin{itemize}
\item Floccuation transition: From heliospheric imaging by the STEREO mission, \citet{DeForestEA16} inferred a transition between 20 and 80 solar radii from striated density structures (streamers) to ``flocculated'' structures that are statistically isotropic in the sky plane.  
This is consistent with the disruption of the streamer structure by Kelvin-Helmholtz dynamics (Fig.~\ref{fig:shear}) and radio observations that show enhanced plasma turbulence levels at distances of 25 to 35 solar radii.
\item The large velocity contrasts between adjacent flux tubes that are necessary for this process are observed in imaging of the outer corona \citep{DeForestEA18}. 
\item From in situ observations, switchbacks are found to have long ($>$10:1) aspect ratios, oriented near the local Parker spiral direction \citep{Horbury20,Laker21}.
From 3D simulation of Kelvin-Helmholtz structures \citep{RuffoloEA20}, field-reversed regions are indeed elongated (Fig.~\ref{fig:shearsim}).
\item The distributions of $B_R$ as observed in situ by PSP and in MHD simulations are similar \citep{RuffoloEA20}
\item The transverse components of the solar wind velocity are found to be bounded by the Alfv\'en speed, which is a natural result of the equipartition of magnetic and kinetic energies in mixing layer dynamics \citep{RuffoloEA20}. 
\item A reduced number of switchbacks (in the sense of Alfv\'{e}nic reversals in sign of the ambient solar wind field) have been observed in sub-Alfv\'enic solar wind \citep{Badman2025,BandyopadhyayEA22sub,PecoraEA22sb,Akhavan-Tafti2024}, which is consistent with switchback generation predominantly at and beyond the Alfv\'en critical zone (note, however, the results of \citealt{Sioulas2024}, discussed further in Sect.~\ref{sec: conclusions}).
\item Many more current sheets have been found inside than outside switchbacks \citep{huang2023structure}. Indeed, current sheets and reconnection jets are common in, and even used as indicators of, Kelvin-Helmholtz structures \citep{Eriksson16,Kieokaew21}, so their presence inside switchbacks is to be expected from the formation mechanism of shear-driven dynamics.

\end{itemize}

\paragraph{Advantages and Limitations}
A key prediction by \citet{RuffoloEA20}, with regard to this mechanism for switchback generation, was as follows: ``As [PSP] perihelia move closer and then enter the Alfv\'en critical zone, we expect to observe a further increase in both the mean magnetic field and the amplitude of broadband turbulence.  
As we move closer to the region of ``striations'', the more random fluctuations seen due to rollups should
give way to
more organized patterns of near-radially aligned flux tubes.
...  
Approaching these more organized magnetic structures, 
we expect the frequency of switchbacks to decrease ...'' 
Indeed, a reduced number of switchbacks (reversals in the ambient magnetic field direction) have been reported in the solar wind inside the Alfv\'en critical zone, i.e., in the solar wind that is sub-Alfv\'enic
\citep{BandyopadhyayEA22sub,PecoraEA22sb,Akhavan-Tafti2024}, though other proposed mechanisms are also consistent with this observation, and it may just be a consequence of switchbacks creating locally super-Alfv\'enic velocity spikes \citep{Sioulas2024,Badman2025}.  

Another advantage of this mechanism is that it satisfies some other observable signatures in the solar wind, as listed in the previous subsection.  
A limitation of the proposed physical scenario for switchback generation is that there are no observations that serve as a ``smoking gun'' of Kelvin-Helmholtz activity, such as periodic Kelvin-Helmholtz structures, in the solar wind near the Alfv\'en critical zone, though there has been an observation farther from the Sun \citep{Kieokaew21}.
Such periodic structures would be an intermediate stage between the initial instability and the highly nonlinear mixing layer.
This may imply that the mixing layer develops rapidly and/or that periodic structures are difficult to observe as PSP moves through the Alfv\'en critical zone with a very high tangential velocity, making it unlikely to observe repeated structures along one radially oriented shear zone between adjacent flux structures (see Fig.~\ref{fig:shear}).

Another limitation is the lack of spherical polarisation in the early stages of the Kelvin-Helmholtz evolution. \citet{RuffoloEA20} found significant variability in the magnitude of $B$ within their simulation but noted that, at least in patches, it remained fairly constant. They suggested a more uniform distribution of $B$ might occur following substantial mixing brought on by the interaction of different mixing layers, but this idea remains to be tested via simulations. A study of PSP data by \citet{Ruffolo2021} found that local domains of spherical polarisation become less volume filling with distance from the Sun, dropping from $90$\% within 0.2 au (43 $R_\odot$) to $38$\% outside 0.9 au. This suggests the solar wind plasma is becoming more isotropic and mixed with distance and would be consistent with the observed flocculation beyond the Alfv\'{e}n surface where the plasma $\beta$ is of order unity \citep{DeForestEA18}.

\subsection{Magnetic Turbulence from Beam and Velocity Shear Instabilities}\label{sec: shear and beams}

\paragraph{Overview}

The observations from PSP linking the modulation of the fast solar wind with the supergranulation scale of the solar surface magnetic field suggested that interchange reconnection could be an important driver of the fast wind and the source of free energy to drive switchbacks \citep{Bale2021,Bale2023}. Reconnection in macroscale systems is known to be highly bursty in time, with Alfv\'enic outflows being spatially localized (see Sect.~\ref{sec:interchange}).  The high velocity and strong heating in the core of reconnection outflows means that they will have sufficient energy to escape from the solar gravitational potential.  Meanwhile, adjacent regions with low outflow velocities and weak heating will either form a weak outflow or fall back into the chromosphere.  Because the Alfv\'en speed in coronal hole sources falls rapidly with distance typically beyond 1 to 2 solar radii above the surface, the flow shear between adjacent regions can exceed the local Alfv\'en speed and overcome magnetic tension to become unstable to shear-driven instabilities such as the Kelvin-Helmholtz instability (see also Sect.~\ref{sec: KH}).  Further, the bursty nature of reconnection outflows can lead to an environment in which fast flows can overtake slower moving flows \citep{Drake06} and, if the relative beam velocity exceeds the local Alfv\'en speed, potentially cause beam-driven turbulence. Shear-driven instabilities can mix plasmas in adjacent streams, while beam-driven instabilities can drive strong heating as well as magnetic turbulence. The turbulence associated with both classes of instabilities can potentially produce switchback-like signatures that have been explored with particle-in-cell simulations.

\paragraph{Explanation}

Instabilities develop when the total shear velocity exceeds the local Alfv\'en speed.  Due to radial variations in the magnetic field and density, the Alfv\'en speed varies in the outgoing wind. PSP has encountered multiple extended intervals in which the radial velocity falls below the local Alfv\'en speed \citep{Kasper21,Zhao22,Bandyopadhyay22}. These intervals are typically associated with low plasma density and, therefore, an increase in the local Alfv\'en speed, rather than a significant change in the radial plasma velocity. Analysis suggests that these sub-Alfv\'enic intervals exhibit fewer switchbacks \citep{Kasper21,Bandyopadhyay22}, as would be expected if velocity shear instabilities were a significant driver of switchbacks.

{How super-Alfv\'enic beams and shear develop via interchange reconnection coupled with the outward expansion in the solar atmosphere has not received as much attention because of the kinetic nature of such outflows \citep{Drake06}, and the resulting instabilities complicate the effort to model the dynamics. However,} the bursty behavior of reconnection is expected to produce time-variable outflows in which fast-moving outflows overtake slower-moving outflows (radial pressure forces are also expected to amplify velocity differences). As a consequence, outflows from coronal hole reconnection are expected to produce beam-like distributions with velocity separations that {could potentially} exceed the ambient Alfv\'en speed. Thus, both velocity shear and beam-like velocity distributions {are expected to} contain substantial free energy that can potentially drive the switchbacks documented by PSP and other satellites. 

\begin{figure*}
\centering
\includegraphics[width=\textwidth]{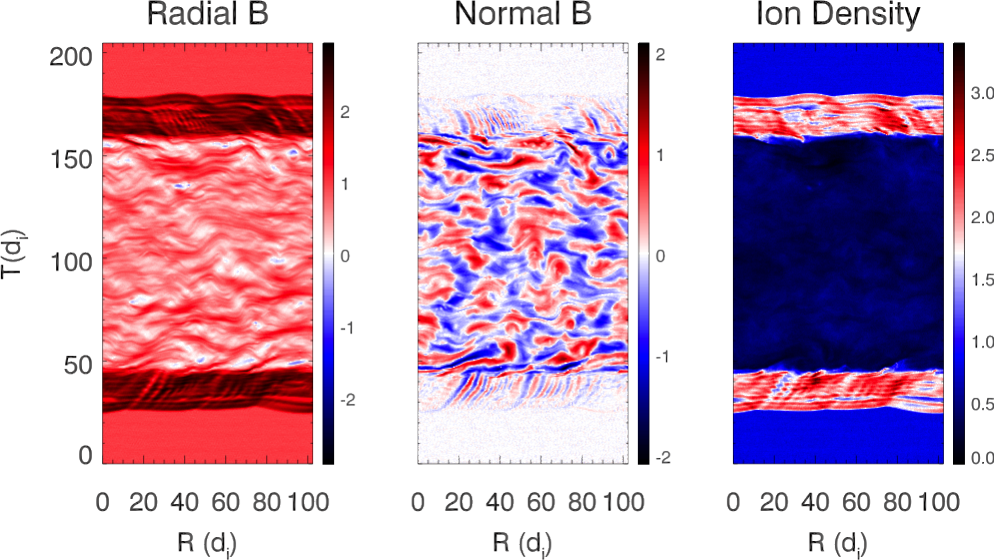}
\caption{Results from a particle-in-cell simulation of the interaction of an outflowing jet and ambient plasma.  The three panels show the radial component of the magnetic field, the normal component, and the ion density, respectively, at a late time. Figure courtesy of M. Swisdak.}
\label{fig:pic_mix}
\end{figure*}

Particle-in-cell (PIC) simulations can be used to explore the development of beam-driven and velocity shear instabilities relevant to coronal heating and the development of solar wind turbulence and switchback formation. Figure~\ref{fig:pic_mix} shows results from a PIC simulation performed with the code {\tt p3d} \citep{Zeiler02}. In each panel, the two outer regions (top and bottom) consist of ambient plasma with zero velocity in a local solar wind frame.  The middle region consists of two populations, half ambient plasma and half plasma flowing at super-Alfv\'enic velocity which is intended to represent a reconnection outflow jet.  (The convecting plasma has velocity $10 \va$ in the simulation shown, but similar -- although somewhat less energetic -- results arise in companion runs with weaker flows.) The result is a system with an ion two-stream system in the center of the domain that is sheared with respect to the ambient plasma at the top and bottom of the simulation domain. The magnetic field is initially uniform and horizontal in the positive radial direction. In Fig.~\ref{fig:pic_mix}, we show three panels from the simulation after turbulence is already well-developed: the radial magnetic field, the magnetic field in the out-of-plane $N$ direction, and the ambient plasma density.   

Two instabilities develop.  First, a Weibel-like instability forms in the central region as the ambient plasma interacts with the super-Alfv\'enic outflowing plasma.  While the initial radial field is of unit strength, the regions of white and blue within the central region correspond to places where the sign of the radial field has reversed, as occurs in switchbacks.  The normal field in the second panel shows that the initially radial field rotates within this region and that the amplitude of this self-generated, turbulent field actually exceeds the initial radial field. Further analysis (not shown) establishes that the perturbations are roughly spherically polarised (i.e., are characterized by constant $B$).  The development of the strong normal magnetic field is interesting in the context of mechanisms for switchbacks because the magnetic field also undergoes a sharp rotation into the normal direction across the switchback boundary. Because it arises from the Weibel instability, it would not appear in an MHD treatment of the problem. 

A second instability, related to the KHI, develops later due to the shear between the central and outer layers.  This is not a pure KHI mode due to the presence of the radial magnetic field (which would inhibit its development) although, in contrast to the Weibel mode, it should be present in MHD simulations as discussed in \S\ref{sec: KH}. 

The development of the Weibel instability in the core of the simulation causes the scattering of the ion beam and a sharp increase in the transverse temperature in this region. The resulting pressure increase causes a rapid expansion of the core in the $T$ direction and drives a shock-like compression of the plasma density at the upper and lower boundaries of the core plasma, as shown in the right panel of Fig.~\ref{fig:pic_mix}. 

\paragraph{Requirements}

Generating switchbacks via the Weibel or KHI mechanism requires: (1) Strong enough interchange {reconnection} 
within coronal holes to drive outflows that can form a wind solution; (2) Nearly continuous interchange reconnection within coronal holes so that outflows can form the space-filling fast wind; and (3) Sufficiently rapid drop in $\va$ with radius so that interpenetrating beams or sheared flows become super-Alfv\'enic. 

\paragraph{Observable Signatures}

In this mechanism, the production of switchbacks is closely tied to the presence of super-Alfv\'enic streams.  Hence, regions with sub-Alfv\'enic flows should exhibit fewer switchbacks.  In addition, the degree of Alfv\'enicity should increase with solar radius as the plasma relaxes from its initially turbulent state.  Finally, the shock-like density enhancements seen in the PIC simulations should be observable at the boundaries of regions with significant switchback activity, and indeed, there is some evidence from PSP observations that  density enhancements bound regions of high-amplitude switchbacks \citep[e.g.,][]{Liu2022}.

\paragraph{Advantages \& Limitations}

Switchback formation in this model follows naturally from the idea that interchange reconnection occurs in the low solar corona and plays an important role in driving the fast wind.  Once that is established, the ensuing mechanisms -- strong, inhomogeneous outflows, a rapid increase of the outflow velocity with altitude, the development of streaming instabilities -- are natural and likely {consequences.} 

One limitation, however, is the current lack of {\it in situ} observational evidence. The development of turbulence and the formation of switchbacks are expected to happen at {distances} inside the perihelion of PSP. In addition, the rapid development may be difficult to observe since PSP moves with a very high tangential velocity in the regions most likely to be of interest.  Hence, direct {\it in situ} observations of switchback formation {by this mechanism} may be unlikely. On the theoretical front, another significant issue with this mechanism lies with the spatial scale of the switchbacks that form. Those that form in the PIC simulations of Fig.~\ref{fig:pic_mix} are at scales ${\sim}10 d_i$, where $d_i$ is the ion-inertial length, while observed switchbacks {are typically orders of magnitude larger} at MHD scales.
While there may exist some process to merge or otherwise grow switchbacks, this has not been unambiguously demonstrated. {So for the moment this mechanism is limited to forming switchbacks at smaller scales and faster time frames than are typically measured by PSP.}

\subsection{Switchbacks from the Merging of Flux Ropes}
\label{sec: merging}

\paragraph{Overview of the Mechanism}
\cite{Drake2021} and \cite{agapitov2022} showed that flux ropes injected into the solar wind can evolve into switchback-like structures through multiple mergings. They evolve during propagation in the solar wind, with the diminishment of the magnetic island structure and the evolution of a switchback topology with a dominant axial magnetic field.

\paragraph{Explanation}

Magnetic flux ropes are known to be a product of reconnection processes in the solar corona or shear flow instabilities in the young solar wind. Cuts through flux ropes (in both observations and simulations) reveal that a strong axial magnetic field is wrapped by magnetic flux, and the radial magnetic field exhibits the characteristic reversal documented in switchback observations in the solar wind \citep[e.g.,][]{Drake2021}. The flux-rope model maintains the direction of the electron strahl with respect to the local magnetic field, as seen in the data. However, the strong magnetic island (wrapped) magnetic field structure, aspect ratio close to one, and lower-compared-to-the-original-perturbation Alfv\'enicity do not allow the direct association of such small-scale flux ropes with switchbacks observed by PSP. The evolution of flux ropes in the solar wind has been explored with 2-D reconnection PIC simulations from an initial state with a band of reversed radial magnetic flux sandwiched within a uniform magnetic field \citep{Drake2021,agapitov2022}. {Although the initial state of this mechanism is not representative of the typical uni-directional field within coronal holes, the late-phase evolution of the subsequently formed flux ropes was the goal of these studies.} The magnetic structure of the resulting flux rope evolves through multiple mergers and reveals signatures that are consistent with switchback observations: a sharp rotation of the ambient solar wind radial magnetic field into the azimuthal direction; weak in-plane magnetic fields within the structure with a local reversal of the radial magnetic field component; and a nearly constant total magnetic field with modest dips at the edges of the structure \citep{Drake2021}. Surface reconnection processes during merging lead to a weak heating of plasma inside switchbacks, which is reported from PSP observations \citep{Larosa2021}.

\begin{figure}
\includegraphics[width=0.99\textwidth]{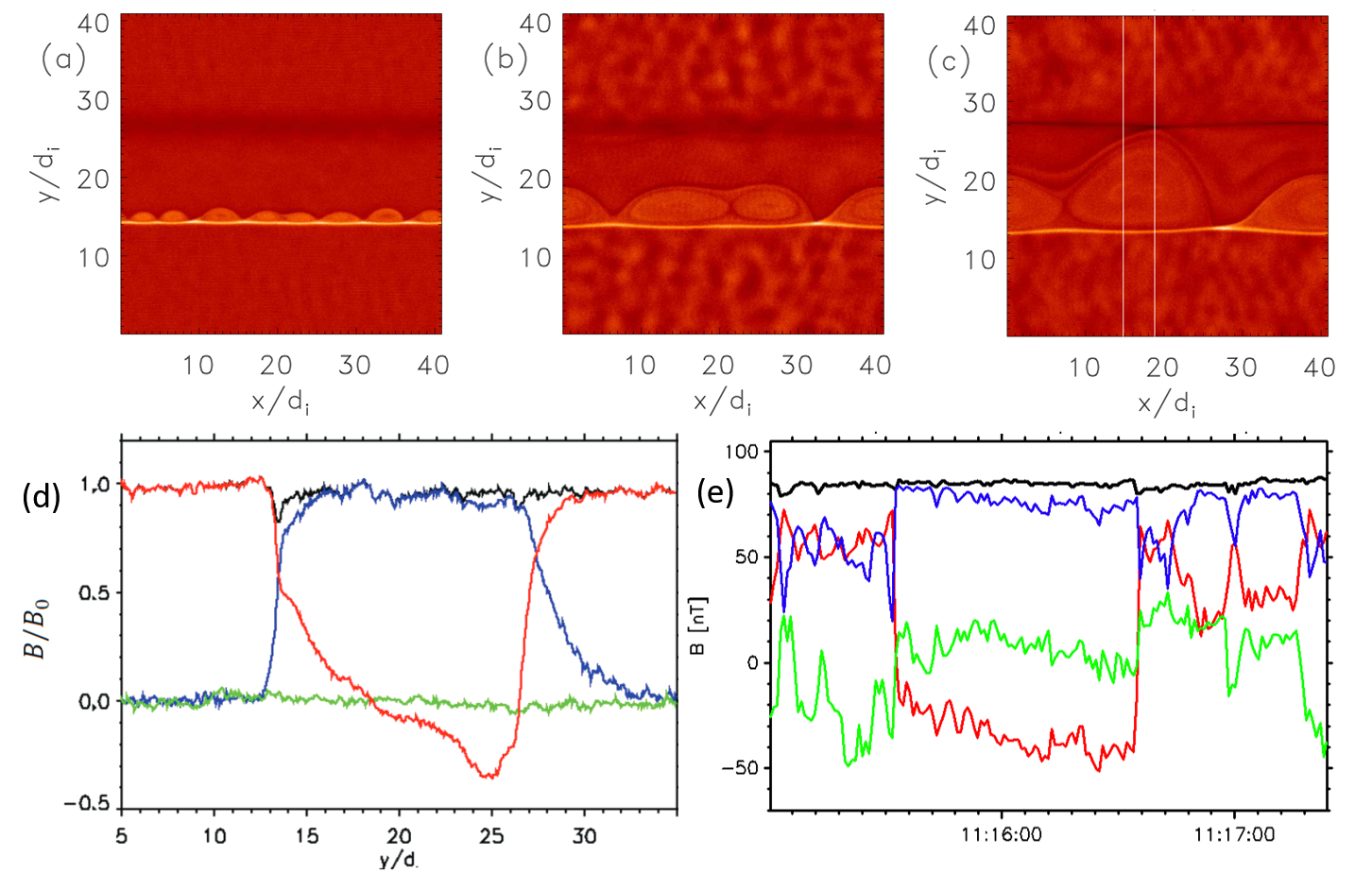}
\centering
    \caption{The formation of the flux rope within the ambient solar-wind magnetic field at various times in the simulations of \cite{Drake2021}: $\Omega_{ci}t = 36$ in (a), 180 in (b), and 292 in (c). Shown is the out-of-plane current $J_{ez}$. In (a) and (e) cuts of the magnetic fields ($B_x$ in red, $B_y$ in green, $B_z$ in blue, and $|B|$ in black) across the large flux rope in panel (c). The cuts are along the white lines in (c). The cut in (a) corresponds to the midplane of the island where $B_y \sim 0$ while that in (e) is offset from the center line where $B_y < 0$. Figure reproduced with permission from \cite{Drake2021}, copyright by ESO}    \label{fig:issi_psp_mering_marc}
\end{figure}

Flux ropes in reconnecting current sheets (within the solar wind or solar corona) typically first form at small spatial scales as current sheets narrow \citep[e.g.,][]{Biskamp1986,Drake06,Bhattacharjee2009}. Small flux ropes then undergo mergers that lead to larger flux ropes. Large current layers can produce a wide distribution of flux rope sizes \citep{fermo2010}. Statistical models of the size distribution of flux ropes suggest that the size distribution of large flux ropes falls off exponentially and there is some observational support for this behavior \citep{fermo2010}. Flux ropes injected into or formed within the solar wind could undergo merging as they propagate away from the Sun. Normally, the magnetic islands that form during merging in a reconnecting current layer become larger, but their aspect ratio (of order unity) does not change since they expand into the region upstream of the current layer as a result of their internal magnetic tension. However, in the case of flux ropes propagating in a unidirectional magnetic field, the merging process should lead to flux rope elongation \citep{agapitov2023}. As revealed in the data, the radial magnetic field within the switchback (the magnetic field that wraps the flux rope in this model) is typically smaller than that of the ambient solar wind. This means that the tension force that tries to make the flux rope round is much weaker than the corresponding backward-acting tension force of the solar wind magnetic field. As a consequence, the merged flux rope becomes significantly longer and only modestly wider in the normal direction \citep{agapitov2022}, consistent with the high aspect ratio of the switchbacks measured in the solar wind \citep{Horbury20}. The merging process also leads to some reduction in the amplitude of the magnetic field wrapping the flux rope while leaving the axial field relatively unchanged. Thus, highly elongated flux ropes with weak wrapping magnetic fields might be a consequence of flux rope merging as the structures propagate outward in the solar wind.

\paragraph{Requirements.} 

The generation of switchbacks from flux ropes requires a population of seed small-scale flux ropes at heliocentric distances below 25-30 $R_\odot$ (outside of that radius, switchback topology is mostly formed). The generation of magnetic flux ropes during reconnection or as a nonlinear stage of shear flow instabilities is a well-established process for doing this in the solar corona and young solar wind. The connection of the fine structure of the coronal magnetic field and the switchback series observed in the magnetically conjugated locations by PSP \citep{Bale2023} supports that small-scale interchange reconnection can seed the conditions favourable for flux rope generation, such as localized plasma jets (Sect.~\ref{sec:interchange}, \ref{sec: KH}, \ref{sec: shear and beams}).

\paragraph{Observable Signatures} 
Expectations from this model of switchback generation due to the merging of small-scale flux ropes include (1) Distribution of the Alfv\'enicity in a range from 0.2--0.3 to 0.8--1 (this accords with switchback observations; \citealp{agapitov2023}); (2) An approach to Alfv\'enic conditions suppresses full merging of the flux rope and leads to a composite structure inside switchbacks, in which they can contain substructures with slightly different parameters (magnetic field rotation angle, plasma density, Alfv\'enicity, plasma temperature) separated by current sheaths (containing plasma density and temperature enhancement) \citep{agapitov2022}; (3) Closed boundaries of switchbacks (presumably tangential discontinuities) \citep{bizien2023} separating plasma with different parameters; (4) Decay of the wrapped magnetic field inside switchbacks with increase in heliocentric distance. 

\paragraph{Advantages \& Limitations} 
The mechanism of switchback generation from a series of small-scale flux ropes {potentially}
explains: the topology of magnetic field perturbations inside switchbacks, the sharp rotation of the magnetic field at the switchback boundary \citep[e.g.,][]{Larosa2021, agapitov2022} with no plasma flow across the boundary \citep{bizien2023}, the enhancement of plasma temperature inside switchbacks \citep{Larosa2021}, and the observed lower Alfv\'enicity \citep{agapitov2023} inside switchbacks relative to perturbations in the background solar wind.

The main limitations are that the flux rope and current layer configurations so far tested are highly idealised. It remains to be shown that the merging dynamics plays out in more realistic solar wind configurations and beyond 2.5 dimensions. For instance, it is now becoming clear that flux ropes are unlikely to be injected into the solar wind directly by interchange reconnection (see Sect.~\ref{sec:interchange}), but Kelvin-Helmholtz (Sect.~\ref{sec: KH}) and beam instabilities (Sect.~\ref{sec: shear and beams}) could provide a plausible route to forming flux ropes in situ. Their subsequent mergers may then form radially elongated switchbacks via this mechanism. However, disentangling which of these mechanisms is more important for switchback formation is likely to be difficult. 

{Furthermore, the double current sheet initial conditions used in \citet{Drake2021,agapitov2022} are not representative of the uni-directional open field of coronal holes (away from the heliospheric current sheet). However, in the later phases of the evolution the current layers become localised to one side of the flux ropes, which may survive longer in practice if the structure moves at the Alfv\'{e}n speed in the solar wind frame and maintains approximate spherical polarisation. Indeed, \citet{Shi24} presented a cylindrically symmetric analytical model for a flux rope topology where the flux rope closed upon itself. No current sheets were required in the solution, and spherical polarisation across the structure meant the flux rope survived against parametric decay for some time. However, again this was a modeler-designed structure rather than one formed dynamically. The main challenge to this mechanism remains in the origin of flux ropes in the first place.}

\section{Conclusions}\label{sec: conclusions}

\subsection{Summary}

This review has been structured to provide a detailed description of a wide range of different mechanisms that have been proposed to explain switchback observations. Of course, it is extremely unlikely that all of these proposed mechanisms are actually involved in the generation of switchbacks in the solar wind, although it is plausible that some combination of them could be playing important roles in switchback formation. In this conclusion, we attempt to provide a broader overview of the mechanisms,
categorizing the different ideas  based on their required `ingredients'.  
In an attempt to provide useful input to the community, we then 
provide a number of specific suggestions for observational studies that could be used to distinguish mechanisms, thereby progressing the field. 

First, it is worth noting that PSP's recent explorations of regions inside the Alfv\'en surface, of heliocentric radius $R_{\rm A}$, have already started yielding constraints. Of particular importance is the observation that switchbacks as we define them in this article (i.e. involving a magnetic deflection beyond $90^\circ$) are less often observed in sub-Alfv\'enic flows \citep{BandyopadhyayEA22sub,PecoraEA22sb,Akhavan-Tafti2024}. The detailed consequences of this observation remain controversial because Alfv\'enic field reversals (switchbacks) naturally have an associated super-Alfv\'enic jet, which will always make the flow locally super-Alfv\'enic even if it is embedded in a background sub-Alfv\'enic stream \citep{Sioulas2024,Badman2025}. Separating this effect  to capture an accurate picture of sub-Alfv\'enic-switchback prevalence is difficult and will require further orbits to build up good statistics, but these initial observations do suggest that switchbacks are lower in amplitude and less numerous in sub-Alfv\'enic regions. This likely rules out ideas that form true switchbacks (i.e., field reversals) in the low corona, requiring instead that they develop as the wind flows outwards. Of course, this does not invalidate the importance of low-coronal physics: there is building observations of statistical correlations between some solar features and switchbacks properties/distributions \citep{Fargette2021,Raouafi2023,Kumar2023,LeeJ24}, which highlights that the low-coronal physics must in some way create the conditions
necessary to develop switchbacks at higher altitudes (e.g., flow shear, Alfv\'enic fluctuations, etc.). But, if the result holds up with future observations, a paucity of switchbacks below the Alfv\'{e}n point will nonetheless be constraining -- indeed, their lower prevalence has likely
 ruled out some previously proposed mechanisms, e.g., via direct injection of switchback field reversals into the wind by interchange reconnection \citep[e.g.][]{2023SSRv..219....8R}.

The emerging picture is thus that processes at the solar surface, as we discuss in Sect.~\ref{sec: ex situ}, lead to conditions that cause switchback reversals to develop around or beyond the Alfvén radius, $R_{\rm A}$. 
Possible mechanisms to create such reversals are discussed in Sect.~\ref{sec: in situ}, each requiring different `ingredients' from lower altitudes in order to operate. These ingredients include:
\begin{description}
    \item[\bf{Alfv\'en waves/fluctuations.}] For the mechanisms of Sect.~\ref{sec: expanding aws} and Sect.~\ref{sec: stream shear}, the key ingredient is some form of Alfv\'enic fluctuation released into the base of open coronal hole flux tubes.
    These propagate upwards and become switchbacks, either due to expansion (Sect.~\ref{sec: expanding aws}) or with the help of velocity shear{/speed gradients} (Sect.~\ref{sec: stream shear}).
    \item[\bf{Velocity shear.}] Several mechanisms, Sect.~\ref{sec: stream shear}, Sect.~\ref{sec: KH}, and Sect.~\ref{sec: shear and beams}, rely on 
    the influence of velocity shear. This can be either small-scale, between neighbouring streams, or on larger, global scales. This shear can then
    cause instabilities (Sects.~\ref{sec: KH} and \ref{sec: shear and beams}) or directly fold up the magnetic field.
    \item[\bf{Compressive fluctuations.}] The mechanism of Sect.~\ref{sec: wkb fast waves} relies on the production of fast waves in the lower corona, which then evolve and grow in amplitude to become switchbacks.
    \item[\bf{Beams.}] In the mechanism of Sect.~\ref{sec: shear and beams}, particle beams play an important role, causing kinetic instabilities that could evolve into switchbacks. 
    \item[\bf{Flux ropes.}] The mechanism of Sect.~\ref{sec: merging} relies on the presence and interaction of small-scale flux ropes within the solar wind.
\end{description}
A full theory then links the generation of one or more of these ingredients from a low coronal source with one of the in situ mechanisms, thereby providing a full pathway to switchback formation. Possible coronal sources that could provide the above ingredients include:
\begin{description}
    \item[\bf{Sources of Alfv\'en waves/fluctuations.}] Convective motions (Sect.~\ref{sec: swirls}), as well as interchange reconnection both directly (Sect.~\ref{sec:interchange}) and indirectly through jets at various scales (Sect.~\ref{sec:untwisting}) launch shear and torsional Alfv\'{e}n waves and more generally Alfv\'{e}nic fluctuations. Open-field reconnection and quasi-2D {turbulence} 
    can further add to and evolve these perturbations as they propagate into the solar wind (Sect.~\ref{sec:turbulence}).    
    \item[\bf{Sources of velocity shear.}] On small scales interchange reconnection either in quasi-continuous low energy bursts (Sect.~\ref{sec:interchange}) or as part of more impulsive small-scale jets (e.g., jetlets ; see Sect.~\ref{sec:untwisting}) could provide fast field-aligned outflows that lead to velocity shear between adjacent solar-wind flux tubes. On global scales, velocity shear arises naturally through differences in expansion and heating between open flux rooted in different regions of coronal holes. 
    \item[\bf{Sources of compressive fluctuations.}] Interchange reconnection (Sect.~\ref{sec:interchange}), jets (Sect.~\ref{sec:untwisting}) and surface motions (Sect.~\ref{sec:turbulence}) could potentially produce fast waves and/or non-linear perturbations with some degree of compressibility.
    \item[\bf{Sources of beams.}] Interchange reconnection again (Sect.~\ref{sec:interchange}), has been put forward as the main source of high energy particle beams (Sect.~\ref{sec: shear and beams}). 
    \item[\bf{Sources of flux ropes.}] Although interchange reconnection was initially put forward as a mechanism for injecting flux ropes into the solar wind, further simulation work has cast doubt on this scenario (Sect.~\ref{sec:interchange}). {If present, flux ropes are likely then a}
    secondary effect, brought on by in situ evolution of instabilities or waves (Sects.~\ref{sec: expanding aws}, \ref{sec: KH}, \ref{sec: shear and beams}).
\end{description}

\begin{figure*}
\centering
\includegraphics[width=\textwidth]{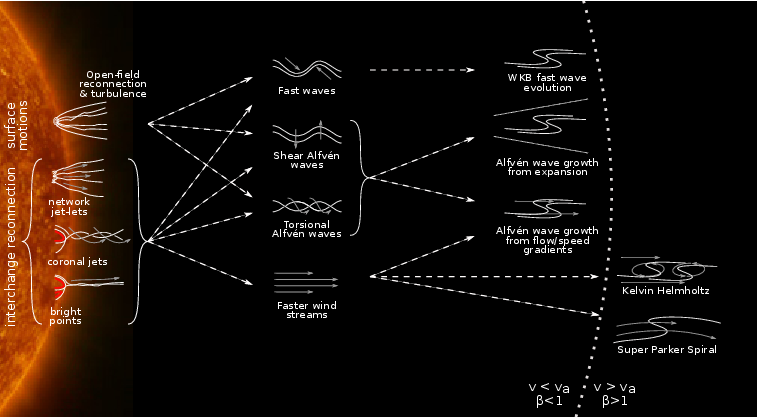}
\caption{Schematic summary of the different mechanisms and phenomena potentially involved in switchback formation and in what cases they feed into each other.} 
\label{fig:con_schem}
\end{figure*}

These ideas are illustrated graphically in Fig. \ref{fig:con_schem}, showing how the different low-coronal processes of Sect.~\ref{sec: ex situ} cause solar wind conditions that evolve into field reversals via the mechanisms of Sect.~\ref{sec: in situ}.

\subsection{An Observational and Theoretical Program}

At this stage, it is fair to say that the community lacks clear answers to many key questions surrounding switchback generation. How can we move beyond this state of affairs? 

One path is theory -- 
models need to be refined and studied further to better understand their predictions. For low-coronal mechanisms, 
there could be important features that feed directly into 
a later in situ process, producing observable consequences at larger altitudes. Examples of such features include the proportion of impulsive versus continuous energy release or latitudinal/longitudinal or helical asymmetries of the fluctuations. 
It is also important to better constrain which of the in situ `ingredients' listed are robustly produced by different
low-coronal processes. For the in situ mechanisms, more theory and simulation work is 
needed to make more detailed predictions of specific switchback distributions, correlations, and features.

The other complementary path to a better understanding
of switchback generation is through observations. With additional statistics from PSP's upcoming 
perihelia, we will start to obtain robust 
statistics of many sub-Alfv\'enic regions for the first 
time, providing an unparalleled opportunity 
to test predictions of the different theoretical ideas laid out above. With this in mind and with the goal of spurring 
further progress in this area, below 
we lay out various possible observational tests that might constrain switchback formation mechanisms. 
We particularly focus on ideas that could rule out or lend support to a particular proposed mechanism rather than those that would support 
a variety of different ideas.
Some of these ideas could be tested with current data while 
some require further statistics from within $R_{\rm A}$.

\begin{description}
    \item[\bf{Connection to the low corona:}] Further statistical and case studies should be conducted to assess the connection between impulsive solar events (e.g., small filament eruptions, jets, jetlets, spicules) and switchback properties and distributions in situ \citep[e.g.][]{Fargette2021,Raouafi2023,Kumar2023,LeeJ24}. This should be augmented by further simulation work assessing how waves, beams and flows launched by such impulsive events propagate into the solar wind \citep[e.g.,][]{Shoda2021,Wyper2022,Bale2023,Touresse2024}. Such studies should also address whether switchbacks are typically due to a particular scale of impulsive event, and if so, what becomes of perturbations launched at larger or smaller scales. \\
    \item[\bf{3D structure:}] By essence, in situ observations only correspond to time sequences of measurements as the spacecraft transits through a switchback structure. The determination of 3D spatial geometry of a switchback is thus convoluted with its time evolution. Analytical and numerical models should aim to provide the 3D structure of switchbacks \citep[e.g.,][]{Shi24}. Fine features and substructures observed within switchbacks (e.g., boundary properties, variations within their core) have not yet been extensively analysed in conjunction with models, nor used as constraints on the latter. This information should, however, be exploited in the future to discriminate between the different formation mechanism scenarios.   
    \\
    \item[\bf{Switchback distribution:}] With lower perihelion passes of PSP, uncertainties in the back mapping of the solar wind to the surface should be further reduced. Future observational studies should look to improve upon the assessment that switchback patches have surface footprints close to supergranular scales \citep{Bale2021} and aim to assess from where, within given supergranules, the peak in switchback activity maps. This will indicate whether the patch structure is the result of a propagation process, e.g., amplifying fluctuations via rapid flux tube expansion above the network in the low corona (Sect.~\ref{sec: expanding aws}), or the result of a spatially and/or temporally localised source region (e.g., network jetlets or coronal interchange reconnection, Sects.~\ref{sec:interchange}, \ref{sec:untwisting}). This should be contrasted with simulation studies of different coronal wave/interchange reconnection scenarios, as it is likely a combination of both scenarios.\\
    \item[\bf{Radial dependence:}] With more perihelion passes of PSP, we will be able to better understand the radial dependence of the switchback prevalence. As discussed above, recent measurements indicate full switchback reversals become less common in the sub-Alfvénic wind (for  $R<R_{\rm A}$), but it remains unclear whether $R_{\rm A}$ represents a sharp transition with fundamentally different properties or a more continuous change over which fluctuations grow. The former would be 
    more consistent with some of the mechanisms invoking shear or beams as an additional ingredient to create instabilities or stretch fields (see Sect.~\ref{sec: stream shear}, Sect.~\ref{sec: KH}, and Sect.~\ref{sec: shear and beams}, and e.g., Figs.~\ref{fig:schwadron2021fig01} and \ref{fig:shear}). The latter would be more consistent with theories related to fluctuation growth (Sect.~\ref{sec: expanding aws} and Sect.~\ref{sec: wkb fast waves}). Theories should aim to produce more detailed predictions of switchback prevalence with radius both in the sub- and super-Alfv\'{e}nic regimes.\\
    \item[\bf{Spherical polarization and Alfv\'enicity:}] Switchbacks are often observed to be highly Alfv\'enic and spherically polarized, but those mechanisms based on beams or shear-flows (Sects.~\ref{sec: stream shear}, \ref{sec: KH}, and \ref{sec: shear and beams}) or flux rope merging (Sect.~\ref{sec: merging}) suggest that this property should be a consequence of post-formation evolution, rather than occurring at their initial formation stage\footnote{At least, it 
    has not yet been shown clearly in simulations that Alfv\'enic, constant-$B$ states arise directly.}. These mechanisms thus predict that, somewhere within or nearby $R_{\rm A}$, we might observe that switchbacks are seeded by less Alfv\'enic structures; i.e., we should observe field reversals with larger $\delta B$ and lower normalized cross helicity. This is at odds with the expanding-Alfv\'en-wave scenario (Sect.~\ref{sec: expanding aws}), which predicts that fluctuations produce switchbacks while remaining highly Alfv\'enic. \\
    \item[\bf{Flow energy:}] As for the previous point, flow-based generation mechanisms predict that energy in flows is converted into magnetic fluctuation energy in order to generate switchbacks. Thus, an interesting test could be to study whether those regions (or particular PSP encounters) with a larger relative large-scale flow shear compared to the magnetic energy also exhibit a larger proportion of large switchbacks. Particularly interesting would be whether a radial dependence were observed in such a correlation. A similar test using the energy of beams would be of interest for assessing aspects of the mechanism of Sect.~\ref{sec: shear and beams}.\\
    \item[\bf{Angular distribution:}] {In this review we have focused mechanisms capable of producing switchbacks with deflections greater than $90^\circ$, but observations show no existence of an obvious cut off in deflection angle, with a distribution of Alfv\'{e}nic, spherically polarized fluctuations present down to small deflection angles \citep[e.g.][]{DudokdeWit2020,Badman2025}. What determines the cut off between switchbacks and low amplitude background turbulent fluctuations, or indeed whether they are significancy different is not yet understood. A continuous distribution of deflections is both consistent with a spectrum of initial deflection angles (below $90^\circ$, Sect. \ref{sec: ex situ}), and the continuous action of in situ mechanisms upon these initial fluctuations (e.g. Sect. \ref{sec: expanding aws}, \ref{sec: wkb fast waves}, \ref{sec: stream shear}, \ref{sec: KH}). So future theoretical and simulations work should aim to give detailed statistical predictions of how and if different combinations of mechanisms lead to different distributions of deflection angles and how they compare against observations. }

\end{description}

It is clear there is still much more work to do to understand the origins of switchbacks and to unpick the different competing and overlapping mechanisms that could be at play in their formation. {The purpose of this review was to provide a broad overview of the many mechanisms proposed by the community. By utilizing a fixed format whereby the strengths and weaknesses of each were discussed, both from a theoretical point of view and from their ability to match to observations, the groundwork has been laid to further rule in or out given mechanisms and to characterize which are the most dominant.} As shown by the recommendations above, this will require continued close collaboration between theorists and observers, and a continued effort to bridge the terminology and understanding gap between the solar and heliospheric communities. However, in this time of new and exciting results from PSP as it orbits ever closer to the Sun, we are at the dawn of this new understanding.

\section*{Compliance with Ethical Standards}
The authors declare no conflicts of interest.

\vspace{\baselineskip}
\noindent
{\large\bf Acknowledgments}
The authors thank the International Space Science Institute (ISSI) for hosting the workshop on ``Magnetic Switchbacks in the Young Solar Wind'' (18-22 September 2023). We thank Judy Karpen and Anna Tenerani for their careful reading of the manuscript. P.W. was supported a Leverhulme Trust Research Project grant RPG-2023-288. E.P. was supported by the JET2SB project of l’Agence Nationale de la Recherche (ANR), project ANR-25-CE31-7416, by the Action Thématique Soleil-Terre (ATST) of the PN ASTRO of CNRS/INSU CNRS/INSU co-funded by CNES and CEA and acknowledge financial support from the French national space agency (CNES) through the APR program. D.R. was supported by the National Science and Technology Development Agency (NSTDA) and National Research Council of Thailand (NRCT): High-Potential Research Team Grant Program (N42A650868) and from the NSRF via the Program Management Unit for Human Resources \& Institutional Development, Research and Innovation (B37G660015). M.M. acknowledges DFG grants WI 3211/8-1 and WI 3211/8-2, project number 452856778. M.M. was also supported by the Brain Pool program funded by the Ministry of Science and ICT through the National Research Foundation of Korea (RS-2024-00408396). M.A.T. was supported by NASA contract Nos. NNN06AA01C, 80NSSC20K1847, 80NSSC20K1014, and 80NSSC21K1662. W.H.M. was partially supported PSP Guest Investigator grant 80NSSC21K1765 and the PSP/ISOIS project. D.T. acknowledges funding from the CEFIPRA Research Project No. 6904-2.

\bibliography{bibliography}

\end{document}